\documentclass[aps, prx, twocolumn, superscriptaddress, nofootinbib, floatfix, english]{revtex4-2}
\usepackage[english]{babel}
\usepackage[normalem]{ulem}
\usepackage{ragged2e}         
\usepackage{amsmath}
\usepackage{amssymb}
\usepackage{graphicx}
\usepackage{physics}
\usepackage{algorithm}
\usepackage{algpseudocode}
\usepackage{float}
\usepackage{braket}
\usepackage[colorlinks=true, allcolors=blue]{hyperref}
\usepackage{xcolor}
\usepackage{comment}
\usepackage{booktabs}
\usepackage{makecell}

\newcommand{\pRBM}{p_{\text{\tiny RBM}}}
\newcommand{\PauliSigma}{{\sigma}}

\usepackage{graphicx}
\usepackage{dcolumn}
\usepackage{bm}
\usepackage{academicons}
\usepackage{multirow}
\newcommand{\orcid}[1]{\href{https://orcid.org/#1}{\textcolor[HTML]{A6CE39}{\aiOrcid}}}

\begin{document}

\preprint{AAPM/123-QED}

\title{Solving Classical and Quantum Spin Glasses with Deep Boltzmann Quantum States}

\author{Luca Leone}
\affiliation{Theoretical Physics III, Center for Electronic Correlations and Magnetism, Institute of Physics, University of Augsburg, Augsburg, Germany.}
\affiliation{Department of Physics, University of Milan, Milan, Italy.}
\author{Arka Dutta}%
\affiliation{Theoretical Physics III, Center for Electronic Correlations and Magnetism, Institute of Physics, University of Augsburg, Augsburg, Germany.}
\author{Markus Heyl}
\affiliation{Theoretical Physics III, Center for Electronic Correlations and Magnetism, Institute of Physics, University of Augsburg, Augsburg, Germany.}
\affiliation{Centre for Advanced Analytics and Predictive Sciences (CAAPS), University of Augsburg, Universitätsstr. 12a, 86159 Augsburg, Germany}

\author{Enrico Prati}
\author{Pietro Torta}
\affiliation{Department of Physics, University of Milan, Milan, Italy.}

\date{\today}
 
\begin{abstract}

Variational neural network models have achieved remarkable success in solving ground-state problems of quantum many-body systems.
However, addressing classical and quantum spin glasses remains challenging, as disorder and energy frustration give rise to an exponentially large number of local energy minima separated by high-energy barriers, hindering the efficiency of conventional Metropolis-based Monte Carlo methods.
To bridge this gap, we introduce Deep Boltzmann Quantum States, a class of neural quantum states inspired by deep Boltzmann machines that inherit efficient block Gibbs sampling. 
We also propose two key advances in the training algorithm. Firstly, we combine natural-gradient updates with state-of-the-art stochastic optimizers.
Secondly, we gradually tune the hardness of the problem Hamiltonian 
by interpolating from an easy to a hard regime, without the need to closely approximate the instantaneous adiabatic state at intermediate times.
We match the exact solution or the best available estimate for several instances of classical and quantum Ising spin-glass models with infinite-range interactions and hundreds of spins.
We also solve instances of the NP-hard Job Shop Scheduling Problem exceeding the current limitations of quantum annealing hardware.
To summarize, deep neural architectures with efficient global update rules and trained within an annealing-like scheme, provide a powerful framework for solving real-world hard combinatorial optimization and for investigating disordered quantum many-body systems.
\end{abstract}
                              
\maketitle

\section{Introduction}

Spin glasses are of central interest across various fields of science and technology.
Beyond their role in modeling disordered magnetic systems, finding classical spin-glass ground states is equivalent to solving many NP-hard combinatorial optimization problems~\cite{PhysRevLett.35.1792,barahona1982computational,lucas2014}.
Quenched disorder and energy frustration give rise to a rugged energy landscape containing an exponential number of local minima, hierarchically organized and separated by high-energy barriers. These minima act as traps for heuristic algorithms. In particular, sampling from a spin-glass low-temperature equilibrium distribution is notoriously hard~\cite{ben2018spectral,Young1998}, posing severe limitations on the performance of Markov chain Monte Carlo (MCMC) methods with local update rules such as standard Simulated Annealing (SA)~\cite{MPV1987}.

In quantum spin glasses, the interplay of quantum effects, quenched disorder, and frustration gives rise to phases and dynamical regimes with no classical counterparts~\cite{Sachdev_2011,PhysRevB.54.3328,cugliandolo2022quantumglassesreview,PhysRevB.105.L020201}, providing a natural setting to study how these phenomena influence phases of quantum matter and non-equilibrium dynamics.
Despite their fundamental interest in quantum many-body theory, quantum spin glasses remain hard to study even in restricted geometries~\cite{oliveira2008complexityquantumspinsystems}.

State-of-the-art algorithms to solve classical spin glasses include integer linear programming methods (e.g.: branch and cut~\cite{doi:10.1137/1033004}) or MCMC methods with global update rules (e.g.: cluster updates~\cite{PhysRevLett.58.86,PhysRevLett.62.361} or parallel tempering~\cite{Marinari_1992,doi:10.1143/JPSJ.65.1604}), and provably exact solutions can usually be obtained for systems up to about 100 spins.
Instances of quantum spin glasses are often solved via exact diagonalization~\cite{Bernaschi_2024}, which is limited to systems of only a few tens of spins. For moderately larger sizes, approximate methods rely on tensor networks~\cite {PhysRevLett.126.090506}, and on variational or quantum Monte Carlo approaches~\cite{10.1007/978-3-642-83154-6_17,viteritti2025quantumspinglasstwodimensional}.
However, their effectiveness critically depends on the problem geometry and the ability to construct a suitable ansatz.

The use of artificial neural networks (ANNs) to solve spin-glass ground-state problems and related combinatorial optimization tasks has been explored in~\cite{bello2016neural,khalil2017learning,gabor2020insights,he2024quantum}. 
%
In recent years, Neural Quantum States (NQS) have emerged as a powerful variational framework for quantum many-body problems~\cite{Carleo_2017}. Combining the expressive power of ANNs~\cite{hopfield1982neural,Goodfellow-et-al-2016}, with the stochastic reconfiguration (SR) algorithm~\cite{sorella1998green,chen2024empowering}, NQS achieved remarkable success in solving ground-state problems of strongly correlated models, without requiring the design of an ansatz tailored for the problem at hand~\cite{PhysRevX.7.021021, PhysRevB.108.054410}. They also show promising results in simulating quantum dynamics~\cite{Carleo_2018, PhysRevLett.125.100503}.
However, state-of-the-art NQS methods typically rely on MCMC sampling with local Metropolis-like update rules, which notoriously struggle to sample from spin-glass probability distributions.

Quantum Annealing (QA)~\cite{Kadowaki_PRE98, Santoro_SCI02, PhysRevB.93.224431} offers an alternative framework rooted in the principles of adiabatic quantum computation (AQC)~\cite{Albash_RMP18}. Here, a quantum system is initialized in the easily prepared ground state of a non-interacting Hamiltonian, which is then slowly transformed into the spin glass Hamiltonian.
However, physical implementations of QA, such as D-Wave quantum annealers~\cite{Johnson2011}, face limitations due to noise, finite temperature effects, limited qubit connectivity, embedding overhead, and finite annealing time~\cite{QA_industry,minor_embedding}.

Quantum-inspired classical algorithms, such as Simulated Quantum Annealing (SQA), have been developed to leverage simulated quantum fluctuations for enhanced sampling of rugged energy landscapes in classically frustrated models~\cite{Martonak_PRB2002, Heim_2015, CrossonHarrow_PRA2016, BaldassiPNAS2018}.
%
In particular, Ref.~\cite{Hibat_Allah_2021} introduces a variational adaptation of (simulated) quantum annealing, termed Variational Quantum Annealing (VQA), and shows that recurrent neural networks (RNNs) can be optimized via VQA to solve classical spin-glass ground-state problems. 

Here, we propose a distinct approach for applying the NQS framework to solve classical and quantum spin glass models.
We first introduce Boltzmann Quantum States (BQS) -- a new variational ansatz inspired by Boltzmann Machines (BMs)~\cite{Goodfellow-et-al-2016,long2010restricted} -- which is of independent theoretical interest and particularly well suited for glassy systems. Here, the physical system is extended by introducing hidden quantum spins, which can be interpreted as unobserved quantum degrees of freedom. We define a Jastrow-like variational wavefunction on the enlarged system, in which the hidden spins do not contribute directly to the energy, but are rather used to encode quantum correlations among the physical spins~\cite{PhysRev.98.1479,becca2017quantum}.

Remarkably, and in contrast to previous quantum extensions of Boltzmann Machines, the construction of efficient Deep Boltzmann Quantum States (DBQS) arises naturally. These variational states inherit the capacity for efficient global updates through block-Gibbs sampling chains~\cite{Goodfellow-et-al-2016}.

To improve the trainability of our model, we propose two algorithmic enhancements. First, we augment the standard SR algorithm with an advanced stochastic optimizer. Second, we adapt the VQA framework to our approach, incorporating key modifications relative to the original formulation~\cite{Hibat_Allah_2021} — in particular, a weakened adiabaticity condition on the variational manifold. We refer to this scheme as Neural Quantum Annealing (NQA), which proves crucial for solving both classical and quantum spin glasses with hundreds of spins.
We perform a hyperparameter optimization (HPO) using Optuna~\cite{optuna_2019} to avoid fixing relevant hyperparameters to arbitrary values.

We train DBQS with NQA and successfully solve instances of the Sherrington–Kirkpatrick (SK) model for $N=100$ and $N=200$ spins. We then move beyond abstract models and address the real-world NP-hard Job Shop Scheduling Problem (JSSP) in its decision version, which admits an Ising reformulation. We solve several instances that exceed the embedding capabilities of the most recent D-Wave Advantage 2 quantum annealer.
For quantum spin glasses, specifically the transverse-field Sherrington-Kirkpatrick (SK) model, we find evidence that our approach can identify ground states of 100-spin instances within the quantum spin-glass phase.
Overall, this work demonstrates that combining a deep neural network ansatz with efficient global update rules and annealing-inspired training algorithms offers a powerful and flexible framework for addressing classical and quantum spin-glass problems, as well as combinatorial optimization tasks, at scales beyond the limits of current quantum annealing hardware and exact classical solvers.

The work is organized as follows.
In Section~\ref{sec: NQS} we outline NQS for spin models, with an emphasis on ground-state search with SR and advanced stochastic optimizers. Section~\ref{sec:QA} reports on the key differences between AQC and practical QA, which are relevant to describe QA-inspired classical algorithms in Section~\ref{sec:VQA}.
We then briefly recall classical BMs in Section~\ref{sec: classical_RBMS}, to set up the notation utilized in Section~\ref{sec:BQS} to introduce Boltzmann Quantum States, our main theoretical advance.
The algorithmic scheme of NQA is described in Section~\ref{sec:NQA}.
Numerical results are presented in Section~\ref{sec: results}.
Finally, Section~\ref{sec: outlook} provides an outlook and highlights future research directions that naturally stem from this work.

A Python implementation of our algorithm, based on JAX~\cite{JAX} and conveniently packaged in an easy-to-deploy NQA server, is freely available on GitHub~\cite {nqa_server_github}.

\section{Preliminaries}
\subsection{Neural Quantum States}
\label{sec: NQS}

Neural Quantum States (NQS) leverage a variational principle, combined with the expressive power of Artificial Neural Networks (ANNs). The wavefunction of a quantum many-body system is represented by a parameterized ANN, by removing the need for an ad-hoc variational ansatz based on physical intuition.
Originally introduced in Ref.~\cite{Carleo_2017}, NQS have been successfully applied to solve a variety of quantum ground-states \cite{PhysRevX.7.021021, PhysRevB.108.054410, chen2024empowering, chen2025convolutionaltransformerwavefunctions} and there are also promising results for the study of quantum dynamics \cite{Carleo_2017, Carleo_2018, PhysRevLett.125.100503, doi:10.1126/sciadv.abl6850,PhysRevX.14.021029}. 
This section provides the background on NQS for spin--$1/2$ systems and Variational Monte Carlo that are necessary for the remainder of this work. 
For a comprehensive review of the topic, see e.g.\ Ref.~\cite{lange2024nqsrev, schmitt2025simulatingdynamicscorrelatedmatter}.

Formally, we consider a system of $N$ spin--$1/2$ degrees of freedom, whose basis states are labeled by spin configurations $\vb x = (x_1, x_2, \dots, x_N)$, with $x_i = \pm 1$. This basis is often conventionally associated with the eigenbasis of Pauli-$Z$ operators~\cite{ft_nqs}, and the general many-body wavefunction can be expanded as
\begin{equation}
\ket{\psi} = \sum_{\vb x} \psi(\vb x) \ket{\vb x}\,.
\end{equation}
In the NQS framework, the wavefunction is expressed as
\begin{equation}
\psi(\vb x;\boldsymbol{\theta})\equiv  \expval{\vb x|\psi(\boldsymbol{\theta})} = f_{\text{ANN}}(\vb x; \boldsymbol{\theta})\,,
\end{equation}
where $f_{\text{ANN}}(\vb x;\boldsymbol{\theta})$ is the output of an ANN that takes $\vb x$ as an input, and $\boldsymbol{\theta}$ as its set of trainable complex or real parameters. 

Leveraging the Time-Dependent Variational Principle (TDVP), it is possible to map the dynamics of the quantum state to a system of ordinary differential equations for the time evolution of the variational parameters \cite{Carleo_2017},
\begin{equation}
\label{eq:TDVP}
i\frac{d}{dt}\ket{\psi(\boldsymbol{\theta})}\!=\!H\ket{\psi(\boldsymbol{\theta})}\ \!\Longleftrightarrow \ \!\frac{d}{dt}{\boldsymbol{\theta}} \!=\! -i S^{-1}\big(\boldsymbol{\theta}\big) F\big(\boldsymbol{\theta}\big)\,.
\end{equation}
Here, $H$ denotes the system Hamiltonian, $S$ is the Fisher information matrix (also known as the quantum geometric tensor), and $F$ is the force array, defined as
\begin{equation}
\label{eq:SR_1}
\left\{
\begin{aligned}
S_{kl}(\boldsymbol{\theta}) &= \expval{O_k \overline{O}_l}_p - \expval{O_k}_p\expval{\overline{O}_l}_p\,,\\
F_k(\boldsymbol{\theta}) &= \expval{E_\text{loc} \overline{O}_k}_p - \expval{E_\text{loc}}_p\expval{\overline{O}_k}_p\,.
\end{aligned}
\right.
\end{equation}
with $O_k$ and $E_\text{loc}$ representing, respectively, the variational derivatives and the local energy
\begin{equation}
\begin{aligned}
\label{eq:SR_2}
O_k(\vb x;\boldsymbol{\theta}) &= \frac{d}{d{\theta}_k}\log\psi(\vb x;\boldsymbol{\theta})\,,\\
E_\text{loc}(\vb x;\boldsymbol{\theta}) &= \frac{\expval{\vb x|H|\psi(\boldsymbol{\theta})}}{\psi(\vb x;\boldsymbol{\theta})}\,.
\end{aligned}
\end{equation}
In Eq.~\eqref{eq:SR_1} we indicated with $\expval{\cdot}_p$ the expectation value over the probability distribution
\begin{equation}
p(\vb x;\boldsymbol{\theta}) = \frac{|\psi(\vb x;\boldsymbol{\theta})|^2}{\langle \psi(\boldsymbol{\theta})|\psi(\boldsymbol{\theta}) \rangle} =
\frac{|\psi(\vb x;\boldsymbol{\theta})|^2}{\sum_{\vb x'}|\psi(\vb x';\boldsymbol{\theta})|^2}\,.
\end{equation}
Since the analytical computation of expectation values on an NQS is generally intractable, practical implementations of the TDVP rely on Monte Carlo sampling to obtain empirical estimates of the quantities in Eq.~\eqref{eq:SR_1}.
More generally, the expectation value of an operator $O$ can be estimated by computing the empirical average over the Born probability of 
\begin{equation}
\label{eq:local_observables}
O_{\text{loc}}(\vb x; \boldsymbol{\theta}) = \frac{\expval{\vb x|O|\psi(\boldsymbol{\theta})}}{\expval{\vb x|\psi(\boldsymbol{\theta})}}\,,
\end{equation}
which is known as the local operator (or local estimator).

Finally, NQS can be employed for ground-state search by performing imaginary-time evolution, which is formally achieved by redefining $\tau = i t$ in Eq.~\eqref{eq:TDVP}. The repeated application of the imaginary-time evolution operator $U=e^{-\delta \tau H}$ for short $\delta\tau$ exponentially suppresses higher-energy components of an initial state, thereby projecting it onto the ground state, provided the initial state has nonzero overlap with it.
Imaginary-time evolution is formally equivalent to minimizing the variational energy 
\begin{equation}
\label{eq:var_energy}
E(\boldsymbol{\theta})= \frac{\expval{\psi(\boldsymbol{\theta})|H|\psi(\boldsymbol{\theta})}}{\expval{\psi(\boldsymbol{\theta})|\psi(\boldsymbol{\theta})}  }
\end{equation}
via Natural Gradient Descent \cite{rattray1998natural, pascanu2014revisitingnaturalgradientdeep}, an algorithm known as Stochastic Reconfiguration (SR) \cite{sorella1998green} (see Appendix~\ref{sec: app_algorithms}). 

The inversion of the Fisher information matrix $S$ is therefore required (as in Eq.~\eqref{eq:TDVP}), and it represents a computational bottleneck. Since $S$ may not be invertible, the Moore-Penrose pseudo-inverse $S^+$ is often used instead~\cite{Carleo_2017,chen2024empowering,ben2006generalized}, and a small diagonal shift can be applied to the Fisher information matrix to stabilize the numerics.
A variant of SR, known as min--SR, has been recently proposed in Ref. \cite{chen2024empowering}, often offering a significant computational speed-up by requiring the inversion of a smaller matrix. Since such two procedures are formally equivalent, the choice between them is guided only by computational convenience.

Solving real-time TDVP dynamics is numerically challenging, mainly due to numerical instabilities arising from multiple sources~\cite {schmitt2025simulatingdynamicscorrelatedmatter}.
A small integration time step $\delta t$ and a large number of Monte Carlo samples are typically required to accurately solve for real-time dynamics.
In contrast, ground-state search is defined as an optimization problem in Eq.~\eqref{eq:var_energy}. When solved with SR, the objective is only to approximate the asymptotic state of the imaginary-time TDVP equations rather than to follow the complete imaginary-time trajectory.
Hence, to solve the ground-state optimization problem, it is not strictly necessary to use standard SR, and natural gradients can be combined with advanced stochastic optimizers developed for deep learning, which can dramatically speed up convergence.
In this work, we combine natural gradients with momentum~\cite{POLYAK19641,rumelhart1986learning,10.5555/3042817.3043064}, as discussed in Section~\ref{sec:NQA}.
To simplify notation, we shall use SR to denote the generalized scheme for NQS optimization in ground-state search, which may combine standard SR (or min-SR) with advanced optimizers, as detailed in Appendix~\ref{sec: app_algorithms}.

\subsection{Adiabatic Quantum Computation and practical Quantum Annealing}
\label{sec:QA}
Quantum Annealing (QA)~\cite{RevModPhys.80.1061,Finnila_CPL94, Kadowaki_PRE98, Santoro_JPA06, AQCfarhiNEW} conceptually relies on the theory of Adiabatic Quantum Computation (AQC)~\cite{Albash_RMP18}.
In this framework, a quantum system is initialized in the easy-to-prepare ground state of a non-interacting \emph{driver} Hamiltonian, often selected as the transverse-field Hamiltonian 
\begin{equation}\label{eq:Hx}
H_{0} = - \sum_{j=1}^N \PauliSigma^x_j \,,
\end{equation}
where $\PauliSigma^\alpha_j$ ($\alpha=x,y,z$) are Pauli operators.
The system is then slowly driven out of equilibrium by tuning an external coupling $s=s(t)$ in an interpolating Hamiltonian
\begin{equation}\label{eq:Hs}
H(s) = s\, H_\mathrm{T} + (1-s)\, H_{0}  \,.
\end{equation}
The objective is to obtain the ground state of the \emph{target} Hamiltonian $H_\mathrm{T}$, which may describe a quantum many-body system or encode the solution(s) of a hard classical computational problem in its lowest-energy state(s).
In an ideal AQC scenario, this is achieved by implementing adiabatic dynamics that, by definition, follow the instantaneous ground-state of $H(s)$, as the coupling $s(t)$ is slowly varied from $s(0) = 0$ to $s(\tau) = 1$ over a sufficiently long total annealing time $\tau$. The system is implicitly assumed to undergo Schr\"odinger evolution, corresponding to a closed quantum system at zero temperature.

When dealing with classical optimization, the target Hamiltonian is often in the form of an Ising spin glass:
\begin{equation}
\label{eq:ising}
H_\mathrm{T} = H_z =  \sum_{i<j}^{N}  J_{ij} \PauliSigma^z_i \PauliSigma^z_j + \sum_{i=1}^{N} h_i \PauliSigma^z_i\,,
\end{equation}
where the non-zero couplings $J_{ij}$ define an interaction graph among the spin variables and the $h_i$ are local longitudinal fields.
This formulation is equivalent to Quadratic Unconstrained Binary Optimization (QUBO) problems and is precisely the type of models that are natively supported on current superconducting QA hardware (e.g., D-Wave systems~\cite{Johnson2011}).
Several critical technological and scientific tasks can be reformulated in terms of QUBO~\cite{glover2019tutorial} or, equivalently, as an Ising spin glass~\cite{lucas2014}, as this represents the prototype of an NP-hard problem. 

The theoretical AQC framework encounters severe practical limitations.
To avoid the transition to excited higher energy states during the dynamics, a large value of $\tau$ is required by the adiabatic theorem(s)~\cite{Albash_RMP18}. Such value critically depends on the minimum spectral gap $\Delta$ encountered during the annealing. The minimum annealing time $\tau$ required for adiabaticity typically scales as $\Delta^{-2}$~\cite{Albash_RMP18}. However, since the annealing path connects a paramagnetic phase in the $x$ direction to a disordered spin glass phase in the $z$ direction, a first-order quantum phase transition is often present in the thermodynamic limit, where the spectral gap closes exponentially fast with the system size $N$~\cite{Amin_2009}. Hence, the value of $\tau$ required for adiabaticity typically diverges exponentially in $N$, representing a major computational bottleneck for AQC. Additional gap closures, such as perturbative crossings, are often encountered during the dynamics~\cite{PhysRevE.84.061152, pnas.1002116107, Knysh2016}.  

While quantum annealers closely implement Eqs.~\eqref{eq:Hs}--\eqref{eq:ising}, an estimate of the order of magnitude of $\tau$ required for adiabaticity is often unfeasible, and practical implementations are also affected by thermal noise, decoherence effects, and hardware constraints~\cite{QA_industry}.
An additional challenge, related to limited hardware connectivity, is the need to embed the interactions in Eq.~\eqref{eq:ising} into the hardware graph~\cite{minor_embedding}.
In practice, current QA hardware does not implement AQC, but is rather employed as a heuristic sampler of the low-energy states of the Hamiltonian in Eq.~\eqref{eq:ising}. Drawing a sample requires system initialization, quantum evolution, and (destructive) measurement on the final state. The whole process is repeated many times (typically thousands of shots).
Hence, rather than a single zero-temperature adiabatic evolution, quantum annealers perform several shorter evolutions with a finite value of $\tau$ related to device constraints, each of which is often modeled as an open quantum system dynamics~\cite{Albash_RMP18}.

Such a heuristic approach shows promising results in several applications, especially when utilized as a subroutine in a hybrid quantum-classical algorithm~\cite {QA_industry, torta2025quantum} or guided through learning procedures~\cite{schulz2025}, yet no conclusive evidence of quantum speedup~\cite{q_speedup} to solve a class of models in the category of Eq.~\eqref{eq:ising} has been shown to date.
Recent progress~\cite{King2025} has been demonstrated with D-Wave hardware in the regime of fast annealing ($\tau$ of the order of a few $\mathrm{ns}$), where coherent Schr\"odinger dynamics can be realized to simulate the physics of quantum spin glasses on hardware-compatible geometries at the boundary or beyond classical simulations (see also Refs~\cite{tindall2025, mauron2025} for improved classical simulations). The latter regime is opposite to the adiabatic theory, and the implications for solving hard \emph{classical} optimization problems as in Eq.~\eqref{eq:ising} remain unclear.

Finally, we observe that the abstract framework of AQC can, in principle, be adapted to prepare the quantum ground state of any many-body target Hamiltonian. A straightforward example is represented by a quantum Ising spin glass, simply obtained as in Eqs.~\eqref{eq:Hx}--\eqref{eq:ising}, by annealing to a final finite value of transverse field coupling, i.e.\ to $s(\tau)<1$. Even more generally, various choices for $H_0$ and $H_T$ can be envisaged, along with different schedules for the time-dependent couplings. Practical realizations of this broad computational scheme, known as \emph{analog quantum computation}, have been implemented on different experimental platforms, such as Rydberg atom arrays~\cite{Ebadi2022QuantumMIS}.

\subsection{Variational Classical Algorithms Inspired by Quantum Annealing}
\label{sec:VQA}

In this work, we borrow concepts and tools from AQC and QA in a classical-simulation setting. A similar approach has inspired a range of classical quantum-inspired algorithms that perform real- or imaginary-time evolution, such as SQA~\cite{Martonak_PRB2002, Heim_2015, CrossonHarrow_PRA2016, BaldassiPNAS2018} and recent results~\cite{Hibat_Allah_2021, Lami_Torta, Torta_Leone}.
Moreover, the theoretical framework of AQC
has recently been recast into a variational setting~\cite{Bojan_PRL}.
The intuitive idea behind these approaches is that a full Hilbert space description of the time evolution may not be necessary to approximate the final ground state.
In particular, when the final Hamiltonian is classical and with a glassy energy landscape, as is typical in the framework of Eq.~\eqref{eq:ising}, obtaining a peaked wavefunction overlapping only with classical lowest-energy configurations may be both unfeasible and unnecessary. In fact, it would be enough to draw a sufficient number of samples from a wavefunction with a finite overlap with one or more of those classical solutions.
In this heuristic setting, there is no evidence that a full Hilbert space description provides better results than a constrained evolution on a classically simulable variational manifold (see also Fig.~1 in Ref~\cite{Bojan_PRL}).
In contrast, an evolution on a restricted manifold may even improve results in some cases~\cite{Lami_Torta}. 

In principle, variational simulations of QA can be achieved by solving real-time TDVP Eqs.~\eqref{eq:TDVP} for a large annealing time $\tau$, as theoretically characterized ~\cite{Bojan_PRL} and numerically tested within a Matrix Product State~\cite{SCHOLLWOCK201196} manifold. 
However, the numerical integration of TDVP equations is computationally demanding for ANNs and would require small time steps, making it challenging for large system sizes and long annealing times $\tau$.

Variational Quantum Annealing (VQA)~\cite{Hibat_Allah_2021} is a different strategy that relies on iteratively \emph{optimizing} the variational ansatz to approximate the instantaneous adiabatic ground-state on the variational manifold. 
VQA leverages the variational principle to avoid solving real-time TDVP dynamics and may have the potential to mitigate the diverging runtime due to vanishing spectral gaps.
Rather than physically evolving a quantum state, one uses a parametrized wavefunction and Variational Monte Carlo to approximately track the ground-state of $H(s)$ in Eq.~\eqref{eq:Hs} at a series of $N_A+1$ discretized points $0=s_0 < s_1 < \cdots < s_{\tiny{N_A}}=1$ along the annealing schedule.
The algorithm realizes a discrete adiabatic path on the variational manifold by iteratively adjusting the ansatz parameters to minimize the variational energy
\begin{equation}
\label{eq:var_energy_VQA}
E(s_j;\boldsymbol{\theta}) = \frac{\expval{\Psi(\boldsymbol{\theta})|H(s_j)|\Psi(\boldsymbol{\theta})}}{\expval{\Psi(\boldsymbol{\theta})|\Psi(\boldsymbol{\theta})}}\,,
\end{equation}
with $j=0,\dots,N_A$.
Optimizing the ansatz parameters with standard SR would be equivalent to performing repeated imaginary-time evolutions for a set of parametric Hamiltonians $H(s_j)$.

VQA was originally presented in Ref.~\cite{Hibat_Allah_2021}, where the authors devised both a variational version of classical Simulated Annealing and of Quantum Annealing. Both algorithms relied on Recurrent Neural Networks (RNNs) and were tested against standard techniques on several spin glass models.
In this setting, VQA with RNNs appeared less effective than its classical counterpart, and VQA results were only shown for polynomial complexity problems, such as the random Ising chain and the 2D Edwards–Anderson model without longitudinal field.

Thanks to its flexibility, the VQA framework can naturally be viewed as an alternative approach to optimizing any NQS for ground-state search. This is useful when a direct optimization with SR on the target Hamiltonian $H_\mathrm{T}$ is expected to fail due to frustration (as in Eq.~\eqref{eq:ising}), disorder, or other effects. In essence, the hardness of the ground-state problem is tuned from \emph{easy} to \emph{hard} while increasing the parameter $s_j$, and the NQS is warm-started with the result of the previous iteration (see Appendix~\ref{sec: app_algorithms} for details). 

\subsection{Restricted and Deep Boltzmann machines}
\label{sec: classical_RBMS}

In this Section, we briefly revise classical Boltzmann machines (BMs) to establish the notation to introduce
Boltzmann Quantum States.
BMs can be defined as undirected graphical models originally introduced as a general approach to learning probability distributions over binary variables~\cite{Goodfellow-et-al-2016}. 
BMs are energy-based models over a $d$-dimensional binary random vector $\vb x=\{-1,+1\}^d$  with a quadratic energy function~\cite{ft_rbm_1}
\begin{equation}
E(\vb x; \boldsymbol{\theta})=- \vb x^T W \vb x - \vb b^T\, \vb x \,,
\end{equation}
where $W$ and $\vb b$ are, respectively, the real-valued weight matrix and bias vector that we collectively denote with $\boldsymbol{\theta} = (W, \vb b)$, and the corresponding probability distribution is given by~\cite{ft_rbm_2}
\begin{equation}
p(\vb  x; \boldsymbol{\theta}) = Z^{-1} \exp\{\vb x^TW \vb x+\vb b^T\,\vb x\}\,,
\end{equation}
where
\begin{equation}
Z = \sum_{\vb x'} \exp{-E(\vb x'; \boldsymbol{\theta})}\,
\end{equation}
is the partition function.
This class of models becomes particularly useful when hidden (unobserved) variables are included to model higher-order interactions among the visible (observed) variables.
With the inclusion of hidden variables, BMs become universal approximators of probability functions over discrete variables \cite{10.1162/neco.2008.04-07-510} and have been successfully used to solve a variety of machine learning problems \cite{multimodal_bm,HJELM2014245,HE2021101871}. For a more detailed discussion on standard BMs, we recommend Ref.~\cite{Goodfellow-et-al-2016}.

Since BMs are notoriously hard to train and their probability distribution is hard to sample, a more practical and widely adopted alternative is the Restricted Boltzmann Machine (RBM), 
characterized by a single layer of visible variables and a single layer of hidden variables. 
No intra-layer interactions are included, resulting in a bipartite interaction graph in which all interactions occur between visible and hidden units.
Formally, the visible layer consists of $N_v$ visible variables that we collectively denote with $\vb v$, and the hidden layer consists of $N_h$ hidden variables that we denote with $\vb h$.
The energy function of the RBM is given by
\begin{equation}
\label{eq:RBMS}
E(\vb v,\vb h;\boldsymbol{\theta}) = - \vb v^T W \vb h - \vb b^T \vb v - \vb c^T \vb h    \,,
\end{equation}
with trainable parameters $\boldsymbol{\theta} = (W, \vb b, \vb c)$.

The RBM architecture can be generalized by introducing several layers of hidden variables, a model known as Deep Boltzmann Machine (DBM).
In contrast to the RBM energy function, the hidden variables in a DBM interact with each other, albeit in a restricted manner, with only adjacent inter-layer connections and no intra-layer connections.
By denoting the units in the $ l$-th layer with $\vb x^{(l)}$, where layer $0$ corresponds to the visible units and all hidden layers are indexed for $l>0$, the energy function of a DBM with $N_L$ layers reads
\begin{equation}
\label{eq:DBMS}
E(\vb x;\boldsymbol{\theta})\!=\!-\!\sum_{l=0}^{N_L-2} \!\vb x^{(l)T}W^{(l)}\vb x^{(l+1)} \!-\! \sum_{l=0}^{N_L-1}\!  \vb b^{(l)T}\vb x^{(l)}\,\!.
\end{equation}

Efficient training of RBMs and DBMs relies on block Gibbs sampling, enabled by the conditional independence of subsets of variables arising from the bipartite graph structure. This is reviewed in Appendix~\ref{sec: app_BMs}.

\section{Boltzmann Quantum States}
\label{sec:BQS}
In this work, we introduce a novel framework for extending BMs to quantum variables, designed to represent quantum many-body states. We refer to this formulation as Boltzmann Quantum States (BQS). This class of states has a clear physical interpretation: both hidden and visible variables are promoted to quantum spins within an extended Hilbert space. 

Several prior extensions of Boltzmann Machines -- particularly RBMs -- to a quantum setting have been proposed, as briefly reviewed in Appendix~\ref{sec: app_quantum_BMs}.
Our goal is to demonstrate that our approach has two key advantages compared to previous formulations.
Firstly, BQS achieve a substantially higher expressive power, thanks to a natural extension from Restricted Boltzmann Quantum States (RBQS) to deep architectures (DBQS).
Secondly, these new models naturally inherit block Gibbs sampling, allowing efficient Monte Carlo estimation even in their deep formulations, with orders-of-magnitude speedups in the sampling phase.
In addition to these two key advantages, both RBQS and DBQS exhibit a further distinguishing feature: the calculation of the quantum geometric tensor and any local estimator is greatly simplified.
In fact, while the former reduces to the computation of few-point correlators, the latter involves only simple matrix multiplications, providing an additional computational speed-up.

\subsection{General formulation}
\label{sec:general_bqs_formulation}
The definition of BQS is based on the idea of introducing unobserved (or hidden) \emph{quantum} spins and using a Jastrow wavefunction to describe the state of the extended system. Physical observables act non-trivially only on the Hilbert subspace related to physical (or visible) quantum spins. 
Formally, given a system of $N_v$ visible spin--$1/2$ degrees of freedom with a Hamiltonian $H_v$, a Hilbert space $\mathcal{H}_v$, and computational basis $\{\ket {\vb v}\}$, we introduce another system of $N_h$ hidden spin--$1/2$ degrees of freedom, Hilbert space $\mathcal{H}_h$, and computational basis $\{\ket{\vb h}\}$. We obtain an extended system with Hilbert space $\mathcal{H}_{vh}=\mathcal{H}_{v}\otimes\mathcal{H}_{h}$ and computational basis $\{\ket{\vb x}\} = \{\ket{\vb v,\vb h}\}$ on which we consider the Hamiltonian $H_{vh}=H_v\otimes\mathbb{I}_h$.
As standard practice in NQS applications, we choose the Pauli-Z basis representation denoted as $\ket{\vb x}=\ket{x_1,...,x_{N_u}}$, with $N_u=N_v+N_h$ and $x_i\in\{-1;+1\}$ corresponding to spin down and spin up, respectively.
Finally, a Jastrow variational wavefunction describes the state of the complete system, effectively coupling visible and hidden quantum spins:
\begin{equation}
\label{eq:BQS}
\log{\expval{\vb x|\Psi(\boldsymbol{\theta})}}=\sum_{ij} W_{ij}x_ix_j + \sum_i b_i x_i\,,
\end{equation}
with $\boldsymbol{\theta} = (W, \vb b)$, where $W\in\mathbb{C}^{N_u\times N_u}$ and $\vb b\in\mathbb{C}^{N_u}$ are complex variational parameters.
We dub this family of ansatz states as Boltzmann Quantum States (BQS). Note that this wavefunction is not normalized, as commonly done in NQS applications. 

The probability function defined by a BQS is easily derived as
\begin{equation}
\label{eq:BQS_prob}
\log[{p(\vb x;\boldsymbol{\theta})}]\!=\!\sum_{ij}\! 2\Re[W_{ij}]x_ix_j \!+\! \sum_i 2\Re[b_i] x_i\,\!,
\end{equation}
which is the same probability function of a standard Boltzmann machine, with \emph{real} parameters $\boldsymbol{\theta}=(2\Re(W),2\Re(\vb b))$.
%
However, the wavefunction in Eq.~\eqref{eq:BQS} is needed to approximate the ground-state or the dynamics of a quantum system as outlined in Section~\ref{sec: NQS}, and the full set of \emph{complex} variational parameters $\boldsymbol{\theta}=(W,\vb b)$ is optimized to capture quantum correlations and entanglement.
Hence, BQS are more general-purpose machines than standard BMs.

Compared to other architectures, such as Convolutional Neural Networks, BQS are particularly suitable for models defined on a graph without a clear 1D or 2D geometric interpretation, or for models that lack translational invariance, such as spin glasses or disordered models.

By definition of the general BQS variational ansatz in Eq.~\eqref{eq:BQS}, visible and hidden quantum spins are entangled. As a consequence, it is not generally possible to marginalize the hidden spins to obtain a pure state for the visible ones. Nonetheless, the ground-state expectation value of an observable acting on the physical Hilbert space $\mathcal{H}_v$ can be computed on the ground-state of the extended system $\mathcal{H}_{vh}$ where the BQS is defined. This is discussed in Appendix~\ref{sec: app_analytical_res_BQS}, where we also show that restricted and deep BQS architectures reduce to previous proposals of quantum BMs~\cite{Carleo_2017, Carleo_2018} by projecting the hidden units on a specific product state.

\subsection{Deep Boltzmann Quantum States}
\label{sec:RDBQ}
Generic BQS suffer from the same intractability as Boltzmann machines.
Nonetheless, it is straightforward to implement restrictions on the terms in the Jastrow wavefunction in Eq.~\eqref{eq:BQS}. This allows us to define restricted and deep Boltzmann quantum states, both of which enable efficient block Gibbs sampling and other useful properties.

As in the standard RBM, one can restrict the non-zero coefficients $W_{ij}$ to connect only visible and hidden spins organized in two separate layers, leading to the variational wavefunction for the Restricted Boltzmann Quantum State (RBQS)
\begin{equation}
\label{eq:RBQS}
\log\expval{\vb v, \vb h|\Psi(\boldsymbol{\theta})}=\vb v^T W \vb h + \vb b^T \vb v + \vb c^T \vb h\,.
\end{equation}
The same notation as in Eq.~\eqref{eq:RBMS} applies here, where $(\vb v, \vb h)$ now label the computational basis states for the visible and hidden quantum spins.
More generally, in close analogy to the case of DBMs, the spins can be arranged into layers $\vb{x}^{(l)}$, with the zero-th layer containing the visible spins, and all the successive layers including hidden quantum spins. By design, non-zero coefficients $W_{ij}$ only connect spins in adjacent layers, and no intra-layer terms are present. Hence, the resulting ansatz for the Deep Boltzmann Quantum State (DBQS) with $N_L$ layers reads:
\begin{equation}
\label{eq:DBQS}
\log\!\expval{\vb x|\Psi(\boldsymbol{\theta})} \!=\!\!\! \sum_{l=0}^{N_L-2}\!\!\vb x^{(l)T}W^{(l)} \vb x^{(l+1)} \!+\!\!\sum_{l=0}^{N_L-1} \vb b^{(l)T} \vb x^{(l)}\,\!.
\end{equation} 
Note that the DBQS ansatz includes the RBQS as a special case. This is the class of variational wavefunction utilized in our work.

Due to Eq.~\eqref{eq:BQS_prob}, the probability distribution for the RBQS and the DBQS is the same as a canonical RBM and DBM, respectively. Hence, it factorizes conditionally on alternate layers, enabling the same efficient block Gibbs sampling routinely used by the machine learning community. A single Markov chain step updates all spins simultaneously, and numerical experiments show that autocorrelation times remain $\mathcal{O}(1)$ as the system size increases, in stark contrast to the superlinear scaling observed with local Metropolis moves. This fact is shown in Appendix~\ref{sec: app_analytical_res_BQS}.
Consequently, a speedup of at least an order of magnitude is achieved during both training and inference, allowing us to benchmark our methods for system sizes up to hundreds of spins. 

\subsection{Computational advantages}

We now outline the other key conceptual innovations and computational advantages enabled by our formulation.

\paragraph{Analytical variational derivatives}
In a generic BQS formulation, the local derivatives $O_k$ required to estimate the Fisher information matrix (see Eqs.~\eqref{eq:SR_1}--\eqref{eq:SR_2}) reduce to simple products of spin variables and admit an analytical expression
\begin{equation}
\begin{aligned}
&\frac{\partial}{\partial W_{ij}}\log \expval{\vb x | \Psi(\boldsymbol{\theta})} = x_i x_j\,,\\
&\frac{\partial}{\partial B_{i}}\log \expval{\vb x | \Psi(\boldsymbol{\theta})} = x_i\,.
\end{aligned}
\end{equation}
These explicit formulas represent a striking difference compared to NQS implementations based on feed-forward neural networks, which require backpropagation to compute gradients, introducing additional computational complexity and memory overhead. Our approach avoids backpropagation, offering both conceptual simplicity and improved numerical stability.
We note that the empirical Fisher information matrix can be efficiently constructed by computing connected correlation functions through sampling of the BQS wavefunction. For instance,  the term of the $S$ matrix associated with the parameters $W_{ij}$ and $W_{lm}$ reads
\begin{equation}
S_{ij,lm}
= \expval{x_i x_j x_l x_m} - \expval{x_i x_j}\expval{x_l x_m}\,,
\end{equation}
and similar equations can be easily derived for the other terms.

\paragraph{Simple evaluation of local operators}
Evaluating the expectation value of an operator $O$ via Monte Carlo requires computing the local estimator defined in Eq.~\eqref{eq:local_observables}. We show that for DBQS, this reduces to simple matrix multiplications.
In the context of spin models, operators can always be decomposed as a linear combination of Pauli strings, and the same holds for their local operators by linearity.
For a system composed of $N$ spins, a generic Pauli string operator is written as
\begin{equation}
P = \PauliSigma_1^{\alpha_1} \otimes \PauliSigma_2^{\alpha_2} \otimes \cdots \otimes \PauliSigma_N^{\alpha_N}\,,
\end{equation}
where each $\alpha_i \in \{0, x, y, z\}$, and $\PauliSigma_i^{\alpha_i}$ denotes either a Pauli operator acting on spin $i$, or the identity operator $\PauliSigma^0 = I$.
The local estimator of a Pauli string has the general form of
\begin{equation}
\label{eq:Pauli_loc_estimator}
P_\text{loc}(\vb x) = f(\vb x)\frac{\psi(\tilde{\vb x} ;\boldsymbol{\theta})}{\psi(\vb x;\boldsymbol{\theta})} 
\end{equation}
where $f(\vb x)$ is some function of the spin configuration $\vb x$ and $\tilde{\vb x}$ is the configuration connected to $\vb x$ by the Pauli string, i.e. $\ket{\tilde{\vb x}} \propto P\, \ket{\vb x}$. 

In NQS applications based on a feed-forward ANN, computing $P_\text{loc}$ requires one full forward pass (network evaluation)
for both configurations $\vb x$ and $\tilde{\vb x}$.
For a BQS, the situation is radically different: inserting the explicit ansatz of Eq.~\eqref{eq:BQS} into the ratio in Eq.~\eqref{eq:Pauli_loc_estimator} yields a simple exponential of differences of quadratic and linear forms
\begin{equation}
\frac{\Psi(\tilde{\vb x};\boldsymbol{\theta})}{\Psi(\vb x; \boldsymbol{\theta})} = 
e^{\sum_{ij}\! W_{ij}(\tilde x_i\tilde x_j\!-\!x_ix_j) +\! \sum_i\! B_i (\tilde x_i\!-\!x_i)}\,.
\end{equation}
In the specific case of a DBQS, when considering operators that act only on the visible spins and inserting Eq.~\eqref{eq:DBQS}, this expression simplifies even further 
\begin{equation}
\frac{\Psi(\tilde{\vb x};\boldsymbol{\theta})}{\Psi(\vb x;\boldsymbol{\theta})}\!=\!e^{\sum_i \left[(\tilde x_{i}^{(0)}-x_{i}^{(0)})\left(b_{i}^{(0)}+\sum_jW_{{i}j}^{(0)}x_j^{(1)}\right)\right]}\,,
\end{equation}
and can be computed using simple tensor operations that can be easily vectorized, further reducing the time complexity of the computation. 
While the evaluation of local observables involving only the visible spins depends solely on the state of the visible units (layer $l=0$) and the first hidden layer ($l=1$), the full set of hidden units remains necessary for sample generation via Gibbs sampling.

To better illustrate this idea, let us consider the case of the local energy for a spin chain of $N$ sites and the Hamiltonian operator $H=\sum_{i=0}^{N-1}\PauliSigma_i^x$. This operator also represents the only non-diagonal contribution for a standard QA Hamiltonian, as shown below in Eq.~\eqref{eq:Hs}.
In this case, the local energy of a given configuration, as defined in Eq.~\eqref{eq:SR_2}, reads
\begin{equation}
E_\text{loc}(\vb x;\boldsymbol{\theta})=\sum_{i=0}^{N-1} \frac{\psi(T_i(\vb x); \boldsymbol{\theta})}{\psi(\vb x; \boldsymbol{\theta})}\,
\end{equation}
where
\begin{equation}
T_i(\vb x) = (x_0, \dots , x_{i-1}, -x_i, x_{i+1}, \dots , x_{N-1})\,.
\end{equation}
Hence, in a NQS implementation based on a feed-forward neural network, one needs to compute the output of the network for all possible single spin flips, requiring a total of $N+1$ evaluations of the network forward function.
On the other hand, in the case of a DBQS, the local energy is simply given by
\begin{equation}
E_\text{loc}(\vb x;\boldsymbol{\theta})=\sum_{i=0}^{N-1}e^{-2x_i^{(0)}\left(b_i^{(0)}+\sum_jW_{ij}^{(0)}x_j^{(1)}\right)}\,.
\end{equation}
Therefore, all the quantities needed to compute the local energy can be evaluated with a single matrix multiplication. 

\section{Neural Quantum Annealing} 
\label{sec:NQA}
\begin{figure*}[th]
\centering
\includegraphics[width=.9\linewidth]{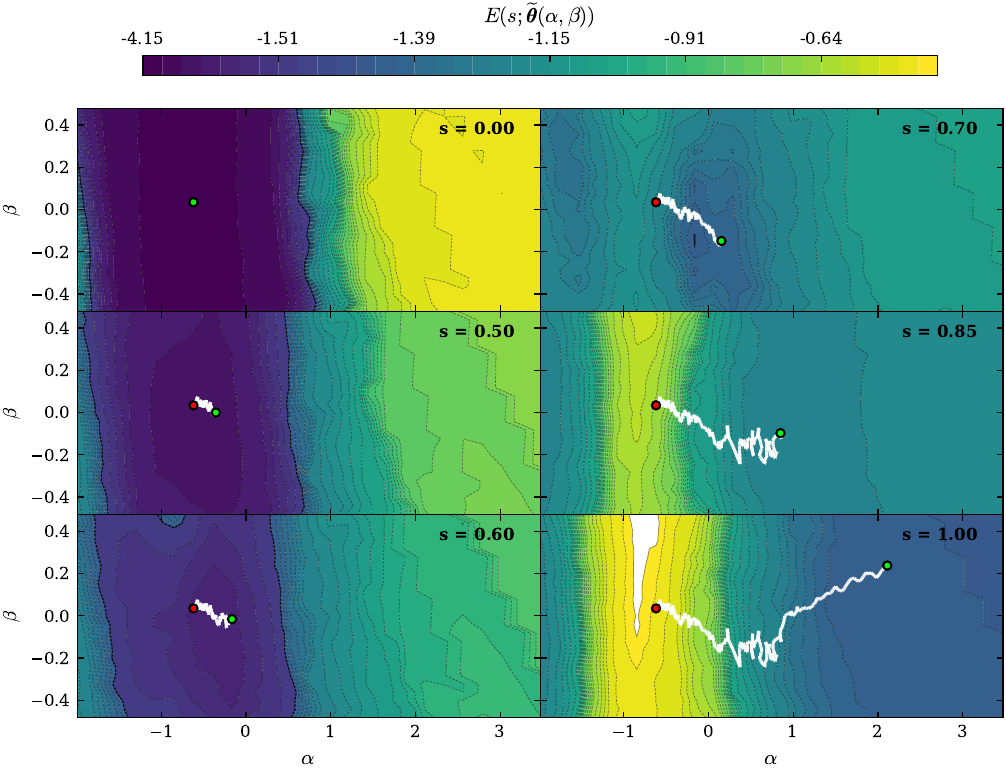}
\caption{
Visualization of the NQA trajectory for a 4-site Ising toy model with all-to-all random couplings. Each subplot shows the projected parameter trajectory  (white line, with red and green dots marking the initial and current parameters) together with the instantaneous energy landscape at the corresponding $s_j$, projected on a plane $\Sigma$ (defined below).
Throughout the variational optimization, the ansatz remains close to the minimum of the instantaneous energy, effectively guiding NQA to the global minimum of the target problem.
Technical details of the plot, following Ref.~\cite{lorch2016visualizing}, are as follows. At each schedule point $s_j$, we record the instantaneous parameters and construct a real-valued data matrix by splitting each complex parameter into its real and imaginary parts and stacking them as separate features, i.e., $X_{j,(k,\mu)}\in\mathbb{R}$, with $k$ spanning the $\boldsymbol{\theta}$ components and $\mu\in\{\mathrm{Re},\mathrm{Im}\}$.
We then project the dynamics onto the affine PCA plane $\Sigma=\{\widetilde{\boldsymbol\theta}(\alpha,\beta)=\bar{\boldsymbol\theta} + \alpha\vb d_1 + \beta \vb d_2\}$ spanned by the first two principal directions $\vb d_{1,2}$ of $X$~\cite{jolliffe2011principal}; $(\alpha,\beta)$ are the in-plane coordinates and $\bar{\boldsymbol\theta}$ is the schedule mean.
}
\label{fig:nqa_visual}
\end{figure*}

Neural Quantum Annealing (NQA) merges the high-level framework of VQA (see Section~\ref{sec:VQA}) with the expressive capabilities and sampling efficiency of DBQS (see Section~\ref{sec:BQS}), iteratively optimized with natural gradients and advanced stochastic optimizers. An illustrative representation of the application of the algorithm is shown in Fig.~\ref{fig:nqa_visual}, where the typical NQA optimization trajectory can be visualized on a reduced-dimensional space for a toy-size classical spin model. 
A preliminary version of our work using a standard cRBM wavefunction (see Ref.~\cite{Carleo_2017} and Appendix~\ref{sec: app_quantum_BMs}) was presented in Ref.~\cite{Torta_Leone}.
In the following, we describe the algorithm, highlighting its new features and the underlying theoretical motivations that distinguish our approach. 
A detailed pseudo-code and additional information on hyper-parameters are reported in Appendix~\ref{sec: app_NQA}. Central to our method is exploiting the scalability and trainability of DBQS, which are initialized in a uniform superposition for hidden and visible spins (see Appendix~\ref{sec: app_analytical_res_BQS} for technical details).

\subsection{DBQS ansatz}
The primary challenge in sampling low-energy configurations of spin glasses lies in their highly rugged energy landscape, which results from energy frustration and competing interactions. It leads to a multitude of local minima that can trap heuristic optimization methods. In particular, Markov Chain Monte Carlo (MCMC) algorithms relying on local update rules are effectively non-ergodic in the spin-glass phase and exhibit exponentially slow mixing due to large free-energy barriers~\cite{mezard1987spin}, necessitating more refined schemes~\cite{Hukushima_1996}.
Since most modern NQS architectures rely on Metropolis-based MCMC (local update rules), they are expected to perform poorly in spin-glass sampling.

In the following, we employ DBQS as a variational ansatz. Unlike other NQS architectures, DBQS enable efficient block Gibbs sampling, which, thanks to its global update rule, is better suited for handling multimodal distributions in glassy energy landscapes.
Moreover, DBQS can encode long-range, non-equivariant correlations on random graphs, providing both the expressivity and the sampling capabilities required for the problem.
This property makes our algorithm viable for large-scale, complex spin-glass problems in which sampling is the primary computational bottleneck.
An alternative approach was proposed in Ref.~\cite{Hibat_Allah_2021}, which developed VQA with recurrent neural network (RNN) wavefunctions with autoregressive sampling. 
Although RNNs yield exact, uncorrelated samples, their ultimate capacity to represent the quantum many-body wavefunction during annealing is not fully settled yet, especially for models on a random graph that lacks any 1D or 2D geometry, as the network must learn non-equivariant, long-range correlations.
In practice, RNNs were found to be more conveniently optimized by thermal annealing~\cite{Hibat_Allah_2021}, at variance with previous studies showing the enhanced effectiveness of \emph{simulated} quantum fluctuations over \emph{simulated} thermal fluctuations~\cite{Kadowaki_PRE98, Santoro_SCI02, Heim_2015, PhysRevB.93.224431, BaldassiPNAS2018} in exploring rugged energy landscapes of classical frustrated models.
Demonstrated results of VQA were restricted to polynomial-complexity cases, such as the random Ising chain and the two-dimensional Edwards-Anderson model without a longitudinal field~\cite{Hibat_Allah_2021}, and recent studies on real-world problems have only focused on the thermal version of the algorithm~\cite{ranabhat2025}.

\subsection{Persistent parallel Gibbs chains}

Building on the sampling capability unlocked by the DBQS ansatz, we run an ensemble of persistent Markov chains throughout the entire variational optimization.
While a straightforward approach would be to generate and thermalize a new set of Markov chains at each annealing step $j=0,\dots, N_A$ in Eq.~\eqref{eq:var_energy_VQA}, this would be a prohibitively slow procedure for high‑dimensional glassy distributions.

Since successive Hamiltonians $H(s_j)$ and $H(s_{j+1})$ change only slightly, it is expected that thermalized chains at step $j$ remain near equilibrium for the new distribution. Consequently, instead of fully re-thermalizing them, one can resume the chains from their last states and apply a few block Gibbs updates to adapt to the updated wavefunction parameters (see Appendix~\ref{sec: app_NQA} for details).

\subsection{SR augmented with momentum}

As anticipated in Section~\ref{sec: NQS}, natural gradients can be combined with advanced stochastic optimizers, enhancing the standard SR algorithm for ground-state search. Rather than a straightforward optimization of the variational energy in Eq.~\eqref{eq:var_energy_VQA}, NQA combines natural gradients with a momentum term to perform the parameter updates. When using momentum~\cite{POLYAK19641,rumelhart1986learning,10.5555/3042817.3043064}, a velocity vector stores an exponentially decaying average of past gradients.
Let us focus on a single annealing step $j$: for a fixed value of \(s_j\), we perform a sequence of $n = 1, \dots, N_{U}$ optimization epochs to iteratively minimize the variational energy in Eq.~\eqref{eq:var_energy_VQA}.
Concretely, at epoch $n$ the updates are
\begin{equation}
\label{eq:SR_momentum}
\boldsymbol{\nu}_{n+1} = \mathbf g_{n}+\mu \, \boldsymbol{\nu}_{n}\,, \qquad 
\boldsymbol{\theta}_{n+1} = \boldsymbol{\theta}_{n} - \eta\,  \boldsymbol{\nu}_{n+1},
\end{equation}
where $\mathbf g_{n}$ is the natural gradient, $\boldsymbol{\nu}$ is the velocity vector, $\eta$ is the learning rate, and $0\le\mu<1$ is the momentum coefficient.  
The velocity term damps oscillations in directions where the gradient sign alternates while accelerating progress along directions with a persistent sign, often yielding faster and more stable convergence than standard gradient descent~\cite{goh2017why}. Crucially, both the optimized variational parameters and the final velocity obtained at annealing step $j$ are carried over to warm start the DBQS ansatz and the optimizer in the next annealing step $j+1$.

\subsection{Relaxed adiabaticity on a manifold}

While in the ideal limit -- infinitely many annealing steps $N_A$, optimization epochs per annealing step $N_U$, and a sufficiently expressive ansatz -- VQA would recover the true adiabatic path in the whole Hilbert space, it is clearly necessary to set a finite computational budget and select an NQS manifold with finite expressivity.
The VQA framework is designed to remain close to the instantaneous variational ground-state, which is defined as the state that minimizes Eq.~\eqref{eq:var_energy_VQA} as $H(s)$ slowly interpolates between $H_0$ and $H_T$. As recently pointed out in~\cite{Bojan_PRL}, it is essential to notice that the instantaneous \emph{variational} ground-state does not need to coincide with the \emph{exact} instantaneous ground-state of $H(s)$ at intermediate points $s_j$.
Hence, the discretized manifold evolution does not, in general, closely approximate the ideal adiabatic dynamics in the full Hilbert space. Nonetheless, if the variational manifold is sufficiently expressive to capture both the initial ground-state of $H_0$ and the target ground-state of $H_T$, this manifold evolution may yield a valid solution -- either the target quantum ground-state, or a wavefunction with finite overlap with classical energy minima of Eq.~\eqref{eq:ising}.

Additionally, even when approximating the adiabatic evolution on the manifold rather than in the whole Hilbert space, it is not necessary to find exact solutions of the intermediate ground-state problems. In fact, even if the ansatz does not fully converge to the instantaneous variational ground-state at some intermediate step, this approximate solution will still serve as a valid warm start for the optimization problem in Eq.~\eqref{eq:var_energy_VQA} at the next annealing step.
Previous numerical errors can be compensated when re-optimizing the ansatz, a feature that does not extend to real dynamics on a quantum annealer or real-time TDVP simulations.
Guided by this observation, we recognize a trade-off in the choice of \(N_A\) and \(N_U\), as the total computational cost scales linearly with their product. Rather than fully optimizing the variational parameters at each annealing step, it is often more efficient to increase the number of annealing steps \(N_A\) while performing only a few parameter updates \(N_U\) per step. In practice, even a single update per step (\(N_U = 1\)) can be sufficient to reach the final target ground state.
To ensure that any residual excited-state components are suppressed at $j=N_A$, the variational parameters can be fine-tuned with a few additional optimization epochs as in Eq.~\eqref{eq:SR_momentum}.

\subsection{Inclusion of a catalyst}

The classical simulation of approximate adiabatic dynamics enables full flexibility in choosing Hamiltonian terms. We test the introduction of a catalyst $H_C$, which is an auxiliary Hamiltonian term vanishing at both $s=0$ and $s=1$:
\begin{equation}\label{eq:Hs_cata}
H(s) = s\, H_\mathrm{T} + A\, (1-s)\, H_{0} + B\, s(1-s)\, H_{C}  \;,
\end{equation}
where $H_0$ is given by Eq.~\eqref{eq:Hx}, and $A>0$ and $B$ are two tunable real hyper-parameters.
By reshaping the instantaneous energy landscape, catalysts can convert a first-order phase transition encountered in real-time dynamics -- with an exponentially small spectral gap -- into a higher-order transition with a polynomially closing gap~\cite{Nishimori_2017, ghosh2024}. Similarly, it may help reduce diabatic errors during variational optimization in a VQA setting.
In this work, we select
\begin{equation}
H_C = \sum_{j=1}^N\sigma_j^y
\end{equation}
as a catalyst, which is a non-stoquastic term that cannot be implemented on current quantum annealers.

\section{Numerical results}
\label{sec: results}

In this Section, we benchmark NQA with DBQS on classical and quantum spin-glass models with long-range interactions, demonstrating its effectiveness for large-scale systems with disorder and energy frustration. We also address an NP-complete real-world task, consisting of the Job Shop Scheduling Problem (JSSP). 
Further details on reproducibility are provided in the Code and Data Availability section.

\subsection{Sherrington–Kirkpatrick model}
We first address the Sherrington–Kirkpatrick (SK) model, a standard benchmark for testing algorithms designed to solve combinatorial optimization problems arising in science and industry, which often admit a reformulation as the energy minimization of similar spin models~\cite{glover2019tutorial, lucas2014}. A notable example is represented by resource allocation and scheduling problems, tackled in Section~\ref{sec: JSSP}. 
The SK model is defined by $N$ spin-$1/2$ degrees of freedom with all-to-all interactions whose coupling constants are sampled independently. 
The Hamiltonian of the model reads
\begin{equation}
\label{eq:SKHamiltonian}
H_\mathrm{T}=\frac{1}{\sqrt{N}}\sum_{i<j}J_{ij}\PauliSigma^z_i\PauliSigma^z_j\,,\ \text{with}\,\ J_{ij}\sim\mathcal{N}(0,1)\,,
\end{equation}
where $\mathcal{N}(\mu,\sigma^2)$ denotes the normal distribution.
Instances of the SK model provide an ideal testbed for 
heuristic algorithms, as the model is fully connected, highly frustrated, and exhibits a hierarchical glassy energy landscape with exponentially many local minima~\cite{parisi1995number,RevModPhys.58.801}.
Our novel DBQS architecture appears well-suited to address real-world NP-hard problems that lack geometric structure and involve infinite-range interactions -- features that are well captured by the SK model.
In contrast, physical models are often characterized by finite dimensionality and symmetries, which can be leveraged to encode efficient inductive biases into NQS ansatz design, as seen for CNNs and convolutional transformers for translationally invariant models, or 1D and 2D RNNs for low-dimensional systems.

We prepared $10$ realizations of the SK model with $N=100$ spins and $10$ realizations with $N=200$ to benchmark our algorithm. Reference ground-state energies for these realizations were computed via $1000$ runs of classical simulated annealing using OpenJij~\cite{Nishimura_OpenJij}. The spin-glass server~\cite{sg_server}, a standard tool for computing exact ground states of relatively small instances, was able to solve only $8$ out of the $10$ instances with $N=100$ spins. In those cases, the results matched the lowest energies obtained with simulated annealing. 
We are testing our methods on an NP-hard problem, where traditional exact solvers already struggle for $ N \approx 100$, while $ N = 200$ is considered an impractically large regime that requires approximate methods. 

For each run, we evaluate two metrics to quantify the quality of the solution, namely the final quantum state residual energy $\varepsilon_Q \equiv (\langle H_\mathrm{T}\rangle - E_0)/|E_0|$, and the residual energy of the best classical state encountered during the simulation $\varepsilon_B \equiv (E_B - E_0)/|E_0|$, where $E_0$ is the exact ground state energy (or the best known empirical estimate).
For a classical spin glass, $\varepsilon_B=0$ is a valid success criterion (a ground state was sampled during the simulation), even if the final wavefunction has not completely converged to that or other ground states.
By construction, $0 \le \varepsilon_B \le \varepsilon_Q$, and $\varepsilon_B = \varepsilon_Q = 0$ signals that the algorithm converges to a state overlapping only with ground-state configuration(s) of a problem instance.

To aggregate results over multiple spin glass instances, we consider the typical (geometric mean) errors $[\varepsilon_Q]_{\rm typ}$ and $[\varepsilon_B]_{\rm typ}$, where $[x]_{\rm typ} \equiv \exp\!\big(\langle \log x\rangle\big)$ 
is the exponential of the average log-error (with a $10^{-10}$ floor for numerical stability) and the average $\langle \cdot \rangle$ is taken over instances.

\begin{figure}[t]
\centering
\includegraphics[width=\linewidth]{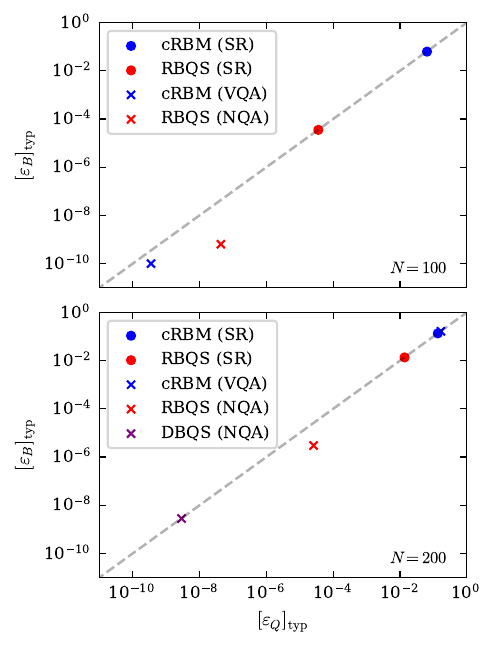}
\caption{Comparison of the typical residual energies on the $10$ realizations of the SK model with $N=100$ (top panel) and $N=200$ (bottom panel). Each point corresponds to a different combination of ansatz and optimization protocol. Only the DBQS results employ a catalyst and complex parameters.}
\label{fig:typical_energies}
\end{figure}

\begin{figure}[t]
\centering
\includegraphics[width=\linewidth]{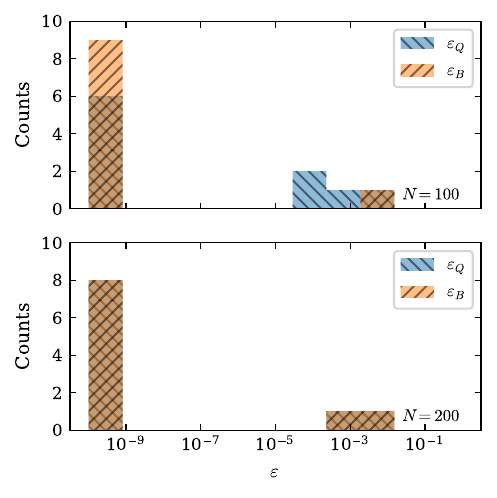}
\caption{Residual energy histograms for the SK model with $N=100$ (top panel) and $N=200$ (bottom panel). Each entry corresponds to a distinct instance tackled by NQA and an RBQS/DBQS ansatz (see main text for details on the variational model). Our setup solves all $N=100$ spin instances and 7 out of 10 $N=200$ spin instances.}
\label{fig:sk_main_1}
\end{figure}

We first test the performance of a cRBM with real parameters (see~\cite{Carleo_2017} and Appendix~\ref{sec: app_quantum_BMs}) optimized via SR (natural gradients with momentum) or VQA, see Appendix~\ref{sec: app_algorithms} for details on the algorithms.
We compare it against RBQS with real parameters, optimized with SR (natural gradients with momentum), or in the framework of NQA (Appendix~\ref{sec: app_NQA}).
The use of real-valued wavefunctions is justified since the QA Hamiltonian in Eq.~\eqref{eq:Hs_cata} is stoquastic for $B=0$, and for a standard driver and target Hamiltonian in Eq.~\eqref{eq:Hx} and Eq.~\eqref{eq:ising}, respectively; hence, its instantaneous ground state can be described by a (positive) real-valued wavefunction.
Sampling is performed using block Gibbs updates for RBQS or single-spin-flip Metropolis-Hastingss chains for cRBM, in both cases kept persistent across annealing steps.
To ensure a fair comparison, we conduct a brief hyperparameter optimization (HPO) using Optuna~\cite{optuna_2019} with a Tree-structured Parzen Estimator (TPE)~\cite{bergstra2011algorithms}. We allocate a fixed amount of $50$ trials to each combination of ansatz and optimization algorithm.
Additional details on HPO and the specific settings used for this benchmark are reported in Appendix~\ref{app:hpo}. 

Figure~\ref{fig:typical_energies} compares the performance of the different setups according to the typical energy errors introduced above.
In the top panel ($N=100$), we observe that the cRBM ansatz fails to find the ground state for any instance when optimized with SR (natural gradients with momentum), as highlighted by large values of both typical metrics. 
The RBQS ansatz with SR solves exactly (i.e.\ $\varepsilon_B=0$) only half of the instances (see Appendix~\ref{app:benchmak_sk} for details).
Note that standard SR optimization is included within our HPO as the limit of zero momentum, which is consistently outperformed by non-zero momentum values.

The results are dramatically improved when using VQA for the cRBM and NQA for the RBQS, respectively solving 10 and 9 benchmark instances.
%
%
These results demonstrate that the guided annealing strategy is significantly more effective in navigating the rugged energy landscape of spin glasses than direct SR optimization. We remark that the number of HPO trials
are set equal for a fair algorithmic comparison.

When scaling up to $N = 200$ spins (bottom panel), the problem hardness increases drastically within our fixed computational budget. The cRBM does not solve any instance, nor does the RBQS optimized with SR. Only the combination of RBQS and NQA manages to solve some of the instances, with typical residual energies of $[\varepsilon_B]_{\rm typ} =2.997\times10^{-6}$ and $[\varepsilon_Q]_{\rm typ} =2.599\times10^{-5}$.
The overall performance on the $N=200$ instances can be improved by including a catalyst and complex-valued variational parameters, which are now needed to describe the non-stoquastic nature of the instantaneous ground state. 
At the same time, we increase the expressivity by selecting a DBQS ansatz with a second hidden layer, optimized within the catalyzed NQA framework, 
this configuration is run with different HPO settings (see Appendix~\ref{app:hpo}), with a budget of 100 trials per instance.
We thereby achieve the highest success rate and accuracy, obtaining the exact ground state in 8 out of 10 instances at $N = 200$, yielding typical residual energies of $[\varepsilon_B]_{\rm typ} =2.807\times10^{-9}$ and $[\varepsilon_Q]_{\rm typ} =2.881\times10^{-9}$. 

Figure~\ref{fig:sk_main_1} presents the residual energy histograms for the benchmark instances with our best performing set-ups, namely RBQS (real parameters, uncatalyzed) for $N=100$ and a two-layer DBQS (complex parameters, with catalyst) for $N=200$. In the latter case, the exact result (i.e.\ $\varepsilon_B=0$) can be obtained even for some instances where the final wavefunction has not fully converged ($\varepsilon_Q>0$), as visible from the right tails of the histogram.
In summary, both the DBQS ansatz and the NQA algorithm emerge as the leading candidates to solve hard classical spin glasses within a given computational budget, according to both residual-energy metrics defined above.


\subsection{The Job Shop Scheduling Problem}
\label{sec: JSSP}

We now test our methods on the Job Shop Scheduling Problem (JSSP), which has relevant applications in logistics, manufacturing, and other domains~\cite{jssp_industry, DACOL2022100249}.
The JSSP involves assigning a set of operations to specific machines while satisfying precedence and resource constraints and minimizing the total processing time, known as timespan. It is an NP-hard problem in the strong sense, and has been empirically shown to be particularly intractable~\cite{dynprog}.

More formally, a JSSP instance is defined by \( N \) jobs, \( \{J_1, \dots, J_N\} \), where each job is defined as a totally ordered sequence of operations. 
Each operation must be executed on a specific machine selected from a set of \( M \) available machines, \( \{M_1, \dots, M_M\} \). The number of operations per job is typically the same for all jobs and is denoted by \( K \). 
Moreover, it is customary to assume \( K = M \), so that all machines are involved exactly once for the operations of any job, in a job-dependent order. 
To be definite, we also set \( N = K = M \), corresponding to a so-called square instance. Our methods can be readily adapted to alternative choices of these hyperparameters, as well as to generalizations of the JSSP, such as the flexible JSSP~\cite{flexibleJSSP_review}.

Each operation is associated with a positive integer processing time \( p_{n,k} \), where \( n = 1, \dots, N \) indexes the job and \( k = 1, \dots, K \) labels the operation within that job.

A valid scheduling of all operations should satisfy two sets of constraints. First, each machine can process only one operation at a time; second, the operations belonging to the same job must be carried out in the prescribed order, each on its preassigned machine. Preemption is not permitted: once an operation starts, it cannot be interrupted and resumed later. The final goal is to obtain a valid schedule that satisfies all constraints and minimizes the timespan \( T \), defined as the total time required to complete all operations across all jobs. In practice, such a schedule is specified by assigning valid integer starting times to each operation.

The decision version of the JSSP formally asks whether a valid schedule exists within a preset value of $T$.
This leads to an NP-complete problem (rather than NP-hard as for the optimization version of the JSSP), which admits a QUBO/Ising reformulation originally presented in Ref.~\cite{Venturelli2015JSSP}.
The timespan can subsequently be lowered in an iterative fashion, e.g.\ by a binary search~\cite{Venturelli2015JSSP}.

To introduce the QUBO model, we need a collective index $i=1,\dots,N\times K$ that spans all operations of all jobs in raw-major order (the first $K$ indices pertain to the operations of the first job). We define the binary variables:
\begin{equation}    x_{i,t} = \begin{cases}
        1 & \text{if operation $i$ starts at time t,}  \\
        0 & \text{otherwise} \,,   \end{cases}
    \label{eq:variablex}   \end{equation}
with the time index $t$ spanning integer values in the range $[0,T-1]$.

As done in~\cite{Venturelli2015JSSP}, we introduce a cost function:
\begin{equation} 
\label{eq:JSSP_cost}
E_{T}(\textbf{x})=  \textbf{x}^\mathsf{T} Q \textbf{x} = \alpha \, \varepsilon_1(\textbf{x}) + \beta \, \varepsilon_2(\textbf{x}) + \gamma \, \varepsilon_3(\textbf{x})\,, 
\end{equation}
where $\textbf{x}$ is a bit string composed of all the $(NK)\times T$ binary variables defined in~\eqref{eq:variablex} in raw-major order (the first $T$ binary variables refer to the possible starting times of the first operation of the first job).

This cost function is composed of three penalty terms, which are weighted by the positive hyperparameters $\alpha\,, \beta\,,\gamma$. A positive value of the cost function indicates that one or more constraints have been violated, whereas the zero-cost minima of the cost function, if they exist, represent valid schedules within the predefined value of $T$, satisfying all constraints~\cite{ft_zero_energy}.

The formal expression of the penalty terms is given by
%
\begin{equation} \varepsilon_1(\textbf{x}) = \sum_{n = 1}^N\left(\sum_{\substack{k_{n-1}<i<k_n\\t+p_i>t'}}x_{i,t}x_{i+1,t'}\right) \label{eq:constraint1} \,; \end{equation}
\begin{equation} \varepsilon_2(\textbf{x}) = \sum_{m=1}^M\left(\sum_{(i,t,j,t')\in R_m}x_{i,t}x_{j,t'}\right) \label{eq:constraint2}\,; \end{equation}
\begin{equation} \varepsilon_3(\textbf{x}) = \sum_{i=1}^{NK}\left(\sum_{t=0}^{T-1}x_{i,t}-1\right)^2 \label{eq:constraint3}\,. \end{equation}
The first two terms formalize the actual constraints of the JSSP. 
In particular, the first term in Eq.~\eqref{eq:constraint1} enforces the proper order of operations within a job, whereas the second one in Eq.~\eqref{eq:constraint2} implies that an operation can be run on a machine only when no other operation is already running on it. Finally, Eq.~\eqref{eq:constraint3} introduces an additional soft constraint that ensures that each operation starts only once: the absence of a starting time or the existence of two (or more) distinct starting times are penalized, resulting in an invalid binary string.
More precisely, in Eq.~\eqref{eq:constraint1}, the index $k_n$ refers to the last operation of job $n$, with the additional definition of $k_0=0$; a cost $\alpha$ is paid whenever an operation starts before the previous one in the same job has been completed.
For the machine constraints in Eq.~\eqref{eq:constraint2}, we define $I_m$ as the set of all the indices of operations that have to be executed on the machine labeled by $m=1,\dots,M$. The set $R_m$ introduced in~\eqref{eq:constraint2} is defined as:
\begin{equation*}
    R_m = A_m \cup B_m\,,
\end{equation*}
where:
\begin{equation*}
      A_m = \{(i,t,j,t') :\;  (i,j) \in I_m \times I_m,\ i \neq j, \  0 < t' - t < p_i \} \,,
\end{equation*}
and
\begin{equation*}
      B_m =\{ (i,t,j,t'):(i,j) \in I_m \times I_m, \ i < j, \ t = t' \} \,.
\end{equation*}

%
The first set $A_m$ introduces a positive penalty cost $\beta$ whenever two operations $i,j$ assigned to the same machine $M_m$ are ordered as $i\to j$, but $j$ starts before $i$ has been completed; in contrast, the second set $B_m$ introduces the same penalty cost whenever such two operations were to start at the same time $t$.

It is worth noting that the fixed timespan $T$ in the decision problem also determines the number of binary variables in the QUBO formulation, which amounts to $NKT$.
Let us consider a square instance. Supposing that the minimum processing time for each operation is one time unit, one necessarily has $T > N$. 
This implies that a lower bound for the Hilbert space dimension is $2^{N^3}$, which is exponentially larger than the scaling $(N!)^N$ associated with a brute-force enumeration of all $N!$ possible schedules on each of the $N$ machines. When dealing with such a large Hilbert space, reducing the computational complexity becomes crucial. Variable pruning techniques achieve this by identifying and eliminating unnecessary binary variables $x_{i,t}$ that would otherwise lead to constraint violations. This approach stems from classical operations research literature~\cite{carlier1990practical, carlier1994adjustment} and has been proposed and implemented to significantly reduce the Hilbert space dimension in a quantum setup~\cite{Venturelli2015JSSP, casati2025iterative}.

\begin{figure*}
\begin{minipage}[c]{0.49\linewidth}
\includegraphics[width=\linewidth]{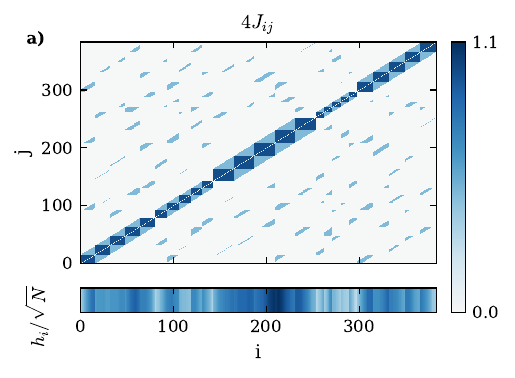}
\end{minipage}
\hfill
\begin{minipage}[c]{0.49\linewidth}
\includegraphics[width=\linewidth]{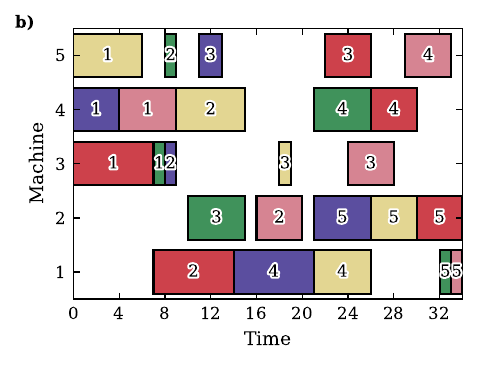}
\end{minipage}%
\caption{\textbf{(a)} Interaction matrix and longitudinal field vector of the Ising problem encoding a typical JSSP square instance. The represented instance has $N=5$ and is displayed after variable pruning. \textbf{(b)} The Gantt diagram of the corresponding optimal schedule. Colors represent operations pertaining to a specific job, whereas the index of each operation indicate their ordering within that job.}
\label{fig:gantt}
\end{figure*}

Here we apply the scheme of Ref.~\cite{casati2025iterative} to perform variable pruning and fix the input value of $T$, subsequently mapping the pruned QUBO matrix to the equivalent Ising spin glass via standard transformations on the binary variables.
The scope of our analysis is to benchmark our methods on realistic problems, beyond unstructured spin-glass models derived from physics.
Indeed, the structure of the matrix $(J_{ij})$ and vector $(h_i)$ is markedly different for a JSSP instance compared to an instance of the SK model, and this may influence the effectiveness of a heuristic algorithm. 
This is conveniently visualized in the left panel of Fig.~\ref{fig:gantt}, showing the interaction matrix and longitudinal field vector of the Ising formulation for a $N=5$ JSSP square instance (after variable pruning).

To tackle challenging models with a large number of spins, we define a batch of $10$ square instances with $N=5$. We select two instances for each value of $p_\mathrm{max}\in \{4,\dots,8\}$ and sample the duration of each operation $p$ uniformly in $\{1,\dots,p_\mathrm{max}\}$, corresponding to a number of spins in a range between $N=300-400$ after variable pruning.
We specifically defined our dataset by only selecting instances that cannot be embedded on the novel D-Wave Advantage 2 solver (within a $5$-minutes time limit). 

We use our best-performing setup, which has been tested on large-scale $N=200$ SK instances: a DBQS with two layers and complex parameters, trained with a catalyst in the NQA framework. We fix the hyperparameters as $\alpha=\beta=\gamma=1$.

For the decision-version JSSP problem formalized by Eq.~\eqref{eq:JSSP_cost}, the ground state energy is known a priori and equals zero~\cite{ft_zero_energy}. 
We run an HPO with Optuna (details in Appendix~\ref{app:hpo}), allocating a budget of 200 maximum trials per instance,
automatically halted when the first valid solution is found.
Within the allocated computational budget, we solve exactly 8 out of 10 instances, while in the remaining two instances, only one constraint is not satisfied. 
The right panel of Fig.~\ref{fig:gantt} shows an example of a Gantt diagram that represents the optimal schedule of the corresponding JSSP instance.



These results demonstrate that our methods are applicable to realistic QUBO instances with hundreds of spins, a regime beyond the embedding capabilities of state-of-the-art quantum annealing hardware.




\subsection{Transverse-field Sherrington-Kirkpatrick model}

Building on the demonstrated performance for disordered classical systems, we now aim to address challenging quantum spin-glass problems. In particular, we focus on the transverse-field Sherrington-Kirkpatrick model
\begin{equation}
\label{eq:TFSKHamiltonian}
H_\mathrm{T}=\frac{1}{\sqrt{N}}\sum_{i<j}J_{ij}\PauliSigma^z_i\PauliSigma^z_j+g\sum_{i=1}^N\PauliSigma_i^x\,,
\end{equation}
with the coupling matrix elements $
J_{ij}\sim\mathcal{N}(0,1)$ drawn independently from a standard Gaussian distribution of unit width.

The transverse-field SK  model has been studied extensively from theoretical~\cite{PhysRevB.41.4858,PhysRevB.52.384,Rieger} and computational~\cite{10.5555/3159044} perspectives, yet finding the ground state for instances beyond the limits of exact diagonalization remains challenging to date. 
Quantum fluctuations lead to a markedly different low-energy manifold from the classical case, making this model an ideal testbed for our methods in the quantum regime.

As we will demonstrate, our approach can provide access to the ground state of instances of the transverse-field SK model.
In the following, we focus on a regime of weak quantum fluctuations with a small value of transverse field coupling, ensuring that the target Hamiltonian in Eq.~\eqref{eq:TFSKHamiltonian} is deep in the spin-glass phase.
We first benchmark our methods against exact diagonalization on ten exactly solvable instances with $N=16$, and then tackle the same disorder realizations with $N=100$ spins used in the classical setting. All simulations are run for the target Hamiltonian with transverse-field coupling $g=0.1$.  

For each instance, we conduct a hyperparameter optimization (HPO) using a complex-parametrized DBQS ansatz with two hidden layers and varying numbers of units per layer, optimized via the catalyzed NQA protocol.
We allocate a budget of 100 trials per instance (details about the HPO are reported in Appendix~\ref{app:hpo}).


\begin{figure*}[t]
\centering
\includegraphics[width=\linewidth]{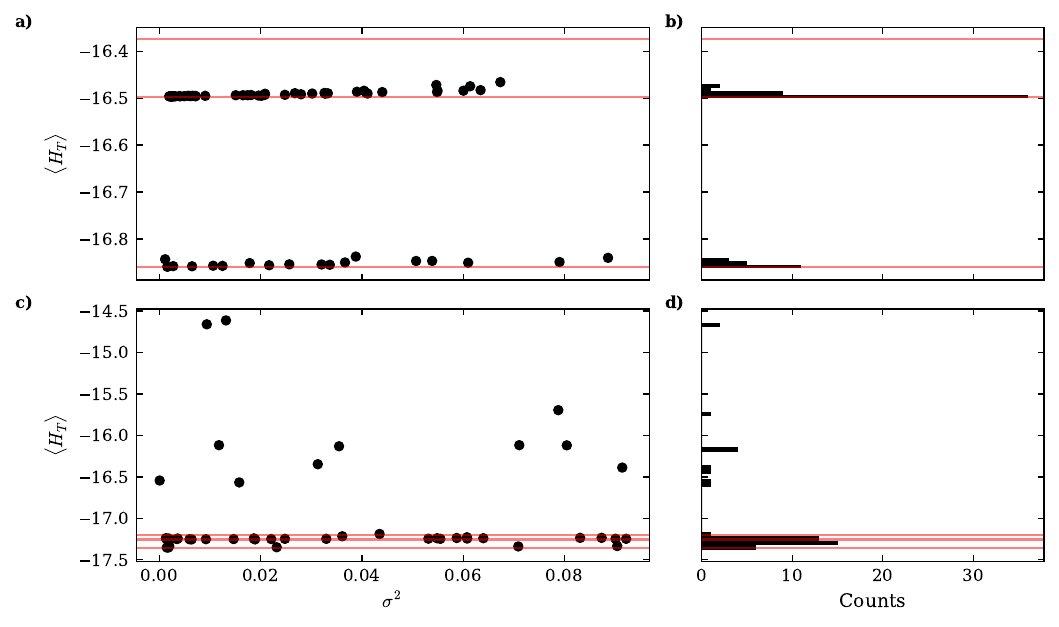} 
\caption{
Convergence analysis for two representative $N=16$ instances of the transverse-field SK  model at $g=0.1$.
The results are obtained by optimizing a two-layer DBQS via catalyzed NQA.
\textbf{(a)} and \textbf{(c)} Final variational energy $\expval{H_T}$ versus final energy variance $\sigma^2$. \textbf{(b)} and \textbf{(d)} Histogram of the final energies over the HPO trials.
The red lines indicate the exact energies of the three lowest eigenstates, for comparison.
Monte Carlo error bars are smaller than the symbol size and thus omitted.
}
\label{fig:tfk_16}
\end{figure*}

\begin{figure}[t]
\centering
\includegraphics[width=\linewidth]{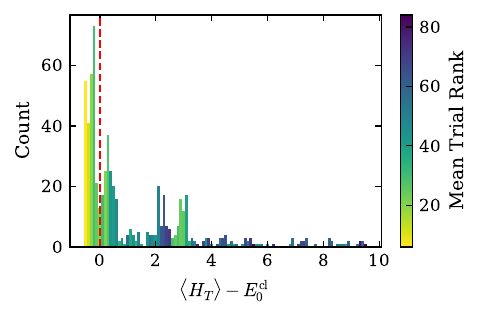} 
\caption{
Aggregated results for the $N=100$ instances of the transverse-field SK model at $g=0.1$.
We plot a histogram of the final energies obtained via the HPO, subtracting the corresponding classical energy. For each instance, we rank the HPO runs by final energy, with rank 1 corresponding to the best result.
Colors indicate the average rank of the trials in the corresponding bin. The best runs for each instance typically capture quantum effects, as shown in the leftmost bins below the classical energy threshold (vertical dashed line).
}
\label{fig:tfk_100}
\end{figure}

\begin{figure}[t]
\centering
\includegraphics[width=\linewidth]{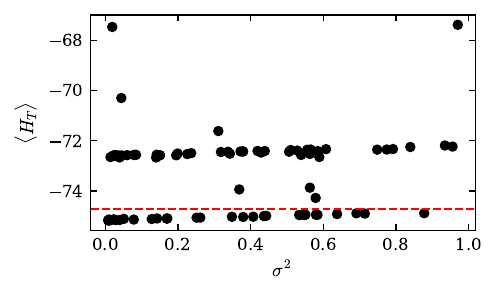} 
\caption{
Convergence analysis for a representative $N=100$ instance of the transverse-field SK  model at $g=0.1$. The best HPO trials lie below the classical ground-state energy (horizontal dashed line), indicating that quantum effects are captured and suggesting convergence to the quantum ground state or to low-lying eigenstates.
Monte Carlo error bars are smaller than the symbol size and thus omitted.
}
\label{fig:tfk_100_i_6}
\end{figure}

We first validate this approach for $N=16$.
On this benchmark, the best trials yield typical errors of $[\varepsilon_Q]_{\rm typ}=3.15\times10^{-5}$ for the ground state energy. 
For every trial $i$, we record the final energy $\expval{H_T}_{(i)}$, and the final energy variance $\sigma^2_{(i)}$.
When visualized in the $(\expval{H_T},\sigma^2)$ plane, for 9 out of 10 instances, the data points corresponding to the best trials do not scatter randomly, but organize in a cluster that is well-separated from the other points. This is shown in Fig.~\ref{fig:tfk_16} (top) for a specific instance, and it can be explained as multiple trials converging to the same eigenstate with different small residual errors.
In the remaining instance, we do not observe a well-separated cluster due to the presence of multiple eigenstates very close to the ground state, see Fig.~\ref{fig:tfk_16} (bottom). Remarkably, even in this case, the best data points closely approximate the actual ground state energy.

In the large-scale regime with $N=100$, a direct comparison with exact results is unavailable. Nonetheless, we observe that the best trial energy typically lies below the ground-state energy of the corresponding classical system. 
Indeed, in the spin glass phase (small $g$), the classical ground-state energy (at $g=0$) constitutes a reliable upper bound on the true quantum ground-state energy, since the expectation value of the transverse-field term vanishes on any classical state.

Hence, our algorithm can consistently capture quantum effects, as demonstrated in Fig.~\ref{fig:tfk_100} by aggregating the best HPO runs over all instances.
Moreover, for each instance, the best trials tend to converge on similar energies, suggesting convergence to the quantum ground state or, at least, to nearby low-energy eigenstates.
Fig.~\ref{fig:tfk_100_i_6} showcases the best HPO trials for a typical instance, whereas detailed results for all instances are reported in Appendix~\ref{app:tfsk_extra}.

Although it is generally impossible to determine whether the identified low-energy states are the true ground state of the problem, this is not a disadvantage of our method, which rather sets a novel benchmark in this large-scale regime of quantum spin glasses.

Overall, these results demonstrate that the combination of DBQS, NQA, and statistical HPO analysis constitutes a viable framework for studying large-scale quantum spin glasses.

\section{Discussion and Outlook}
\label{sec: outlook}

In this work, we introduced Deep Boltzmann Quantum States (DBQS), a class of neural quantum states (NQS) that offers significant computational advantages, particularly enabling efficient sampling via block Gibbs chains.
Our results provide evidence that DBQS constitute an effective variational ansatz for large-scale classical and quantum spin-glass problems. When combined with neural quantum annealing (NQA) and automated hyperparameter optimization (HPO), our variational ansatz is an effective solver for classical and quantum ground-states in many-body models with disorder and for realistic optimization problems. 


On classical Sherrington-Kirkpatrick (SK) instances up to $N=200$, NQA attains exact or near-exact energies, whereas on transverse-field SK, final variational solutions across HPO trials form well-separated clusters in the energy–variance plane, consistent with convergence to distinct eigenstates.
Finally, on QUBO encodings of the Job Shop Scheduling Problem, we reach exact solutions for instances that exceed current quantum-annealing hardware constraints.

At the same time, some practical limitations remain. Performance depends on a set of hyperparameters related to the variational model, the annealing schedule, the optimizer, and the Gibbs chains. However, HPO constitutes an effective automated tool for collecting statistics across hyperparameter realizations, as common practice in deep learning applications. Thanks to the efficiency of DBQS, the computational cost of a full HPO can be comparable to that of a single variational optimization run for other large-scale NQS, such as transformers.
More broadly, our numerical approach is heuristic: at present, it provides no theoretical guarantees on time-to-solution, and optimal settings can vary across instance families.
Nonetheless, the consistent empirical performance across classical, quantum, and real-world benchmarks indicates that high-quality, frequently optimal solutions can be obtained in practice with moderate computational effort, positioning our method as a promising candidate for challenging optimization and disordered many-body problems.

Several directions could further improve robustness and scalability. A natural extension is to move from fixed annealing schedules to adaptive schedules guided by local geometric signals on the variational manifold, such as the Fubini–Study distance or Fisher curvature.
A promising conceptual advance would integrate population-annealing variants that exchange persistent Gibbs chains across distinct annealing schedule points, further mitigating slow mixing and mode collapse, similarly to the renowned parallel-tempering algorithms~\cite{Marinari_1992,doi:10.1143/JPSJ.65.1604}.
On the modeling side, it seems promising to explore architectural generalizations of DBQS, including a convolutional structure, attention mechanisms, and graph-coupled layers.
Finally, the same framework is immediately applicable beyond SK-type benchmarks to broader classes of classical or quantum ground-state problems with spin-glass features, including recently explored amorphous Rydberg atom arrays~\cite{brodoloni2025spinglassquantumphasetransition}.

To conclude, deep neural architectures with efficient global update rules, trained within an annealing-inspired protocol, provide a flexible and scalable framework for both real-world hard combinatorial optimization and the study of disordered quantum many-body systems.

\section*{Code and Data Availability}
The data that support the findings of this article are openly available~\cite{zenodo_repo}.
All code to reproduce our experiments, along with scripts to generate figures and tables, is publicly available at~\cite{nqa_server_github} in the \texttt{Paper/} directory.

\begin{acknowledgments}
We thank Thomas Giudici for his help in generating the JSSP instances, and Bojan \v{Z}unkovi\v{c}, Marcus Kollar, and Arnab Das for insightful discussion. 
We thank the maintainers of open-source projects that we relied on for baselines and validation, in particular OpenJij~\cite{Nishimura_OpenJij} and the spin-glass server~\cite{sg_server}. We also acknowledge the broader open-source ecosystem and the scientific Python stack, specifically JAX~\cite{JAX} and Optuna~\cite{optuna_2019}, which enabled our implementation and experiments.
Experiments were run on a Linux cluster with CUDA-enabled GPUs, on a single A100 GPU. 
We acknowledge the high-performance computing resources provided by the LiCCA HPC cluster at the University of Augsburg, cofunded by the DFG (Project ID 499211671).
This work was partially supported by the Italian Space Agency (ASI) (project CQES grant N.
2023-46-HH.0) and the 
German Research Foundation (Project No. 492547816; TRR 360, Project C4). 
This project has received funding from the European Research Council (ERC) under the European Union’s Horizon 2020 research and innovation programme (grant agreement No. 853443).
\end{acknowledgments}

\appendix
\section{Algorithmic details on NQS optimization for ground-state search}
\label{sec: app_algorithms}

\subsection{Natural gradients and advanced optimizers}
In Section~\ref{sec: NQS}, we introduced the real and imaginary-time TDVP and defined the Fisher information matrix. A noteworthy observation is that imaginary-time TDVP is formally equivalent to the optimization of the cost function in Eq.~\eqref{eq:var_energy} via natural gradient descent, a seminal algorithm known as Stochastic Reconfiguration (SR) \cite{sorella1998green}, recently improved by min-SR~\cite{chen2024empowering}.
This equivalence follows from the definition of natural gradient, obtained by applying the inverse Fisher information matrix to the standard gradient:
\begin{equation}
\label{eq:natural_gradient}
\vb g_{\boldsymbol{\theta}} = S^{-1}(\boldsymbol{\theta}) \boldsymbol{\nabla}_{\boldsymbol{\theta}}\,.
\end{equation}
Indeed, a first-order numerical integration of the imaginary-time TDVP equations leads to an iterative parameter update
\begin{equation}
\boldsymbol{\theta}(\tau+\delta \tau) = \boldsymbol{\theta}(\tau) - \delta\tau S^{-1}\big(\boldsymbol{\theta}(\tau)\big)F\big(\boldsymbol{\theta}(\tau)\big)\,,
\end{equation}
which corresponds to a Natural Gradient Descent step with step size $\delta \tau$ since 
\begin{equation}
\vb g_{\boldsymbol{\theta}} E(\boldsymbol{\theta}) = S^{-1}(\boldsymbol{\theta}) \boldsymbol{\nabla}_{\boldsymbol{\theta}} E(\boldsymbol{\theta}) = S^{-1}(\boldsymbol{\theta}) F(\boldsymbol{\theta})\,.
\end{equation}

As briefly mentioned in Section~\ref{sec: NQS}, natural gradients can be combined with advanced stochastic optimizers developed for deep learning.
While this strategy has not been explored in the NQS literature to our knowledge, it has been studied in the broader machine learning community.  For example, works such as~\cite{khan2018fastscalablebayesiandeep, martens2020optimizingneuralnetworkskroneckerfactored} precondition the stochastic gradient with an (approximate) inverse Fisher information matrix and apply momentum before updating parameters. Ref.~\cite{li2025acceleratednaturalgradientmethod} further proposes a unified framework for accelerated natural gradient descent.

In practice, an optimizer maintains an auxiliary state $\alpha$, and given the (natural) gradient, returns both the parameter update and the new auxiliary state.
As stated in Section~\ref{sec:NQA}, in this work, we combine SR with momentum, where the auxiliary state is a velocity vector storing an exponentially decaying average of past natural gradients.

Algorithm \ref{algo:NQS+SR} presents a high-level overview of NQS optimization for ground-state search, potentially combining natural gradients with an advanced optimizer.
As remarked in Section~\ref{sec: NQS}, we collectively refer to all these variants as SR in the remainder of the text.

\begin{algorithm}[H]
\caption{SR for ground-state search}
\label{algo:NQS+SR}
\begin{algorithmic}[1]
\Require
Number of training epochs $N_\varepsilon$;
Optimizer $\text{opt}(\vb g,\alpha)$;
Variational ansatz $\psi(\vb x;\boldsymbol{\theta})$.
\Ensure 
Optimized parameters $\boldsymbol{\theta}$

\State Initialize parameters $\boldsymbol{\theta} \gets \boldsymbol{\theta}_0$
\State Initialize auxiliary state of the optimizer $\alpha \gets \alpha_0$
\State \textbf{for} $n = 0$ \textbf{to} $N_{\varepsilon}-1$ \textbf{do} 
\State \quad Sample configurations $\{\vb x\}$ from $p(\vb x)\propto |\psi(\vb x;\boldsymbol{\theta})|^2$
\State \quad Compute the natural gradient $\vb g \gets \vb g_{\boldsymbol{\theta}} E(\boldsymbol{\theta})$
\State \quad Compute the updates for the parameters and the new optimizer state $(\delta\boldsymbol{\theta},\alpha)\gets \text{opt}(\vb g,\alpha)$
\State \quad Update the variational parameters $\boldsymbol{\theta} \gets \boldsymbol{\theta} + \delta \boldsymbol{\theta}$
\State \textbf{end for} 
\State \Return $\boldsymbol{\theta}$
\end{algorithmic}
\end{algorithm}

\subsection{Variational Quantum Annealing pseudo-code}

As explained in Section~\ref{sec:VQA}, an alternative approach to ground-state optimization with NQS relies on interpolating from an \emph{easy} to a \emph{hard} regime by tuning a parameter in the Hamiltonian.
The resulting high-level implementation of VQA for optimizing a variational ansatz (typically an NQS) is illustrated in Algorithm~\ref{algo:VQA_optimization}.
We note that, in Ref.~\cite{Hibat_Allah_2021}, the parametric Hamiltonian is defined slightly differently compared to Eq.~\eqref{eq:Hs}, precisely as $H(s)=H_\mathrm{T}+(1-s)H_0$, with $H_0=-B^x\sum_i \sigma_i^x$ and $B^x\gg1$. Although Ref.~\cite{Hibat_Allah_2021} uses Adam and simple gradients to train RNN wavefunctions, it is possible to extend this approach to other NQS architectures and to natural gradients.

\begin{algorithm}[H]
\caption{VQA for ground-state search}
\label{algo:VQA_optimization}
\begin{algorithmic}[1]
\Require 
Number of warm-up steps $N_{W}$;
Number of annealing steps $N_{A}$;
Number of updates per annealing step $N_{U}$;
Optimizer $\text{opt}(\vb g,\alpha)$ (Adam and simple gradients in~\cite{Hibat_Allah_2021});
Variational ansatz $\psi(\vb x;\boldsymbol{\theta})$ (RNN in~\cite{Hibat_Allah_2021}).
\Ensure  Optimized parameters $\boldsymbol{\theta}$
\State Initialize parameters $\boldsymbol{\theta} \gets \boldsymbol{\theta}_0$
\State Initialize auxiliary state of the optimizer $\alpha \gets \alpha_0$
\State $H(s)\gets H_0$
\For{$n = 0$ \textbf{to} $N_{W}-1$}
\State Sample configurations $\{\vb x\}$ from $p(\vb x)\propto |\psi(\vb x;\boldsymbol{\theta})|^2$
\State Compute (natural) gradients of the variational energy $\vb g\gets {\boldsymbol\nabla}_{\boldsymbol{\theta}} E(s;\boldsymbol{\theta})$
\State Compute the updates for the parameters and the new optimizer state $(\delta\boldsymbol{\theta},\alpha)\gets \text{opt}(\vb g,\alpha)$
\State Update the variational parameters $\boldsymbol{\theta} \gets \boldsymbol{\theta} + \delta \boldsymbol{\theta}$
\EndFor
\For{$j = 1$ \textbf{to} $N_{A}$} 
\State $s\gets j/N_A$ 
\State $H(s)\gets s H_T + (1-s) H_0$
\For{$n = 0$ \textbf{to} $N_{U}-1$}
\State Sample configurations $\{\vb x\}$ from $p(\vb x)\propto |\psi(\vb x;\boldsymbol{\theta})|^2$
\State Compute (natural) gradients of the variational energy $\vb g\gets {\boldsymbol\nabla}_{\boldsymbol{\theta}} E(s;\boldsymbol{\theta})$
\State Compute the updates for the parameters and the new optimizer state $(\delta\boldsymbol{\theta},\alpha)\gets \text{opt}(\vb g,\alpha)$
\State Update the variational parameters $\boldsymbol{\theta} \gets \boldsymbol{\theta} + \delta \boldsymbol{\theta}$
\EndFor
\EndFor
\State \Return $\boldsymbol{\theta}$
\end{algorithmic}
\end{algorithm}

\section{Standard analytical results on Boltzmann Machines}
\label{sec: app_BMs}
In Section~\ref{sec: classical_RBMS}, we introduced the RBM as the simplest efficiently trainable version of generic BMs.
Starting from Eq.~\eqref{eq:RBMS}, the joint probability function over visible and hidden variables reads
\begin{equation}
p(\vb v,\vb h;\boldsymbol{\theta}) = 
Z^{-1} \exp\{\vb v^T W \vb h + \vb b^T \vb v + \vb c^T \vb h\}\,,
\end{equation}
where
\begin{equation}
Z = \sum_{\vb v',\vb h'} \exp{-E(\vb v',\vb h';\boldsymbol{\theta})} \,. 
\end{equation}
This partition function $Z$ has been proven to be intractable~\cite{long2010restricted}, implying that the explicit computation of a normalized marginal probability distribution for the visible units is also intractable. 
Nonetheless, it is possible to exactly trace out the hidden variables to obtain the unnormalized marginal distribution 
\begin{equation}
\label{eq:p_rbm}
\begin{aligned}
&\pRBM(\vb v; \boldsymbol{\theta}) = \sum_{\vb h} p(\vb v,\vb h;\boldsymbol{\theta})\\
&=Z^{-1} \exp{\vb b^T \vb v}\prod_{j=0}^{N_h}\cosh\left\{c_j+\sum_{i=0}^{N_v}v_i W_{ij}\right\}\,.
\end{aligned}
\end{equation} 
The inability to compute the exact partition function does not constitute a serious limitation, as multiple methods have been devised to approximate it~\cite{Goodfellow-et-al-2016}, and its explicit computation is often unnecessary in practice. For instance, Monte Carlo methods circumvent the problem of computing the partition function by computing unweighted averages over the samples drawn from the model.

The key property of RBMs that enables efficient sampling and training stems from the bipartite structure of their interaction graph. Indeed,  conditioned on the hidden units, the visible units are mutually independent, and vice versa.
This fact allows us to explicitly compute the normalized conditional probabilities 
\begin{equation}
\begin{aligned}
&p(\vb v|\vb h;\boldsymbol{\theta}) = \frac{p(\vb v,\vb h;\boldsymbol{\theta})}{p(\vb h;\boldsymbol{\theta})}\\
&=\frac{1}{Z} \frac{\exp{\vb c^T\vb h}}{p(\vb h;\boldsymbol{\theta})} \exp\{\vb b^T\vb v+\vb v^TW\vb h\} \\
&=\frac{1}{Z'}\exp\{\vb b^T\vb v+\vb v^TW\vb h\} \\ 
&=\frac{1} {Z'}\prod_{i=1}^{N_v}\exp\left\{\left(b_i + \sum_{j=1}^{N_h} W_{ij} h_j\right) v_i\right\} \\
&\equiv  \frac{1} {Z'}\prod_{i=1}^{N_v}\tilde p(v_i|\vb h; \boldsymbol{\theta})\,.
\end{aligned}
\end{equation}
Here, the hidden variables are treated as constants, as we are conditioning on their values. Hence, all terms not explicitly depending on visible units are included in a new constant $Z'$.
The conditional probability distribution is factorized over the visible units, and the single terms can be easily normalized since computing the partition function in this case involves a sum over two configurations only. The normalized conditional probabilities read:
\begin{widetext}
\begin{equation}
\begin{aligned}
p(v_i|\vb h;\boldsymbol{\theta})&=\frac{\tilde p(v_i|\vb h;\boldsymbol{\theta})}{\tilde{p}(v_i=+1|\vb h;\boldsymbol{\theta})+\tilde{p}(v_i=-1|\vb h;\boldsymbol{\theta})}\\
&=\frac{\exp\left\{\left(b_i + \sum_{j=1}^{N_h} W_{ij} h_j\right) v_i\right\}}{\exp\left\{\left(b_i + \sum_{j=1}^{N_h} W_{ij} h_j\right)\right\}+\exp\left\{-\left(b_i + \sum_{j=1}^{N_h} W_{ij} h_j\right)\right\}}\\
&=\exp\left\{\left(b_i + \sum_{j=1}^{N_h} W_{ij} h_j\right) (v_i-1)\right\}\sigma\left[2\left(b_i + \sum_{j=1}^{N_h} W_{ij}h_j\right)\right]\,,
\end{aligned} 
\end{equation}
\end{widetext}
where $\sigma(x)$ denotes the logistic sigmoid function.
The normalized probability for the $i$-th visible unit to have value $+1$, conditioned on all hidden variables, simplifies to
\begin{equation}
\label{eq:condprob_RBM_1}
p(v_i=+1|\vb h;\boldsymbol{\theta}) = \sigma\left[2\left(b_i + \sum_{j=1}^{N_h} W_{ij}h_j\right)\right]\,,
\end{equation}
and a similar expression holds for the value $-1$.
Since the difference between visible and hidden units is only formal, the conditional probability for the hidden units given the visible ones can be computed in the same way:
\begin{equation}
\label{eq:condprob_RBM_2}
p(h_j=+1|\vb v;\boldsymbol{\theta}) = \sigma\left[2\left(c_j + \sum_{i=1}^{N_v} W_{ij}v_i\right)\right]\,.
\end{equation}
The mutual independence of visible units conditioned on hidden units, and vice versa, enables efficient block Gibbs sampling of the joint probability function $p(\vb v, \vb h;\boldsymbol{\theta})$.
By starting from a random configuration, a Markov chain can be constructed with the update rule that, at each iteration, all visible units are simultaneously sampled conditioned on the hidden units, and then all hidden units are sampled conditioned on the visible units. 
Moreover, these samples can be drawn via exact sampling of the conditional probability in Eqs.~\eqref{eq:condprob_RBM_1} and~\eqref{eq:condprob_RBM_2}.
Such an alternating conditional sampling efficiently generates samples from the joint probability distribution $p(\vb v, \vb h; \boldsymbol{\theta}$) \cite{Casella01081992,Goodfellow-et-al-2016}, which are necessary both during training and inference.

During training, the RBM parameters $\boldsymbol{\theta}$ are typically optimized by minimizing the negative log-likelihood using variants of gradient descent. The gradient of such objective function can be expressed in terms of expectation values and two-point correlations between hidden and visible units~\cite{Amin_2018}, which are efficiently estimated via Gibbs sampling.
At the inference stage, the trained model would be used to compute the expectation values of functions of the visible units $f(\bf v)$ over the marginal probability distribution $p_\text{\tiny RBM}(\vb v; \boldsymbol{\theta})$ in Eq.~\eqref{eq:p_rbm}, which cannot be explicitly normalized. This expectation value, denoted as $\langle\cdots \rangle_\text{\tiny RBM}$, is equal to the expectation value of the same function over the joint probability $p(\vb v, \vb h; \boldsymbol{\theta}$):
\begin{equation}
\begin{aligned}
&\expval{f(\vb v)}_\text{\tiny RBM} = \frac{\sum_{\vb v} p(\vb v) f(\vb v)}{\sum_{\vb v} p(\vb v)} \\
&= \frac{\sum_{\vb v, \vb h} p(\vb v, \vb h) f(\vb v)}{\sum_{\vb v, \vb h} p(\vb v, \vb h)} = \expval{f(\vb v)}_{p(\vb v, \vb h)}\,.
\end{aligned}
\end{equation}
This allows us to efficiently estimate it by performing block Gibbs sampling and using the normalized conditional probability functions in Eqs.~\eqref{eq:condprob_RBM_1} and~\eqref{eq:condprob_RBM_2}.

The more general case of DBMs in Eq.~\eqref{eq:DBMS} is similarly treated.
The joint probability function is given by
\begin{equation}
p(\vb x;\boldsymbol{\theta}) =\frac{1}{Z}e^{\sum_{l=0}^{N_L-1} \vb x^{(l)T}W^{(l)}\vb x^{(l+1)} + \sum_{l=0}^{N_L}  \vb b^{(l)T}\vb x^{(l)}}\,.
\end{equation}
where
\begin{equation}
Z = \sum_{\vb x'} \exp{-E(\vb x'; \boldsymbol{\theta})}  \,.
\end{equation}

The computation of the marginal probability function for the visible units, as done for the RBM in Eq.~\eqref{eq:p_rbm}, is now intractable even in its unnormalized version.
Nonetheless, similarly to the case of RBMs, the variables can be rearranged into a bipartite graph, with odd layers on one side and even layers on the other.
This immediately implies that when we condition on the units in the even layers, the variables in the odd layers become independent, and vice versa. More precisely, it is enough to condition on the adjacent layers, and the explicit computation of the normalized conditional probability leads to 
\begin{equation}
\begin{aligned}
\label{eq:condprob_DBM}
&p(x^{(l)}_i=+1|\vb x^{(l-1)},\vb x^{(l+1)};\boldsymbol{\theta})=\\ 
& \ \ \sigma\!\left[2\!\left(\!\sum_{j} W^{(l-1)}_{ji} x^{(l-1)}_j \!+\!\sum_{j} W^{(l)}_{ij} x^{(l+1)}_j \!+\!b^{(l)}_{i}\!\right)\!\right]\!\,,
\end{aligned}
\end{equation}
with slightly different expressions for the first and last layers.
Analogously to the RBM case, this conditional independence enables efficient block Gibbs sampling by updating all units in only two iterations. Firstly, one can update all the units in the even layers (including the $l=0$ visible layer) as a block, conditioned on the units in the odd layers, and, secondly, all the hidden units in the odd layers conditioned on the units in the even ones. The samples can be drawn via exact sampling of the conditional probability in Eq.~\eqref{eq:condprob_DBM}.

\section{Previous quantum extensions of Boltzmann Machines}
\label{sec: app_quantum_BMs}
Several extensions of BMs and specifically RBMs to quantum spin systems have been previously proposed. 
For quantum systems, the fundamental object is not the probability function, but rather the wavefunction, which is a complex-valued function representing the probability amplitudes. 

\subsection{pmRBM and cRBM}
\label{sec:carleo_RBM}
A conceptually straightforward extension proposed in Ref.~\cite{torlai2018neural} and known as pmRBM uses two independent RBMs to model the phase and the modulus of the wavefunction.
The wavefunction is expressed in the spin eigenbasis, which is commonly associated with the eigenstates of the Pauli-Z operators:
\begin{equation}
\psi(\vb x;\boldsymbol{\theta})=\sqrt{\pRBM^\mathrm{m}(\vb x;\boldsymbol{\theta}_\mathrm{m})}\exp\{i\,\pRBM^\mathrm{p}(\vb x;\boldsymbol{\theta}_\mathrm{ p})\}\,,
\end{equation}
where $\pRBM(\vb x;\boldsymbol{\theta})$ denotes the RBM unnormalized marginal probability distribution (see Appendix~\ref{sec: app_BMs}). Here, $\boldsymbol{\theta}=(\boldsymbol{\theta}_m,\boldsymbol{\theta}_p)$ are real parameters, and $\vb x$ is a collective label for a spin eigenstate in the Pauli-Z eigenbasis. 
Current literature suggests that the artificial separation of modulus and phase information is suboptimal, making the learning problem more challenging, regardless of the specific ANN model used. Almost all recent NQS applications~\cite{schmitt2025simulatingdynamicscorrelatedmatter} utilize ANNs parameterized by complex variational parameters to output a complex number representing the wavefunction amplitude.
This approach was proposed in the context of RBMs in Ref.~\cite{Carleo_2017}, which is based on the idea of promoting the probability distribution defined by an RBM to quantum amplitudes by introducing complex parameters. 
The resulting ansatz, commonly referred to as cRBM, reads
\begin{equation}
\label{eq:crbm_wavefunction}
\psi(\vb x;\boldsymbol{\theta})= 
e^{\sum_{i=0}^{N_v} b_i x_i }\prod_{j=0}^{N_h}\cosh{\left\{ c_j + \sum_{i=0}^{N_v} x_i W_{ij}\right\}} \,,
\end{equation}
which is formally equivalent to the unnormalized marginal probability distribution of an RBM (see Appendix~\ref{sec: app_BMs}), 
with the only difference that $\boldsymbol{\theta} = (W, \vb b, \vb c)$ are now complex parameters.

In combination with the standard SR algorithm, this ansatz has been successfully used to find a variety of quantum ground-states with remarkable precision \cite{Carleo_2017, PhysRevB.107.165149, 10.21468/SciPostPhys.12.5.166, Nomura_2021} and in the context of NQA in Ref. \cite{Torta_Leone}.
However, the possibility of efficient block Gibbs sampling is lost in the cRBM, so one must utilize Metropolis-Hastingss sampling with local update rules, which has significantly longer autocorrelation times, making it slower and more computationally expensive.

\subsection{DBM wavefunctions}
A possible quantum extension of DBMs based on the same principle of the cRBM was introduced in Ref.~\cite{Carleo_2018} and reads
\begin{equation}
\label{eq:dbm_wavefunction}
\psi(\vb x; \boldsymbol{\theta})\! =\! \sum_{\substack{\vb x^{(l)}\\l>0}}e^{\sum_{l=0}^{N_L-1} \vb x^{(l)T}W^{(l)}\vb x^{(l+1)} + \sum_{l=0}^{N_L}  \vb b^{(l)T}\vb x^{(l)}}.
\end{equation}
The original reference also provides an interesting constructive approach that does not require a variational optimization of the DBM parameters. 
However, due to the intractability of the summation in Eq.~\eqref{eq:dbm_wavefunction} and the restrictive conditions under which efficient block Gibbs sampling of the model is possible, this variational ansatz has only been applied to small-scale benchmarks.

\subsection{Quantum Boltzmann Machines}
Another possible extension, dubbed Quantum Boltzmann Machine (QBM), was proposed in Ref.~\cite{Amin_2018}. The units of the BM are promoted to quantum spins, and the state of the system is described as the exponential of a transverse-field Ising Hamiltonian
\begin{equation}
H_\text{QBM} = -\sum_a \Gamma_a \PauliSigma_a^x - \sum_a b_a \PauliSigma_a^z - \sum_{ab}w_{ab}\PauliSigma^z_a\PauliSigma^z_b
\end{equation}
yielding the density matrix operator
\begin{equation}
\rho = \frac{1}{Z}\exp\{-H_\text{QBM}\}\,.
\end{equation}
The presence of non-commuting terms poses significant challenges in training the model, as detailed in the original reference~\cite{Amin_2018}, which applies this framework to address classical machine learning tasks. Later, Ref.~\cite{patel2024quantumboltzmannmachinelearning} used a simplified version of QBMs without hidden units to learn the ground-states of quantum systems through a hybrid quantum-classical method inspired by the variational quantum eigensolver.

\section{Analytical results for Boltzmann Quantum States}
\label{sec: app_analytical_res_BQS}
In this Section, we report additional results on BQS and their relationship to the cRBM introduced in~\cite{Carleo_2018} and described in Appendix~\ref{sec: app_quantum_BMs}.
We adopt the same notation as in Section~\ref{sec:BQS}.

\subsection{Ground-state structure of the extended Hilbert space and expectation values}
The BQS formalism describes both visible and hidden spins as quantum spins in an extended Hilbert space $\mathcal{H}_{vh}=\mathcal{H}_{v}\otimes\mathcal{H}_{h}$.
Although the two Hamiltonians $H_v$ and $H_{vh}=H_v\otimes\mathbb{I}_h$ share the same spectrum, the ground-state of the extended system can have a non-trivial entanglement structure.
Specifically, if $H_v$ has a degenerate ground-state, then there exist ground-states of $H_{vh}$ that are entangled across the visible/hidden partition, rather than simple tensor products of two pure states.
Let $H_v$ be a Hamiltonian with a $d$-fold degenerate ground-state, let $\{\ket{\Psi_v^{a}}\}_{a=1}^d$ be a set of orthonormal ground states, and let $\{\ket{\Phi_v^{b}}\}_{b=1}^{\dim\mathcal{H}_v-d}$ be a set of orthonormal higher energy eigenstates so that $\{\ket{\Psi_v^{a}},\ket{\Phi_v^{b}}\}$ forms a complete orthonormal basis of $\mathcal H_{v}$.
Then, any state $\ket{\psi_{vh}}$ of the extended system can be expanded as
\begin{equation}
\begin{aligned}
\label{eq:energy_basis_expansion}
\ket{\psi_{vh}(G,E)} = &\sum_{a,\vb h} G_{a \vb h}\ket{\Psi_v^a}\otimes\ket{\vb h} 
\\+ &\sum_{b,\vb h} E_{b \vb h}\ket{\Phi_v^b}\otimes\ket{\vb h}\,,
\end{aligned}
\end{equation}
and its energy expectation value is
\begin{widetext}
\begin{equation}
\frac{\expval{\psi_{vh}(G,E)|H_{vh}|\psi_{vh}(G,E)}}{\expval{\psi_{vh}(G,E)|\psi_{vh}(G,E)}}=\frac{\sum_{a,\vb h}|G_{a\vb h}|^2 \expval{\Psi_v^a|H_v|\Psi_v^a}+\sum_{b,\vb h}|E_{b\vb h}|^2 \expval{\Phi_v^b|H_v|\Phi_v^b}}{\sum_{a,\vb h}|G_{a\vb h}|^2 +\sum_{b,\vb h}|E_{b\vb h}|^2}\,.
\end{equation}
\end{widetext}
Hence, by the variational principle, the ground states of $H_{vh}$ take the form
\begin{equation}
\label{eq:general_gs_hv}
\ket{\Psi_{vh}(G)}=\sum_{a,\vb h} G_{a\vb h}\ket{\Psi_v^a}\otimes\ket{\vb h}\,,
\end{equation}
and variational energy minimization in the extended Hilbert space can generally yield entangled states that do not decompose into the tensor product of a state of the visible spins and a state of the hidden spins, provided the variational ansatz can describe entanglement between the two partitions.

Indeed, if we denote by $\Psi(\vb v, \vb h; \boldsymbol{\theta})=\expval{\vb v, \vb h|\Psi(\boldsymbol{\theta})}$ the (converged) BQS ansatz written in the computational basis $\{\ket{\vb x}\} = \{\ket{\vb v,\vb h}\}$, we can formally trace out the hidden degrees of freedom. This generally yields a mixed state for the visible spins:
\begin{equation}
\label{eq: reduced_rho_visible}
\rho_v = \sum_{\vb v,\vb v',\vb h} \Psi(\vb v, \vb h; \boldsymbol{\theta})\overline{\Psi(\vb v', \vb h; \boldsymbol{\theta})} \ket{\vb v}\bra{\vb v'}\,.
\end{equation}
The summation over $\vb h$ required to compute the matrix elements is tractable only in the case of the RBQS ansatz, as an analytical expression can be obtained for the reduced density operator
\begin{widetext}
\begin{equation}
\begin{aligned}
\rho_v^\text{RBQS} &= \sum_{\vb v,\vb v'} \exp{\left[\sum_i b_i v_i - \overline{b}_i v'_i\right]}\prod_j\cosh\left\{\sum_i (W_{ij}v_i -\overline{W}_{ij}v_i') + c_j-\overline c_j\right\} \ket{\vb v}\bra{\vb v'}\,.
\end{aligned}
\end{equation}
\end{widetext}

On the contrary, and similarly to the classical case, this summation cannot be performed analytically for the general DBQS ansatz.
Nonetheless, in practical applications, we typically need to compute the ground-state expectation value of operators $O_v$ acting solely on visible spins.
After a variational optimization, a DBQS converges to a close approximation of some ground state of the extended system in the form of Eq.~\eqref{eq:general_gs_hv}.
Hence, the required expectation value can be more conveniently estimated on the extended system by computing the expectation value of the operator $O_v\otimes\mathbb{I}_h$, using standard Monte Carlo sampling on the final variational state.

We can now derive a formal expression of such ground-state expectation values on the states in Eq.~\eqref{eq:general_gs_hv}. By means of the Schmidt decomposition, any such state can be rewritten as
\begin{equation}
\ket{\Psi_{vh}(G)}=\sum_{n} \widetilde{G}_{n}\ket{\widetilde{\Psi}_v^n}\otimes\ket{\widetilde{\vb h}^n}\,,
\end{equation}
where the coefficients \(\widetilde G_n\) are obtained from the singular
value decomposition \(G=U\widetilde G
V^\dagger\). The Schmidt vectors are given by
\begin{equation}
\ket{\widetilde{\Psi}_v^n}
=
\sum_a U_{an}\ket{\Psi_v^a},
\qquad
\ket{\widetilde{\vb h}^{\,n}}
=
\sum_{\vb h} V_{\vb h n}^*\ket{\vb h},
\end{equation}
and \(\widetilde G_n=\widetilde G_{nn}\).

Using such expansion immediately exposes that the ground-state expectation value for the operator $O_v\otimes\mathbb{I}_h$ on a general ground state of the extended system is
\begin{equation}
\begin{aligned}
&\frac{\expval*{\Psi_{vh}(G)|O_v \otimes \mathbb{I}_h|\Psi_{vh}(G)}}{\expval*{\Psi_{vh}(G)|\Psi_{vh}(G)}}\\
&=\frac{\sum_{n,m} \widetilde{G}^*_{n} \widetilde{G}_{m}\expval*{\widetilde{\Psi}_v^n|O_v|\widetilde{\Psi}_v^m}\expval*{\widetilde{\vb h}^n|\widetilde{\vb h}^m}}{\sum_{n}|\tilde{G}_{n}|^2}\\
&=\sum_n p_n \expval*{\widetilde{\Psi}_v^n|O_v|\widetilde{\Psi}_v^n}\,,
\end{aligned}
\end{equation}
where $p_n=|\widetilde{G}_n|^2/(\sum_m |\widetilde{G}_m|^2)$.
Clearly, this expression simplifies if the relevant observable $O_v$ acts as a scalar in the ground-state eigenspace.
Unsurprisingly, the same result would be found by first tracing out the hidden units and then computing the expectation value $\expval{O_v} = \tr{\rho_v O_v}$ on the reduced density matrix for the visible units in Eq.~\eqref{eq: reduced_rho_visible}.

\subsection{Relation to the cRBM architecture and universality of DBQS}
\label{sec:extracting_pure_states}
Here, we show that applying a projection operator to the hidden partition collapses the visible spins into a pure state. 
Then, we demonstrate that specific projections reduce the RBQS and DBQS ansatzes to the cRBM and DBM wavefunctions of Refs.~\cite{Carleo_2017,Carleo_2018}, establishing that the DBQS framework generalizes these earlier models.

Given a pure state of the hidden spins
$\ket{\chi_h}$, 
applying the projection operator 
$P_\chi = \mathbb{I}_v \otimes \ket{\chi_h}\bra{\chi_h}$
to a generic state of the extended system $\ket{\psi_{vh}}$ yields
\begin{equation}
\begin{aligned}
&(\mathbb{I}_v\otimes\ket{\chi_h}\bra{\chi_h})\ket{\psi_{vh}}\\
&=\sum_{\vb v,\vb h}\ket{\vb v,\vb h}\bra{\vb v,\vb h}(\mathbb{I}_v\otimes\ket{\chi_h}\bra{\chi_h})\sum_{\vb v',\vb h'}\ket{\vb v',\vb h'}\expval{\vb v',\vb h'|\psi_{vh}}\\
&=\sum_{\vb v,\vb h,\vb v',\vb h'}\ket{\vb v,\vb h}\expval{\vb v|\vb v'}\expval{\vb h|\chi_h}\expval{\chi_h|\vb h'}\expval{\vb v',\vb h'|\psi_{vh}}\\
&=\sum_{\vb v,\vb h,\vb h'}\ket{\vb v,\vb h}\expval{\vb h|\chi_h}\expval{\chi_h|\vb h'}\expval{\vb v,\vb h'|\psi_{vh}}\\
&=\sum_{\vb v}\left[\sum_{\vb h'}\expval{\chi_h|\vb h'}\expval{\vb v,\vb h'|\psi_{vh}}\right]\ket{\vb v}\otimes\sum_{\vb h}\expval{\vb h|\chi_h}\ket{\vb h}\\
&=\sum_{\vb v}\left[\sum_{\vb h'}\expval{\chi_h|\vb h'}\expval{\vb v,\vb h'|\psi_{vh}}\right]\ket{\vb v}\otimes\ket{\chi_h}\,.
\end{aligned}
\end{equation}
Now consider the special case where \(\ket{\psi_{vh}} = \ket{\Psi_{RBQS}}\) is an (unnormalized) RBQS wavefunction as in Eq.~\eqref{eq:RBQS}, and 
\[\ket{\chi_h} = \ket{\rightarrow_h} = 
\sum_{\vb h} \ket{\vb h}\] is the (unnormalized) uniform superposition over all hidden configurations. Then, the projected state becomes
\begin{equation}
\begin{aligned}
&(\mathbb{I}_v\otimes\ket{\rightarrow_h}\bra{\rightarrow_h})\ket{\Psi_{RBQS}}\\
&=
\sum_{\vb v}\left[\sum_{\vb h'}\exp\left\{\vb v^T W \vb h' + \vb b^T \vb v + \vb c^T \vb h'\right\}\right]\ket{\vb v}\otimes\ket{\rightarrow_h}\\
&=\!\sum_{\vb v}\!\left[\!e^{\sum_{i=0}^{N_v} b_i v_i}  \prod_{j=0}^{N_h}\cosh\!{\left\{\! c_j \!+\! \sum_{i=0}^{N_v} v_i W_{ij}\!\right\}}\right]\!\ket{\vb v}\!\otimes\!\ket{\rightarrow_h}\!\,.
\end{aligned}
\end{equation}
Hence, the state in which the visible spins are found exactly matches the cRBM wavefunction introduced in Ref.~\cite{Carleo_2017}.
Similarly, projecting a DBQS wavefunction with two hidden layers onto the state \(\ket{\rightarrow_h}\) for the hidden units yields the DBM wavefunction of Ref.~\cite{Carleo_2018}. 
This establishes that both the cRBM and DBM can be viewed as specific subclasses of the RBQS and DBQS ansatzes, respectively.
Remarkably, since the DBM wavefunction family has been proven to be a universal approximator of quantum states~\cite{gao2017efficient}, the DBQS inherits this expressive power, while potentially representing complex quantum states more effectively.

\subsection{Dataset--free initialization of the DBQS parameters}
\label{sec:dbqs_init}

Typical uses of DBMs in machine learning rely on greedy layer‑wise pretraining~\cite{Goodfellow-et-al-2016}. 
In the NQA setting such pretraining yields no benefit: the first few optimization steps rapidly overwrite any pre‑trained weights, effectively undoing the layer‑wise initialization long before the algorithm approaches the hard part of the evolution.
To our knowledge, however, dataset‑free initialization of DBMs has not been systematically explored.
Motivated by Ref.~\cite{yasuda2025dataset}, which shows that Xavier initialization~\cite{pmlr-v9-glorot10a} improves training of canonical RBMs, we adopt a Xavier‑normal initialization scheme for the variational parameters: all biases are set to zero, and the real and imaginary parts of the weights connecting layer $(l)$ and layer $(l+1)$ are sampled independently from a normal distribution with $0$ mean and variance $a/(N_{(l)}+N_{(l+1)})$, where $N_{(l)}$ is the number of units in layer $l$ and $a$ is a gain factor (here fixed at $a=10^{-2}$).

If the initial DBQS parameters are drawn from $\mathcal{N}(0,\sigma^2)$, the ensemble‑averaged log‑amplitude $\mathbb{E}[\log\psi(\vb x)]$ is exactly zero for any spin configuration $\ket{\vb x}$, and the variance scales linearly with $\sigma$. Hence, for sufficiently small gain factor, the untrained network already provides a reasonable representation of the uniform superposition state for all spins (visible and hidden). This is chosen as the initialization of the variational DBQS wavefunction.

\subsection{Comparison of block Gibbs chains and Metropolis-Hastings chains}
\label{sec: autocorr_time}
In Markov Chain Monte Carlo (MCMC) sampling, successive samples are typically correlated because each new sample is generated from the previous one, rather than drawn independently from the target distribution. This autocorrelation reduces the efficiency of the sampling procedure by decreasing the number of effectively independent samples obtained from the chain. The degree of this dependence can be quantified using the integrated autocorrelation time (IAT), which provides a measure of the number of iterations required, on average, to produce an effectively independent sample.

Let $\{\vb x_t\}_{t=0}^{N-1}$ denote a stationary Markov chain, and consider a real-valued function of the chain state $f(\vb x)$.
The autocorrelation function at lag $t$ is defined as
\begin{equation}
\label{eq:ac_lag}
\rho_f(t) = \frac{\mathbb{E}[(f(\vb x_0)-\mu_f)(f(\vb x_t)-\mu_f)]}{\sigma_f^2}
\end{equation}
where $\mu_f$ and $\sigma_f^2$ denote the mean and variance of $f(x)$ on the chain.
The IAT can be estimated by
\begin{equation}
\tau_f = 1+2\sum_{t=1}^{+\infty}\rho_f(t)
\end{equation}
In practice, the infinite sum is truncated at a lag where $\rho_f(t)$ becomes negligible.
When considering multiple functions, one should consider the highest IAT across all functions to ensure reliable statistical estimates.
In the context of spin systems,
the functions of interest are typically the individual spins: $f_i(\vb x)=x_i$.

For MCMC algorithms employing local update rules, such as the single-spin flip Metropolis-Hastings (MH) chain, the autocorrelation time often scales at least linearly with the system size.
Specifically, if only a fraction $\alpha$ of the $N_s$ spins is updated in each iteration, then roughly $1/\alpha$ iterations are required for every spin to have an opportunity to be updated once, corresponding to one sweep over the system.
This local nature of the dynamics can lead to large autocorrelation times.
In Metropolis-Hastings algorithms, rejected proposals further increase correlations because the chain remains in the same state after a rejection, whereas Gibbs updates are always accepted by construction.

\begin{figure}[t]
\centering
\includegraphics[width=0.45\textwidth]{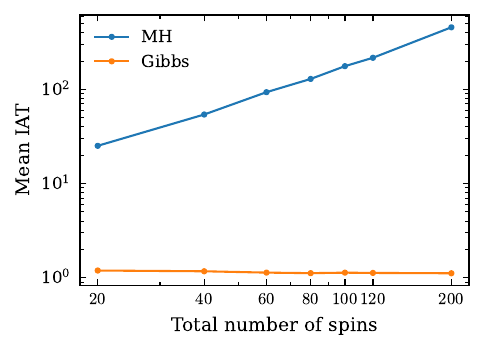}
\caption{Comparison of the magnetization IAT between the single-spin flip Metropolis-Hastings chain and the block Gibbs chain, in a log-log scale plot. Results are averages over $10$ realizations of initialized DBQS ansatzes with $3$ hidden layers and a hidden-unit density of $1$ per layer.
}
\label{fig:ac_times}
\end{figure}

To draw a quantitative comparison, we run both Gibbs and MH chains across several DBQS realizations as the system size increases. We measure the IAT for the magnetization along $z$ (on the computational basis) for each spin, and take the maximum value.
Figure~\ref{fig:ac_times} shows the maximum IAT for untrained DBQS with $3$ hidden layers and a hidden unit density per layer of $1$.
For each system size, we average over $10$ realizations of the dataset-free initialization described in the previous section.
As expected, we find that block Gibbs chains systematically converge faster and have autocorrelation times of order $O(1)$, whereas single-spin flip MH chains show IAT values of $O(N)$. Similar results are observed empirically during training of the variational DBQS wavefunction.

\section{Algorithmic details on Neural Quantum Annealing}
\label{sec: app_NQA}

This section provides supplementary technical details on NQA. In particular, we specify the exact implementation of our code and list all tunable hyperparameters, commenting on their effect on algorithmic behavior and computational cost. From a broad algorithmic perspective,
NQA is an advanced Monte Carlo method: its runtime scales polynomially, and it does not guarantee finding the exact solution in a strict mathematical sense. Despite the lack of performance guarantees, for many spin glass problems described in Section~\ref{sec: results}, it successfully finds the best or best-known solution, or a good approximation. As for any Monte Carlo algorithm, the quality of this approximation can be systematically improved through careful hyperparameter tuning.

\subsection{Algorithmic overview}
\label{sec:NQA_update}
As discussed in Section~\ref{sec:NQA}, NQA combines standard SR with momentum acceleration and persistent Gibbs chains enabled by our novel DBQS architecture.
The update step is shown in Algorithm~\ref{algo:NQA_step}, and recursively executed throughout the optimization. It comprises three actions:
\begin{enumerate}
\item \textbf{Sampler update}: Advance the persistent chains with a few block-Gibbs updates and draw a set of configurations;
\item \textbf{Natural-gradient evaluation}: Compute the natural gradient;
\item \textbf{Parameter update}: Update the variational parameters and the optimizer state.
\end{enumerate}

This update step is embedded into the full NQA optimization showcased in Algorithm~\ref{algo:NQA_optimization}, which comprises the following actions:
\begin{enumerate}
\item \textbf{Initialization:} The parameters of the DBQS are drawn from the Xavier-normal distribution as described in Section~\ref{sec:dbqs_init}. An ensemble of block-Gibbs chains is then initialized and thermalized.
\item \textbf{Pre-training:} The ansatz is first trained to approximate the ground state of the initial Hamiltonian at $s_0=0$.  When the initial Hamiltonian is chosen as $H_0=-\sum_i \sigma_i^x$, this step can be skipped because, as argued in Section~\ref{sec:dbqs_init}, the Xavier initialization already yields an accurate approximation of its ground state.
\item \textbf{Iterative annealing:} For each intermediate annealing point $s_j$, we perform a few NQA update steps. In practice, a single step for each value of $s_j$ ($j=1,\dots, N_A-1$) is usually sufficient to keep the state close to the variational adiabatic path and successfully approximate the final target state.
\item \textbf{Fine-tuning:} Finally, additional update steps are carried out at $s_{N_A}=1$ in order to suppress any residual excited-state components of the wavefunction.
\end{enumerate}

\begin{algorithm}[H]
\caption{NQA\_UpdateStep}
\label{algo:NQA_step}
\begin{algorithmic}[1]
\Require 
Variational parameters $\boldsymbol{\theta}$;
Persistent Gibbs chains $\mathcal{C}$;
Optimizer state $\alpha$.
\Ensure Updated parameters $\boldsymbol{\theta}$; Updated Gibbs chains $\mathcal{C}$; Updated optimizer state $\alpha$.
\State Update the Gibbs chains $\mathcal{C}$ by applying $N_\text{sweep}$ block Gibbs updates, and sample a set of configurations
\State Compute the natural gradient $\vb g \gets \vb g_{\boldsymbol{\theta}} E(\boldsymbol{\theta})$
\State Compute the updates for the parameters and the new optimizer state $(\delta\boldsymbol{\theta},\alpha)\gets \text{opt}(\vb g,\alpha)$
\State Update the variational parameters $\boldsymbol{\theta} \gets \boldsymbol{\theta} + \delta \boldsymbol{\theta}$
\State \Return $\boldsymbol{\theta}$, $\mathcal{C}$, $\alpha$
\end{algorithmic}
\end{algorithm}

\begin{algorithm}[H]
\caption{NQA for ground-state search}
\label{algo:NQA_optimization}
\begin{algorithmic}[1]
\Require 
Number of warm-up steps $N_{W}$;
Number of annealing steps $N_{A}$;
Number of updates per annealing step $N_{U}$;
Number of fine-tuning steps $N_{F}$;
Optimizer $\text{opt}(\vb g,\alpha)$ (natural gradients with momentum);
Variational ansatz $\psi(\vb x;\boldsymbol{\theta})$ (DBQS).
\Ensure  Optimized parameters $\boldsymbol{\theta}$
\State Initialize parameters $\boldsymbol{\theta} \gets \boldsymbol{\theta}_0$ \Comment{Initialization}
\State Initialize auxiliary state of the optimizer $\alpha \gets \alpha_0$
\State Initialize and thermalize Gibbs Chains $\mathcal{C}$
\State $H(s)\gets H_0$
\For{$n = 0$ \textbf{to} $N_{W}-1$}\Comment{Pretraining}
\State $\boldsymbol{\theta}, \mathcal{C}, \alpha \gets \textrm{NQA\_UpdateStep}(\boldsymbol{\theta}, \mathcal{C}, \alpha)$ 
\EndFor
\For{$j = 1$ \textbf{to} $N_{A}-1$}\Comment{Iterative annealing}
\State $s\gets j/N_A$
\State $H(s)\gets s H_T + (1-s) A H_0 + s(1-s)BH_C$
\For{$n = 0$ \textbf{to} $N_{U}-1$}
\State $\boldsymbol{\theta}, \mathcal{C}, \alpha \gets \textrm{NQA\_UpdateStep}(\boldsymbol{\theta}, \mathcal{C}, \alpha)$ 
\EndFor
\EndFor
\State $H(s)\gets H_T$
\For{$n = 0$ \textbf{to} $N_{F}-1$}\Comment{fine-tuning}
\State $\boldsymbol{\theta}, \mathcal{C}, \alpha \gets \textrm{NQA\_UpdateStep}(\boldsymbol{\theta}, \mathcal{C}, \alpha)$ 
\EndFor
\State \Return $\boldsymbol{\theta}$
\end{algorithmic}
\end{algorithm}

\subsection{NQA Hyperparameters}

In the following, we enumerate the tunable hyperparameters of NQA, organizing them into five groups based on their role in the algorithm.

\subsubsection{Annealing Hyperparameters}
These parameters characterize the annealing schedule and the parametric Hamiltonian $H(s)$ in Eq.~\eqref{eq:Hs_cata}.
\begin{itemize}
\item Number of annealing steps $N_A$: it fixes the number of discretized values of the annealing parameter along the schedule, precisely given by $0=s_0 < s_1 < \cdots < s_{\tiny{N_A}}=1$.
\item Number of warm-up update steps $N_W$: number of \texttt{\small NQA\_UpdateSteps} to perform for $s_0=0$ to prepare the initial ground-state.
\item Number of updates steps per annealing step $N_U$: number of \texttt{\small NQA\_UpdateSteps} performed for each intermediate value of $s_j$, for $j=1,\dots,N_A-1$. Using a number of updates per step equal to one is sufficient for most classical and quantum applications, by selecting a sufficiently large number of annealing steps $N_A$, as anticipated in Section~\ref{sec:NQA}.
\item Number of fine-tuning update steps $N_F$: number of \texttt{\small NQA\_UpdateSteps} performed at $s_{N_A}=1$. Although this final refinement does not seem to yield any performance improvement on classical problems, it empirically improves the performance for quantum ground states.
\item Annealing field scale $A>0$: multiplicative prefactor of the initial Hammiltonian $H_0$. Typically, during the annealing dynamics, the system undergoes a phase transition (a crossover, for finite systems) from a paramagnetic to a spin glass phase (additional phase transitions, such as perturbative crossings, may also occur). Heuristically, NQA is expected to work better when such a transition occurs somewhat close to $s\approx1/2$. Indeed, if the transition happens for $s\approx 0$, there is almost no difference between NQA and standard NQS optimization on the target problem Hamiltonian, since the hard regime is immediately reached. On the contrary, when the transition happens for $s\approx 1$, the variational ansatz may not have enough time to relax on the target ground-state in the hard regime, and computational resources may be wasted in the easy regime. Correctly tuning $A$ such that the transition occurs neither too soon nor too late during the annealing can be achieved automatically through hyperparameter optimization. 
\item Catalyst field scale $B$: multiplicative prefactor of $H_C$, included for the same reason as the annealing field scale. If we explicitly fix $B=0$, no catalyst is utilized.
\end{itemize}

\subsubsection{Optimizer Hyperparameters}
As stated in the main text, we use natural gradients with momentum.
Hence, the optimizer hyperparameters are the learning rate $\eta>0$, momentum $0\le\mu<1$, and diagonal shift $\lambda_d$. The diagonal shift does not appear to be critical for the algorithm's performance. On the contrary, selecting optimal learning rate and momentum values is crucial and should be done automatically via hyperparameter optimization, as the optimal values strongly depend on the target problem and on the NQS architecture.

\subsubsection{DBQS Hyperparameters}
\begin{itemize}
\item Number of hidden layers $N_L$: number of hidden layers in the DBQS architecture defined in Section~\ref{sec:BQS}.
\item Hidden-units density per layer $\alpha$: ratio between the number of hidden spins in each hidden layer (assumed to be layer-independent) and the number of visible (physical) spins in the visible layer. 
\item Parameters type (real or complex): whether to use real or complex variational parameters $\boldsymbol{\theta}$. Real parameters can be used for stoquastic Hamiltonians as the instantaneous ground state can be represented by real positive amplitudes. Using real parameters enforces this inductive bias, reduces the ansatz's capacity, and may yield lower computational cost and runtime. 
\end{itemize}

\subsubsection{Block Gibbs sampler Hyperparameters}
\begin{itemize}
\item Number of samples $N_\text{samples}$: number of samples to generate to compute observables, such as the local energy.
\item Number of chains $N_\text{chains}$: number of persistent block Gibbs chains to run in parallel. We recommend having one chain per sample, if possible, as this yields a faster runtime and guarantees that the samples are uncorrelated.
\item Number of thermalization steps $N_\text{therm}$: number of Gibbs chain steps to thermalize the chains at the beginning of the algorithm. (Recommended values may be $100-1000$).
\item Number of sweep updates $N_\text{sweeps}$: number of block Gibbs updates to perform in between drawing successive samples from a given chain. (Recommended $10-100$).
\end{itemize}


\subsection{Computational complexity}
The computational complexity of NQA is linear in the number of calls of the \texttt{\small NQA\_UpdateStep} function, which amounts to $N_W +  N_F + N_AN_U$.
The first two terms in the sum are usually negligible, as is the computational cost of the rest of the code; hence, the overall computational cost is essentially proportional to $N_AN_U$ times the computational cost of a \texttt{\small NQA\_UpdateStep} function call.
The latter comprises three contributions; we will now analyze them one by one.

First, propagating each persistent block Gibbs chain for $N_{\text{sweeps}}$ sweeps through a DBQS with $N_{L}$ hidden layers requires $O(N_{\text{sweeps}}N_L)$ matrix multiplications of matrices of size $O(N_v)$ times a vector of size $O(N_v)$. This operation has computational complexity of $O(N_v^2)$, for a total complexity of $O(N_{\text{sweeps}}N_L N_v^{2})$, and updating all the chains costs $O(N_{\text{samples}}N_{\text{sweeps}}N_L N_v^{2})$.

Secondly, the complexity of computing the natural gradients via SR scales as $O(N_{\text{params}}^{3})$ or via min--SR $O(N_\textrm{params}N_\textrm{samples}^2 + N_\textrm{samples}^3)$~\cite{chen2024empowering} for the matrix pseudoinverse.
Since a DBQS has number of parameters $O(N_{\text{spins}}^{2} N_{\text{layers}})$, the SR route costs $O(N_{\text{spins}}^{6} N_{\text{layers}}^{3})$, and min--SR reduces this to $O(N_{\text{spins}}^{2} N_{\text{layers}} N_{\text{samples}}^{2} + N_{\text{samples}}^{3})$. 
There is also the calculation of the logarithmic derivatives, which is just the computation of averages over the previously generated samples, at a cost of $O(N_\textrm{params}N_\textrm{samples})$. Concerning the local energies, as mentioned in the main text, they involve a matrix multiplication of $O(N_v^2)$, and a scalar product $O(N_v)$.
Total cost via min--SR is $O(N_\textrm{params}N_\textrm{samples}^2 + N_\textrm{samples}^3)$

Thirdly, evaluating the local energy and related observables is $O(N_{\text{spins}}^{2})$, while the subsequent momentum--SGD update of the parameters is negligible. 

Collecting all terms, the per-step complexity for the practical (min--SR) implementation is
\begin{equation}
O(N_{\text{spins}}^{2} N_{\text{layers}} (N_{\text{samples}}^{2}+N_{\text{sweeps}}) + N_{\text{samples}}^{3}),
\end{equation}
which remains tractable for the modest $N_{\text{samples}}, N_{\text{sweeps}} \ll N_{\text{spins}}^{2}$ typically required in practice.

With carefully chosen settings, NQA can provide exact solutions for instances of spin glasses with hundreds of spins
on a consumer laptop with a single NVIDIA RTX3060m GPU.


\begin{table}[h]
\centering
\caption{HPO for the SK benchmark with cRBM and RBQS.}
\begin{tabular}{llll}
\toprule
\textbf{Hyperparameter} & \textbf{Type} & \textbf{Sampling} & \textbf{Range} \\
Learning rate                    & float & log     & $[10^{-4}, 1]$ \\
Momentum                         & float & linear  & $[0.0, 0.9]$ \\
\label{tab:sk_hpo_1}
\end{tabular}

\vspace{0.5ex}
\footnotesize \emph{Note:} “log” indicates log-uniform sampling for hyperparameters spanning multiple orders of magnitude.
\end{table}

\begin{table}[h]
\centering
\caption{Fixed settings for the SK benchmark with cRBM and RBQS.}
\begin{tabular}{ll}
\toprule
\textbf{Setting} & \textbf{Value} \\
Number of warm-up update steps                 & 1 \\
Number of update steps per annealing step      & 1 \\
Number of fine-tuning steps                     & 1 \\
Number of thermalization steps                 & 1024 \\
Number of annealing steps          & $100000$ \\
Catalyst field scale (if present)  & $1$ \\
Annealing field scale              & $1$ \\
Number of hidden layers            & $1$ \\
Unit density per layer             & $1$ \\
Number of samples                  & $128$ \\
Number of sweep updates            & $128$ \\
Diagonal shift                     & $10^{-4}$ \\
\label{tab:sk_hpo_2}
\end{tabular}
\end{table}

\begin{table}[h]
\centering
\caption{Search spaces for the SK benchmark with DBQS ansatz.}
\begin{tabular}{llll}
\toprule
\textbf{Hyperparameter} & \textbf{Type} & \textbf{Sampling} & \textbf{Range} \\
Number of annealing steps        & int   & log     & $[10^{3}, 10^{6}]$ \\
Annealing field scale            & float & log     & $[0.1, 10]$ \\
Catalyst field scale             & float & log     & $[0.1, 10]$ \\
Number of hidden layers          & int   & linear  & $[1, 3]$ \\
Unit density per layer           & float & log     & $[0.25, 4]$ \\
Number of samples                & int   & log     & $[8, 128]$ \\
Number of sweep updates            & int   & log     & $[4, 128]$ \\
Learning rate                    & float & log     & $[10^{-4}, 1]$ \\
Momentum                         & float & linear  & $[0.0, 0.9]$ \\
Diagonal shift                   & float & log     & $[10^{-9}, 1]$ \\
\label{tab:sk_hpo_3}
\end{tabular}

\vspace{0.5ex}
\footnotesize \emph{Note:} “log” indicates log-uniform sampling for hyperparameters spanning multiple orders of magnitude. The number of Markov chains is set equal to the number of samples.
\end{table}

\begin{table}[h]
\centering
\caption{Fixed settings for the SK benchmark with DBQS ansatz.}
\begin{tabular}{ll}
\toprule
\textbf{Setting} & \textbf{Value} \\
Number of warm-up update steps                 & 1 \\
Number of update steps per annealing step      & 1 \\
Number of fine-tuning steps                     & 1 \\
Number of thermalization steps                 & 128 \\
\label{tab:sk_hpo_4}
\end{tabular}
\end{table}

\begin{table}[h]
\centering
\caption{HPO for the JSSP benchmark.}
\begin{tabular}{llll}
\toprule
\textbf{Hyperparameter} & \textbf{Type} & \textbf{Sampling} & \textbf{Range} \\
Number of annealing steps        & int   & log     & $[10^{4}, 10^{6}]$ \\
Annealing field scale            & float & log     & $[0.1, 10]$ \\
Catalyst field scale             & float & log     & $[0.1, 10]$ \\
Unit density per layer           & float & log     & $[0.25, 4]$ \\
Number of samples                & int   & log     & $[8, 128]$ \\
Number of sweep updates            & int   & log     & $[4, 128]$ \\
Learning rate                    & float & log     & $[10^{-3}, 1]$ \\
Momentum                         & float & linear  & $[0.0, 0.9]$ \\
Diagonal shift                   & float & log     & $[10^{-9}, 1]$ \\
\label{tab:jssp_hpo_1}
\end{tabular}

\vspace{0.5ex}
\footnotesize \emph{Note:} “log” indicates log-uniform sampling for hyperparameters spanning multiple orders of magnitude. The number of Markov chains is set equal to the number of samples.
\end{table}

\begin{table}[h]
\centering
\caption{Fixed settings for the JSSP benchmark.}
\begin{tabular}{ll}
\toprule
\textbf{Setting} & \textbf{Value} \\
Number of hidden layers                        & 2 \\
Number of warm-up update steps                 & 10 \\
Number of update steps per annealing step      & 1 \\
Number of fine-tuning steps                     & 10 \\
Number of thermalization steps                 & 128 \\
\label{tab:jssp_hpo_2}
\end{tabular}
\end{table}

\begin{table}[h]
\centering
\caption{Search spaces for N = 16 transverse-field SK benchmark HPO.}
\begin{tabular}{llll}
\toprule
\textbf{Hyperparameter} & \textbf{Type} & \textbf{Sampling} & \textbf{Range} \\
Number of annealing steps        & int   & log     & $[10^{3}, 10^{6}]$ \\
Unit density per layer           & float & log     & $[0.25, 4]$ \\
Number of sweep updates            & int   & log     & $[4, 128]$ \\
Number of samples            & int   & log     & $[128, 1024]$ \\
Learning rate                    & float & log     & $[10^{-4}, 1]$ \\
Momentum                         & float & linear  & $[0.0, 0.9]$ \\
\end{tabular}

\vspace{0.5ex}
\footnotesize \emph{Note:} “log” indicates log-uniform sampling for hyperparameters spanning multiple orders of magnitude. The number of Markov chains is set equal to the number of samples.
\label{tab:tfsk_hpo_16_v}
\end{table}

\begin{table}[h]
\centering
\caption{Fixed settings for the N = 16 transverse-field SK benchmark.}
\begin{tabular}{ll}
\toprule
\textbf{Setting} & \textbf{Value} \\
Number of warm-up update steps & $10$ \\
Number of update steps per annealing steps & $1$ \\
Number of fine-tuning steps & $100$ \\
Number of thermalization steps & $128$ \\
Annealing field scale & $1$ \\
Catalyst field scale & $1$ \\
Diagonal shift & $0.001$
\end{tabular}
\label{tab:tfsk_hpo_16_c}
\end{table}

\begin{table}[h]
\centering
\caption{Search spaces for N = 100 transverse-field SK benchmark HPO.}
\begin{tabular}{llll}
\toprule
\textbf{Hyperparameter} & \textbf{Type} & \textbf{Sampling} & \textbf{Range} \\
Number of annealing steps        & int   & log     & $[10^{3}, 10^{6}]$ \\
Unit density per layer           & float & log     & $[0.25, 4]$ \\
Number of sweep updates            & int   & log     & $[4, 128]$ \\
Learning rate                    & float & log     & $[10^{-4}, 1]$ \\
Momentum                         & float & linear  & $[0.0, 0.9]$ \\
\end{tabular}

\vspace{0.5ex}
\label{tab:tfsk_hpo_v}
\end{table}

\begin{table}[h]
\centering
\caption{Fixed settings for the N = 100 transverse-field SK benchmark.}
\begin{tabular}{ll}
\toprule
\textbf{Setting} & \textbf{Value} \\
Number of warm-up update steps & $10$ \\
Number of update steps per annealing steps & $1$ \\
Number of fine-tuning steps & $100$ \\
Number of thermalization steps & $128$ \\
Number of samples
      & 128   \\
Annealing field scale & $1$ \\
Catalyst field scale & $1$ \\
Diagonal shift & $0.001$
\end{tabular}
\label{tab:tfsk_hpo_c}
\end{table}

\section{Additional information on the hyperparameter optimization (HPO)}
\label{app:hpo}

Identifying optimal hyperparameters for NQA is challenging, as the hyperparameters detailed in Appendix~\ref{sec: app_NQA} are non-trivially correlated in their effect on algorithmic performance. The optimal choice depends on the problem at hand, and as for most machine learning algorithms, there is no universal set of hyperparameters that works every time for every problem.
To find the optimal hyperparameters for a given problem, we use Optuna~\cite{optuna_2019}, a framework for black box optimization commonly used in the machine learning community for hyperparameter tuning.
The specific algorithms we use are the Tree-structured Parzen Estimator (TPE)~\cite{bergstra2011algorithms} and the Covariance Matrix Adaptation Evolution Strategy (CMA-ES)~\cite{hansen2016cma}, depending on the computational budget and the parallelization capabilities available.



We recommend setting a fixed budget for the total number of trials and sampling the first $\sim80\%$ with CMA-ES and the remaining $\sim20\%$ with TPE. This will generate sufficient statistics to support claims of optimality with actual data.
Moreover, we recommend setting a wall-clock time limit per trial to prevent the HPO from selecting hyperparameter configurations that lead to ever-increasing computational complexity.

\subsection{HPO for SK benchmarks}
The HPO for the SK benchmark instances with cRBM and RBQS ansatzes are performed by only optimizing the learning rate and momentum, as shown in Table~\ref{tab:sk_hpo_1}, whereas all other hyperparameters are fixed as reported in Table~\ref{tab:sk_hpo_2}. Here, no maximum time limit per run is set, as each run requires the same runtime within this simplified HPO.
The HPO for the DBQS ansatz is reported in Table~\ref{tab:sk_hpo_3}, with fixed settings in Table~\ref{tab:sk_hpo_4}. Here, we implement a 2-hour wall-clock time limit per trial to avoid computationally inefficient hyperparameter combinations; successful runs usually take $\leq30$ minutes of wall-clock time on our hardware.
In both cases, new trial settings are sampled via TPE.
Note that the RBQS can function with a significantly smaller number of sweep updates than the cRBM; the large number reported in Table~\ref{tab:sk_hpo_2} was chosen to give the cRBM and RBQS ansatz the same HPO setting to obtain a fair comparison.

\subsection{HPO for JSSP benchmarks}
The ranges and settings for the HPO studies on the JSSP benchmark instances are reported in Table~\ref{tab:jssp_hpo_1} and Table~\ref{tab:jssp_hpo_2}.
New trial settings are sampled via CMA-ES; here, we implement a 4-hour wall-clock time limit per trial to prune computationally inefficient hyperparameter combinations. 
Successful runs take $5-50$ minutes of wall-clock time on our hardware.

\subsection{HPO for transverse-field SK benchmarks}
The ranges and settings for the HPO studies on the transverse-field SK benchmark instances with $N=16$ are reported in Table~\ref{tab:tfsk_hpo_16_v} and Table~\ref{tab:tfsk_hpo_16_c}, and those for the $N=100$ benchmarks are reported in Table~\ref{tab:tfsk_hpo_v} and Table~\ref{tab:tfsk_hpo_c}.
The first 100 trials are sampled via CMA-ES, and the last 50 via TPE.
Here, we implement a 1-hour wall-clock time limit per trial to prune computationally inefficient hyperparameter combinations.

\section{Additional data on the SK benchmark}
\label{app:benchmak_sk}

\begin{figure*}[ht]
\centering
\includegraphics[width=.9\linewidth]{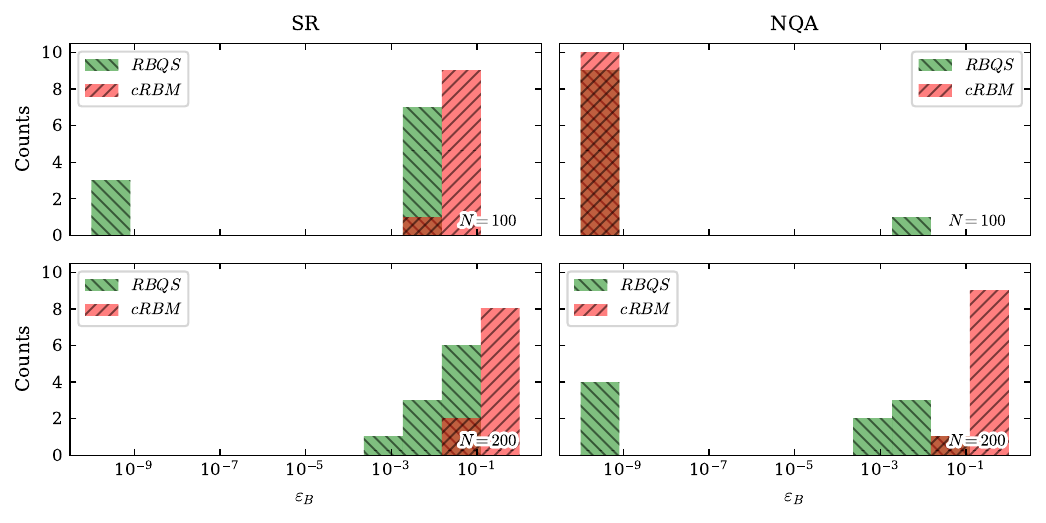} 
\caption{Residual energy error histograms for each combination of variational ansatz and optimization strategy on the benchmark instances with $N=100$ (top panels) and $N=200$ (bottom panels) spins. Left panels show the comparison of the final energies of real parameterized cRBM and RQBS ansatz when optimized via SR. The right panels showcase the same comparison when the ansatzes are optimized via NQA.
The leftmost bars correspond to instances solved exactly.
}
\label{fig:sk1}
\end{figure*}

\begin{figure*}[ht]
\centering
\includegraphics[width=.9\linewidth]{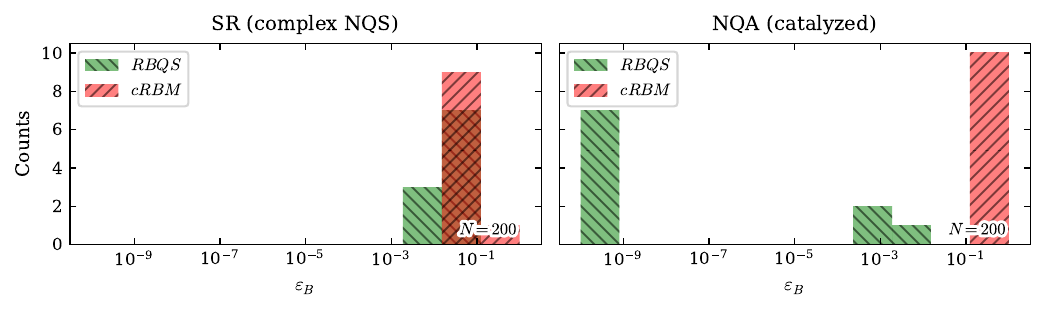} 
\caption{Residual energies histogram for each combination of variational ansatz and optimization strategy for the 10 realizations of the SK model with $N=200$ spins. 
Left panel shows the comparison of the final energies of complex parameterized cRBM and RQBS ansatz when optimized via SR. Right panel showcases the same comparison when the ansatz are optimized via NQA, with the inclusion of a catalyst.
Leftmost bars corresponds to instances solved exactly.
}
\label{fig:sk2}
\end{figure*}

Table~\ref{tab:typical-energies} reports the typical energy errors on the $N=200$ benchmarks for all the combinations explored. Figure~\ref{fig:sk1} and Figure~\ref{fig:sk2} show the energy error histograms for the set-ups not shown in the main text.

\begin{table}[ht]
\centering
\caption{Typical energy errors on the $N=200$ SK benchmark.}
\renewcommand{\arraystretch}{1.2}
\begin{tabular}{|c|c|c|}
\hline
\textbf{Method} & $[\varepsilon_B]_{\rm typ}$ & $[\varepsilon_Q]_{\rm typ}$ \\\hline
cRBM-SR            & $1.353\times10^{-1}$ & $1.360\times10^{-1}$ \\\hline
cRBM-SR-cata       & $9.661\times10^{-2}$ & $9.685\times10^{-2}$ \\\hline
cRBM-NQA           & $1.653\times10^{-1}$ & $1.687\times10^{-1}$ \\\hline
cRBM-NQA-cata      & $3.207\times10^{-1}$ & $3.280\times10^{-1}$ \\\hline
RBQS-SR            & $1.362\times10^{-2}$ & $1.371\times10^{-2}$ \\\hline
RBQS-SR-cata       & $1.693\times10^{-2}$ & $1.752\times10^{-2}$ \\\hline
RBQS-NQA           & $2.997\times10^{-6}$ & $2.599\times10^{-5}$ \\\hline
RBQS-NQA-cata      & $1.256\times10^{-8}$ & $8.938\times10^{-6}$ \\\hline
\textbf{DBQS-NQA-cata}      & $2.807\times10^{-9}$ & $2.881\times10^{-9}$ \\\hline
\end{tabular}
\label{tab:typical-energies}
\end{table}


\section{Additional information on the transverse-field SK benchmark}
\label{app:tfsk_extra}

\subsection{Additional results on the benchmark}
\label{app:tfsk_add_bench}

The energies obtained for the $N=16$ benchmark are reported in Table~\ref{tab:tfsk_hpo_16}. 
Table~\ref{tab:tfsk_hpo_100} shows the results for our benchmark instances with $N=100$ spins.
Additional figures showing the data for all the examined instances are available at~\cite{zenodo_repo} or~\cite{nqa_server_github} in the \texttt{Paper/Figures/FiguresExtra}.
To improve the accuracy of the reported energies, the final state obtained via NQA is resampled to generate $65536$ samples, which are then used to compute the reported energies.
Reported uncertainties are the usual Monte Carlo error $\delta=\sqrt{\frac{\rm Var}{N_{\rm samples}}}$.

\begin{table}[h]
\centering
\caption{Comparison between classical ground-state energy and the best energy found via NQA for the $N=100$ benchmark.}
\renewcommand{\arraystretch}{1.2}
\begin{tabular}{|c|c|c|c|c|}
\hline
Instance & classical energy & NQA \\ \hline
0 & -75.2018 & -75.7082 $\pm$ 0.0005 \\ 
\hline
1 & -71.8734 & -71.6367 $\pm$ 0.0005 \\ 
\hline
2 & -75.0439 & -75.4007 $\pm$ 0.002 \\ 
\hline
3 & -75.3180 & -75.8620 $\pm$ 0.0006 \\ 
\hline
4 & -79.6096 & -80.0250 $\pm$ 0.0003 \\ 
\hline
5 & -73.0136 & -73.5789 $\pm$ 0.0007 \\ 
\hline
6 & -74.6966 & -75.1646 $\pm$ 0.0004 \\ 
\hline
7 & -72.8547 & -73.1961 $\pm$ 0.003 \\ 
\hline
8 & -71.9467 & -72.4551 $\pm$ 0.0005 \\ 
\hline
9 & -73.3053 & -73.7775 $\pm$ 0.001 \\ 
\hline
\end{tabular}
\label{tab:tfsk_hpo_16}
\end{table}

\begin{table}[h]
\centering
\caption{Comparison between ED and NQA for the $N=16$ benchmark.}
\renewcommand{\arraystretch}{1.2}
\begin{tabular}{|c|c|c|c|c|}
\hline
Instance & ED & NQA \\ \hline
0 & -16.8595 & -16.8589 $\pm$ 0.0002 \\ 
\hline
1 & -17.0621 & -17.0600 $\pm$ 0.0002 \\ 
\hline
2 & -19.8560 & -19.8557 $\pm$ 0.0001 \\ 
\hline
3 & -18.45559 & -18.4555 $\pm$ 0.0001 \\ 
\hline
4 & -17.3565 & -17.3515 $\pm$ 0.0002 \\ 
\hline
5 & -23.21253 & -23.21250 $\pm$ 0.00007 \\ 
\hline
6  & -18.7474 & -18.7471 $\pm$ 0.0002 \\ 
\hline
7 & -18.6813 & -18.6807 $\pm$ 0.0002 \\ 
\hline
8 & -16.8657 & -16.8652 $\pm$ 0.0001 \\ 
\hline
9 & -18.5272 & -18.5186 $\pm$ 0.0003 \\ 
\hline
\end{tabular}
\label{tab:tfsk_hpo_100}
\end{table}

\bibliography{sample}

@article{PhysRevLett.35.1792,
  title = {Solvable Model of a Spin-Glass},
  author = {Sherrington, David and Kirkpatrick, Scott},
  journal = {Phys. Rev. Lett.},
  volume = {35},
  pages = {1792},
  year = {1975},
  doi = {10.1103/PhysRevLett.35.1792},
  url = {https://link.aps.org/doi/10.1103/PhysRevLett.35.1792}
}

@article{Carleo_2018,
  title = {Constructing exact representations of quantum many-body systems with deep neural networks},
  author = {Carleo, Giuseppe and Nomura, Yusuke and Imada, Masatoshi},
  journal = {Nature Communications},
  volume = {9},
  number = {1},
  pages = {5322},
  year = {2018},
  publisher = {Nature Publishing Group UK London},
  doi = {10.1038/s41467-018-07520-3},
  url = {http://dx.doi.org/10.1038/s41467-018-07520-3}
}

@article{barahona1982computational,
  title = {On the computational complexity of Ising spin glass models},
  author = {Barahona, F.},
  journal = {Journal of Physics A: Mathematical and General},
  volume = {15},
  pages = {3241},
  year = {1982},
  doi = {10.1088/0305-4470/15/10/028},
  url = {https://doi.org/10.1088/0305-4470/15/10/028}
}

@article{Lucas2014,
  title = {Ising formulations of many NP problems},
  author = {Lucas, Andrew},
  journal = {Frontiers in Physics},
  volume = {2},
  pages = {5},
  year = {2014},
  doi = {10.3389/fphy.2014.00005},
  url = {https://www.frontiersin.org/articles/10.3389/fphy.2014.00005}
}

@article{DACOL2022100249,
title = {Industrial-size job shop scheduling with constraint programming},
journal = {Oper. Res. Perspect.},
volume = {9},
pages = {100249},
year = {2022},
issn = {2214-7160},
author = {G Da Col and E C Teppan}
}

@article{jssp_industry,
  title={Review of job shop scheduling research and its new perspectives under Industry 4.0},
  author={Zhang, Jian and Ding, Guofu and Zou, Yisheng and Qin, Shengfeng and Fu, Jianlin},
  journal={J. Intell. Manuf.},
  volume={30},
  pages={1809--1830},
  year={2019},
  publisher={Springer}
}

@article{flexibleJSSP_review,
title = {The flexible job shop scheduling problem: A review},
journal = {European Journal of Operational Research},
volume = {314},
number = {2},
pages = {409-432},
year = {2024},
issn = {0377-2217},
doi = {https://doi.org/10.1016/j.ejor.2023.05.017},
url = {https://www.sciencedirect.com/science/article/pii/S037722172300382X},
author = {Stéphane Dauzère-Pérès and Junwen Ding and Liji Shen and Karim Tamssaouet}
}

@article{dynprog,
title = {Solving the job-shop scheduling problem optimally by dynamic programming},
journal = {Computers \& Operations Research},
volume = {39},
number = {12},
pages = {2968-2977},
year = {2012},
issn = {0305-0548},
doi = {https://doi.org/10.1016/j.cor.2012.02.024},
url = {https://www.sciencedirect.com/science/article/pii/S0305054812000500},
author = {Joaquim A.S. Gromicho and Jelke J. {van Hoorn} and Francisco Saldanha-da-Gama and Gerrit T. Timmer}
}

@article{Carleo_2017,
  title = {Solving the quantum many-body problem with artificial neural networks},
  author = {Carleo, Giuseppe and Troyer, Matthias},
  journal = {Science},
  volume = {355},
  pages = {602},
  year = {2017},
  doi = {10.1126/science.aag2302},
  url = {http://dx.doi.org/10.1126/science.aag2302}
}

@article{Amin_2018,
  title = {Quantum Boltzmann Machine},
  author = {Amin, Mohammad H. and Andriyash, Evgeny and Rolfe, Jason and Kulchytskyy, Bohdan and Melko, Roger},
  journal = {Physical Review X},
  volume = {8},
  pages = {021050},
  year = {2018},
  doi = {10.1103/PhysRevX.8.021050},
  url = {http://dx.doi.org/10.1103/PhysRevX.8.021050}
}

@article{Amin_2009,
  title = {First-order quantum phase transition in adiabatic quantum computation},
  volume = {80},
  number = {6},
  journal = {Physical Review A},
  publisher = {American Physical Society (APS)},
  author = {Amin, M. H. S. and Choi, V.},
  year = {2009},
  doi = {10.1103/PhysRevA.80.062326},
  url = {http://dx.doi.org/10.1103/PhysRevA.80.062326}
}

@article{King2025,
  author = {Andrew D. King and Alberto Nocera and Marek M. Rams and Jacek Dziarmaga and Roeland Wiersema and William Bernoudy and Jack Raymond and Nitin Kaushal and Niclas Heinsdorf and Richard Harris and Kelly Boothby and Fabio Altomare and Mohsen Asad and Andrew J. Berkley and Martin Boschnak and Kevin Chern and Holly Christiani and Samantha Cibere and Jake Connor and Martin H. Dehn and Rahul Deshpande and Sara Ejtemaee and Pau Farre and Kelsey Hamer and Emile Hoskinson and Shuiyuan Huang and Mark W. Johnson and Samuel Kortas and Eric Ladizinsky and Trevor Lanting and Tony Lai and Ryan Li and Allison J. R. MacDonald and Gaelen Marsden and Catherine C. McGeoch and Reza Molavi and Travis Oh and Richard Neufeld and Mana Norouzpour and Joel Pasvolsky and Patrick Poitras and Gabriel Poulin-Lamarre and Thomas Prescott and Mauricio Reis and Chris Rich and Mohammad Samani and Benjamin Sheldan and Anatoly Smirnov and Edward Sterpka and Berta Trullas Clavera and Nicholas Tsai and Mark Volkmann and Alexander M. Whiticar and Jed D. Whittaker and Warren Wilkinson and Jason Yao and T. J. Yi and Anders W. Sandvik and Gonzalo Alvarez and Roger G. Melko and Juan Carrasquilla and Marcel Franz and Mohammad H. Amin},
  title = {Beyond-classical computation in quantum simulation},
  journal = {Science},
  volume = {388},
  number = {6743},
  pages = {199},
  year = {2025},
  doi = {10.1126/science.ado6285},
  url = {https://www.science.org/doi/10.1126/science.ado6285}
}

@misc{tindall2025,
  title = {Dynamics of disordered quantum systems with two- and three-dimensional tensor networks},
  author = {Tindall, Joseph and Mello, Antonio and Fishman, Matt and Stoudenmire, Miles and Sels, Dries},
  year = {2025},
  eprint = {2503.05693},
  archivePrefix = {arXiv},
  primaryClass = {quant-ph},
  doi = {10.48550/arXiv.2503.05693},
  url = {https://arxiv.org/abs/2503.05693}
}

@misc{mauron2025,
  title = {Challenging the Quantum Advantage Frontier with Large-Scale Classical Simulations of Annealing Dynamics},
  author = {Mauron, Linda and Carleo, Giuseppe},
  year = {2025},
  eprint = {2503.08247},
  archivePrefix = {arXiv},
  primaryClass = {quant-ph},
  doi = {10.48550/arXiv.2503.08247},
  url = {https://arxiv.org/abs/2503.08247}
}

@misc{schulz2025,
  title = {Learning-Driven Annealing with Adaptive Hamiltonian Modification for Solving Large-Scale Problems on Quantum Devices},
  author = {Schulz, Sebastian and Willsch, Dennis and Michielsen, Kristel},
  year = {2025},
  eprint = {2502.21246},
  archivePrefix = {arXiv},
  primaryClass = {quant-ph},
  doi = {10.48550/arXiv.2502.21246},
  url = {https://arxiv.org/abs/2502.21246}
}

@book{mezard1987spin,
  title = {Spin Glass Theory And Beyond: An Introduction To The Replica Method And Its Applications},
  author = {Mezard, M. and Parisi, G. and Virasoro, M. A. and Hopfield, J. J.},
  isbn = {9789813103917},
  series = {World Scientific Lecture Notes In Physics},
  doi = {10.1142/0271},
  url = {https://books.google.it/books?id=DwY8DQAAQBAJ},
  year = {1987},
  publisher = {World Scientific Publishing Company}
}

@article{Hukushima_1996,
  title = {Exchange Monte Carlo Method and Application to Spin Glass Simulations},
  volume = {65},
  issn = {1347-4073},
  doi = {10.1143/JPSJ.65.1604},
  number = {6},
  journal = {Journal of the Physical Society of Japan},
  publisher = {Physical Society of Japan},
  author = {Hukushima, Koji and Nemoto, Koji},
  year = {1996},
  month = {jun},
  pages = {1604-1608}
}

@article{Nishimori_2017,
  title = {Exponential Enhancement of the Efficiency of Quantum Annealing by Non-Stoquastic Hamiltonians},
  volume = {4},
  issn = {2297-198X},
  doi = {10.3389/fict.2017.00002},
  journal = {Frontiers in ICT},
  publisher = {Frontiers Media SA},
  author = {Nishimori, Hidetoshi and Takada, Kabuki},
  year = {2017},
  month = {feb}
}

@article{Ebadi2022QuantumMIS,
  title = {Quantum optimization of maximum independent set using Rydberg atom arrays},
  author = {Ebadi, S. and Keesling, A. and Cain, M. and Wang, T. T. and Levine, H. and Bluvstein, D. and Semeghini, G. and Omran, A. and Liu, J.-G. and Lukin, M. D. and others},
  journal = {Science},
  volume = {376},
  number = {6598},
  pages = {1209-1215},
  year = {2022},
  month = {may},
  doi = {10.1126/science.abo6587},
  publisher = {American Association for the Advancement of Science}
}

@misc{ghosh2024,
  title = {Exponential speed-up of quantum annealing via n-local catalysts},
  author = {Ghosh, Roopayan and Nutricati, Luca A. and Feinstein, Natasha and Warburton, P. A. and Bose, Sougato},
  year = {2024},
  eprint = {2409.13029},
  archivePrefix = {arXiv},
  primaryClass = {quant-ph},
  doi = {10.48550/arXiv.2409.13029},
  url = {https://arxiv.org/abs/2409.13029}
}

@article{PhysRevB.107.165149,
  title = {Ground state search by local and sequential updates of neural network quantum states},
  author = {Zhang, Wenxuan and Xu, Xiansong and Wu, Zheyu and Balachandran, Vinitha and Poletti, Dario},
  journal = {Phys. Rev. B},
  volume = {107},
  issue = {16},
  pages = {165149},
  numpages = {6},
  year = {2023},
  month = {Apr},
  publisher = {American Physical Society},
  doi = {10.1103/PhysRevB.107.165149},
  url = {https://link.aps.org/doi/10.1103/PhysRevB.107.165149}
}

@Article{10.21468/SciPostPhys.12.5.166,
  title = {{Accuracy of restricted Boltzmann machines for the one-dimensional $J_1$-$J_2$ Heisenberg model}},
  author = {Viteritti, Luciano Loris and Ferrari, Francesco and Becca, Federico},
  journal = {SciPost Phys.},
  volume = {12},
  pages = {166},
  year = {2022},
  publisher = {SciPost},
  doi = {10.21468/SciPostPhys.12.5.166},
  url = {https://scipost.org/10.21468/SciPostPhys.12.5.166}
}

@article{10.1162/neco.2008.04-07-510,
  author = {Le Roux, Nicolas and Bengio, Yoshua},
  title = {Representational Power of Restricted Boltzmann Machines and Deep Belief Networks},
  journal = {Neural Computation},
  volume = {20},
  number = {6},
  pages = {1631-1649},
  year = {2008},
  month = {06},
  issn = {0899-7667},
  doi = {10.1162/neco.2008.04-07-510},
  url = {https://doi.org/10.1162/neco.2008.04-07-510}
}

@book{Goodfellow-et-al-2016,
  title = {Deep learning},
  author = {Goodfellow, Ian and Bengio, Yoshua and Courville, Aaron and Bengio, Yoshua},
  year = {2016},
  publisher = {MIT Press, Cambridge},
  url = {http://www.deeplearningbook.org}
}

@inproceedings{long2010restricted,
  author = {Long, Philip M. and Servedio, Rocco A.},
  title = {Restricted Boltzmann Machines are Hard to Approximately Evaluate or Simulate},
  year = {2010},
  isbn = {9781605589077},
  publisher = {Omnipress},
  address = {Madison, WI, USA},
  booktitle = {Proceedings of the 27th International Conference on Machine Learning (ICML)},
  pages = {703-710},
  numpages = {8},
  location = {Haifa, Israel},
  series = {ICML'10},
  url = {https://www.cs.columbia.edu/~rocco/Public/final-camera-ready-icml10.pdf}
}

@misc{patel2024quantumboltzmannmachinelearning,
  title = {Quantum Boltzmann machine learning of ground-state energies},
  author = {Patel, Dhrumil and Koch, Daniel and Patel, Saahil and Wilde, Mark M.},
  year = {2024},
  eprint = {2410.12935},
  archivePrefix = {arXiv},
  primaryClass = {quant-ph},
  doi = {10.48550/arXiv.2410.12935},
  url = {https://arxiv.org/abs/2410.12935}
}

@article{Casella01081992,
  author = {Casella, George and George, Edward I.},
  title = {Explaining the Gibbs Sampler},
  journal = {The American Statistician},
  volume = {46},
  number = {3},
  pages = {167-174},
  year = {1992},
  publisher = {American Statistical Association},
  doi = {10.1080/00031305.1992.10475878},
  url = {https://www.tandfonline.com/doi/abs/10.1080/00031305.1992.10475878}
}

@article{HJELM2014245,
  title = {Restricted Boltzmann machines for neuroimaging: An application in identifying intrinsic networks},
  journal = {NeuroImage},
  volume = {96},
  pages = {245-260},
  year = {2014},
  issn = {1053-8119},
  doi = {10.1016/j.neuroimage.2014.03.048},
  url = {https://www.sciencedirect.com/science/article/pii/S1053811914002080},
  author = {Hjelm, R. Devon and Calhoun, Vince D. and Salakhutdinov, Ruslan and Allen, Elena A. and Adali, Tulay and Plis, Sergey M.}
}

@article{multimodal_bm,
  title = {Multimodal learning with deep Boltzmann machines},
  author = {Srivastava, Nitish and Salakhutdinov, Russ R.},
  journal = {Advances in Neural Information Processing Systems},
  volume = {25},
  year = {2012}
}

@article{HE2021101871,
  title = {A neuron image segmentation method based Deep Boltzmann Machine and CV model},
  journal = {Computerized Medical Imaging and Graphics},
  volume = {89},
  pages = {101871},
  year = {2021},
  issn = {0895-6111},
  doi = {10.1016/j.compmedimag.2021.101871},
  url = {https://www.sciencedirect.com/science/article/pii/S0895611121000197},
  author = {He, Fuyun and Huang, Xiaoming and Wang, Xun and Qiu, Senhui and Jiang, F. and Ling, Sai Ho}
}

@article{chen2024empowering,
  title = {Empowering deep neural quantum states through efficient optimization},
  author = {Chen, Ao and Heyl, Markus},
  journal = {Nature Physics},
  volume = {20},
  number = {9},
  pages = {1476-1481},
  year = {2024},
  publisher = {Nature Publishing Group UK London},
  doi = {10.1038/s41567-024-02566-1},
  url = {https://www.nature.com/articles/s41567-024-02566-1}
}

@article{sorella1998green,
  title = {Green Function Monte Carlo with Stochastic Reconfiguration},
  author = {Sorella, Sandro},
  journal = {Phys. Rev. Lett.},
  volume = {80},
  number = {20},
  pages = {4558-4561},
  year = {1998},
  month = {May},
  publisher = {American Physical Society},
  doi = {10.1103/PhysRevLett.80.4558},
  url = {https://link.aps.org/doi/10.1103/PhysRevLett.80.4558}
}

@article{PhysRevB.108.054410,
  title = {High-accuracy variational Monte Carlo for frustrated magnets with deep neural networks},
  author = {Roth, Christopher and Szab\'o, Attila and MacDonald, Allan H.},
  journal = {Phys. Rev. B},
  volume = {108},
  issue = {5},
  pages = {054410},
  numpages = {12},
  year = {2023},
  month = {Aug},
  publisher = {American Physical Society},
  doi = {10.1103/PhysRevB.108.054410},
  url = {https://link.aps.org/doi/10.1103/PhysRevB.108.054410}
}

@article{PhysRevX.7.021021,
  title = {Quantum Entanglement in Neural Network States},
  author = {Deng, Dong-Ling and Li, Xiaopeng and Das Sarma, S.},
  journal = {Phys. Rev. X},
  volume = {7},
  issue = {2},
  pages = {021021},
  numpages = {17},
  year = {2017},
  month = {May},
  publisher = {American Physical Society},
  doi = {10.1103/PhysRevX.7.021021},
  url = {https://link.aps.org/doi/10.1103/PhysRevX.7.021021}
}

@article{PhysRevLett.125.100503,
  title = {Quantum Many-Body Dynamics in Two Dimensions with Artificial Neural Networks},
  author = {Schmitt, Markus and Heyl, Markus},
  journal = {Phys. Rev. Lett.},
  volume = {125},
  issue = {10},
  pages = {100503},
  numpages = {7},
  year = {2020},
  month = {Sep},
  publisher = {American Physical Society},
  doi = {10.1103/PhysRevLett.125.100503},
  url = {https://link.aps.org/doi/10.1103/PhysRevLett.125.100503}
}

@article{rattray1998natural,
  title = {Natural Gradient Descent for On-Line Learning},
  author = {Rattray, Magnus and Saad, David and Amari, Shun-ichi},
  journal = {Phys. Rev. Lett.},
  volume = {81},
  issue = {24},
  pages = {5461-5464},
  year = {1998},
  month = {Dec},
  publisher = {American Physical Society},
  doi = {10.1103/PhysRevLett.81.5461},
  url = {https://link.aps.org/doi/10.1103/PhysRevLett.81.5461}
}

@misc{lange2024nqsrev,
  title = {From Architectures to Applications: A Review of Neural Quantum States},
  author = {Lange, Hannah and Van de Walle, Anka and Abedinnia, Atiye and Bohrdt, Annabelle},
  year = {2024},
  eprint = {2402.09402},
  archivePrefix = {arXiv},
  primaryClass = {cond-mat.dis-nn},
  doi = {10.48550/arXiv.2402.09402},
  url = {https://arxiv.org/abs/2402.09402}
}

@article{torlai2018neural,
  title = {Neural-network quantum state tomography},
  author = {Torlai, Giacomo and Mazzola, Guglielmo and Carrasquilla, Juan and Troyer, Matthias and Melko, Roger and Carleo, Giuseppe},
  journal = {Nature Physics},
  volume = {14},
  number = {5},
  pages = {447-450},
  year = {2018},
  publisher = {Nature Publishing Group UK London},
  doi = {10.1038/s41567-018-0048-5}
}

@article{Finnila_CPL94,
  title = {Quantum annealing: A new method for minimizing multidimensional functions},
  journal = {Chemical Physics Letters},
  volume = {219},
  number = {5},
  pages = {343-348},
  year = {1994},
  issn = {0009-2614},
  doi = {10.1016/0009-2614(94)00117-0},
  url = {https://www.sciencedirect.com/science/article/pii/0009261494001170},
  author = {Finnila, A. B. and Gomez, M. A. and Sebenik, C. and Stenson, C. and Doll, J. D.}
}

@article{RevModPhys.80.1061,
  title = {Colloquium: Quantum annealing and analog quantum computation},
  author = {Das, Arnab and Chakrabarti, Bikas K.},
  journal = {Rev. Mod. Phys.},
  volume = {80},
  issue = {3},
  pages = {1061-1081},
  numpages = {0},
  year = {2008},
  month = {Sep},
  publisher = {American Physical Society},
  doi = {10.1103/RevModPhys.80.1061},
  url = {https://link.aps.org/doi/10.1103/RevModPhys.80.1061}
}

@article{Kadowaki_PRE98,
  title = {{Quantum annealing in the transverse Ising model}},
  author = {Kadowaki, Tadashi and Nishimori, Hidetoshi},
  journal = {Phys. Rev. E},
  volume = {58},
  issue = {5},
  pages = {5355-5363},
  numpages = {0},
  year = {1998},
  month = {Nov},
  publisher = {American Physical Society},
  doi = {10.1103/PhysRevE.58.5355}
}

@article{Santoro_SCI02,
  author = {Santoro, Giuseppe E. and Marto\v{n}\'{a}k, Roman and Tosatti, Erio and Car, Roberto},
  title = {Theory of Quantum Annealing of an Ising Spin Glass},
  journal = {Science},
  volume = {295},
  number = {5564},
  pages = {2427-2430},
  year = {2002},
  doi = {10.1126/science.1068774},
  url = {https://www.science.org/doi/abs/10.1126/science.1068774}
}

@article{Martonak_PRB2002,
  title   = {Quantum annealing by the path-integral Monte Carlo method: The two-dimensional random Ising model},
  author  = {Marton{\'a}k, Roman and Santoro, Giuseppe E. and Tosatti, Erio},
  journal = {Physical Review B},
  volume  = {66},
  number  = {9},
  pages   = {094203},
  year    = {2002},
  doi     = {10.1103/PhysRevB.66.094203}
}

@article{CrossonHarrow_PRA2016,
  title   = {Simulated Quantum Annealing can be exponentially faster than classical simulated annealing},
  author  = {Crosson, Elizabeth and Harrow, Aram W.},
  journal = {Physical Review A},
  volume  = {93},
  number  = {4},
  pages   = {042307},
  year    = {2016},
  doi     = {10.1103/PhysRevA.93.042307}
}

@ARTICLE{Santoro_JPA06,
  AUTHOR = {Santoro, Giuseppe E. and Tosatti, Erio},
  TITLE = {Optimization using Quantum Mechanics: Quantum annealing through adiabatic evolution},
  JOURNAL = {J. Phys. A: Math. Gen.},
  VOLUME = {39},
  YEAR = {2006},
  PAGES = {R393-R431},
  doi = {10.1088/0305-4470/39/36/R01}
}

@article{AQCfarhiNEW,
  author = {Farhi, Edward and Goldstone, Jeffrey and Gutmann, Sam and Lapan, Joshua and Lundgren, Andrew and Preda, Daniel},
  title = {{A Quantum Adiabatic Evolution Algorithm Applied to Random Instances of an NP-Complete Problem}},
  journal = {Science},
  volume = {292},
  number = {5516},
  pages = {472-475},
  year = {2001},
  doi = {10.1126/science.1057726},
  url = {https://www.science.org/doi/abs/10.1126/science.1057726}
}

@article{Albash_RMP18,
  title = {Adiabatic quantum computation},
  author = {Albash, Tameem and Lidar, Daniel A.},
  journal = {Rev. Mod. Phys.},
  volume = {90},
  issue = {1},
  pages = {015002},
  numpages = {64},
  year = {2018},
  month = {Jan},
  publisher = {American Physical Society},
  doi = {10.1103/RevModPhys.90.015002},
  url = {https://link.aps.org/doi/10.1103/RevModPhys.90.015002}
}

@article{Johnson2011,
  title = {Quantum annealing with manufactured spins},
  author = {Johnson, M. W. and Amin, M. H. S. and Gildert, S. and Lanting, T. and Hamze, F. and Dickson, N. and Harris, R. and Berkley, A. J. and Johansson, J. and Bunyk, P. and Chapple, E. M. and Enderud, C. and Hilton, J. P. and Karimi, K. and Ladizinsky, E. and Ladizinsky, N. and Oh, T. and Perminov, I. and Rich, C. and Thom, M. C. and Tolkacheva, E. and Truncik, C. J. S. and Uchaikin, S. and Wang, J. and Wilson, B. and Rose, G.},
  journal = {Nature},
  volume = {473},
  pages = {194--198},
  year = {2011},
  doi = {10.1038/nature10012}
}

@misc{glover2019tutorial,
  title = {A Tutorial on Formulating and Using QUBO Models},
  author = {Glover, Fred and Kochenberger, Gary and Du, Yu},
  year = {2019},
  eprint = {1811.11538},
  archivePrefix = {arXiv},
  primaryClass = {cs.DS},
  doi = {10.48550/arXiv.1811.11538}
}

@misc{minor_embedding,
  title = {A practical heuristic for finding graph minors},
  author = {Cai, Jun and Macready, William G. and Roy, Aidan},
  year = {2014},
  eprint = {1406.2741},
  archivePrefix = {arXiv},
  primaryClass = {quant-ph},
  doi = {10.48550/arXiv.1406.2741},
  url = {https://arxiv.org/abs/1406.2741}
}

@article{PhysRevE.84.061152,
  title = {Exponential complexity of the quantum adiabatic algorithm for certain satisfiability problems},
  author = {Hen, Itay and Young, A. P.},
  journal = {Phys. Rev. E},
  volume = {84},
  issue = {6},
  pages = {061152},
  numpages = {8},
  year = {2011},
  month = {Dec},
  publisher = {American Physical Society},
  doi = {10.1103/PhysRevE.84.061152},
  url = {https://link.aps.org/doi/10.1103/PhysRevE.84.061152}
}

@article{pnas.1002116107,
  author = {Altshuler, Boris and Krovi, Hari and Roland, J\'{e}r\'{e}mie},
  title = {Anderson localization makes adiabatic quantum optimization fail},
  journal = {Proceedings of the National Academy of Sciences},
  volume = {107},
  number = {28},
  pages = {12446-12450},
  year = {2010},
  doi = {10.1073/pnas.1002116107},
  url = {https://www.pnas.org/doi/abs/10.1073/pnas.1002116107}
}

@article{Knysh2016,
  author = {Knysh, Sergey},
  title = {Zero-temperature quantum annealing bottlenecks in the spin-glass phase},
  journal = {Nature Communications},
  volume = {7},
  number = {12370},
  year = {2016},
  doi = {10.1038/ncomms12370},
  url = {https://www.nature.com/articles/ncomms12370},
  publisher = {Nature Publishing Group}
}

@article{QA_industry,
  doi = {10.1088/1361-6633/ac8c54},
  url = {https://dx.doi.org/10.1088/1361-6633/ac8c54},
  year = {2022},
  month = {sep},
  publisher = {IOP Publishing},
  volume = {85},
  number = {10},
  pages = {104001},
  author = {Yarkoni, Sheir and Raponi, Elena and B{\"a}ck, Thomas and Schmitt, Sebastian},
  title = {Quantum annealing for industry applications: introduction and review},
  journal = {Reports on Progress in Physics}
}

@article{Hibat_Allah_2021,
  title = {Variational neural annealing},
  volume = {3},
  issn = {2522-5839},
  doi = {10.1038/s42256-021-00401-3},
  number = {11},
  journal = {Nature Machine Intelligence},
  publisher = {Springer Nature},
  author = {Hibat-Allah, Mohamed and Inack, Estelle M. and Wiersema, Roeland and Melko, Roger G. and Carrasquilla, Juan},
  year = {2021},
  month = {oct},
  pages = {952-961}
}

@misc{ranabhat2025,
      title={Large-scale portfolio optimization with variational neural annealing}, 
      author={Nishan Ranabhat and Behnam Javanparast and David Goerz and Estelle Inack},
      year={2025},
      eprint={2507.07159},
      archivePrefix={arXiv},
      primaryClass={cond-mat.dis-nn},
      url={https://arxiv.org/abs/2507.07159}, 
}

@Article{Lami_Torta,
  title = {{Quantum annealing for neural network optimization problems: A new approach via tensor network simulations}},
  author = {Lami, Guglielmo and Torta, Pietro and Santoro, Giuseppe E. and Collura, Mario},
  journal = {SciPost Phys.},
  volume = {14},
  pages = {117},
  year = {2023},
  publisher = {SciPost},
  doi = {10.21468/SciPostPhys.14.5.117},
  url = {https://scipost.org/10.21468/SciPostPhys.14.5.117}
}

@INPROCEEDINGS{Torta_Leone,
  author = {Torta, Pietro and Leone, Luca and Casati, Rebecca and Prati, Enrico},
  booktitle = {2024 IEEE International Conference on Quantum Computing and Engineering (QCE)},
  title = {Neural Quantum Annealing for Real-World Quadratic Unconstrained Binary Optimization},
  year = {2024},
  volume = {02},
  pages = {514-515},
  doi = {10.1109/QCE60285.2024.10382}
}

@article{carlier1990practical,
  title={A practical use of {J}ackson's preemptive schedule for solving the job shop problem},
  author={Carlier, Jacques and Pinson, Eric},
  journal={Annals of Operations Research},
  volume={26},
  pages={269--287},
  year={1990},
  publisher={Springer}
}

@article{carlier1994adjustment,
  title={Adjustment of heads and tails for the job-shop problem},
  author={Carlier, Jacques and Pinson, Eric},
  journal={European Journal of Operational Research},
  volume={78},
  number={2},
  pages={146--161},
  year={1994},
  publisher={Elsevier}
}

@article{Bojan_PRL,
  title = {Variational Ground-State Quantum Adiabatic Theorem},
  author = {\v{Z}unkovi\v{c}, Bojan and Torta, Pietro and Pecci, Giovanni and Lami, Guglielmo and Collura, Mario},
  journal = {Phys. Rev. Lett.},
  volume = {134},
  issue = {13},
  pages = {130601},
  numpages = {9},
  year = {2025},
  month = {Mar},
  publisher = {American Physical Society},
  doi = {10.1103/PhysRevLett.134.130601},
  url = {https://link.aps.org/doi/10.1103/PhysRevLett.134.130601}
}

@article{q_speedup,
  author = {R{\o}nnow, Troels F. and Wang, Zhihui and Job, Joshua and Boixo, Sergio and Isakov, Sergei V. and Wecker, David and Martinis, John M. and Lidar, Daniel A. and Troyer, Matthias},
  title = {Defining and detecting quantum speedup},
  volume = {345},
  number = {6195},
  pages = {420-424},
  year = {2014},
  doi = {10.1126/science.1252319},
  publisher = {American Association for the Advancement of Science},
  journal = {Science}
}

@article{BaldassiPNAS2018,
  author = {Baldassi, Carlo and Zecchina, Riccardo},
  title = {Efficiency of quantum vs. classical annealing in nonconvex learning problems},
  volume = {115},
  number = {7},
  pages = {1457-1462},
  year = {2018},
  doi = {10.1073/pnas.1711456115},
  publisher = {National Academy of Sciences},
  journal = {Proceedings of the National Academy of Sciences}
}

@article{Heim_2015,
  title = {Quantum versus classical annealing of Ising spin glasses},
  volume = {348},
  issn = {1095-9203},
  doi = {10.1126/science.aaa4170},
  number = {6231},
  journal = {Science},
  publisher = {American Association for the Advancement of Science (AAAS)},
  author = {Heim, Bettina and R{\o}nnow, Troels F. and Isakov, Sergei V. and Troyer, Matthias},
  year = {2015},
  month = {apr},
  pages = {215-217}
}

@article{PhysRevB.93.224431,
  title = {Quantum annealing speedup over simulated annealing on random Ising chains},
  author = {Zanca, Tommaso and Santoro, Giuseppe E.},
  journal = {Phys. Rev. B},
  volume = {93},
  issue = {22},
  pages = {224431},
  numpages = {6},
  year = {2016},
  month = {Jun},
  publisher = {American Physical Society},
  doi = {10.1103/PhysRevB.93.224431},
  url = {https://link.aps.org/doi/10.1103/PhysRevB.93.224431}
}

@article{SCHOLLWOCK201196,
  title = {The density-matrix renormalization group in the age of matrix product states},
  journal = {Annals of Physics},
  volume = {326},
  number = {1},
  pages = {96-192},
  year = {2011},
  note = {January 2011 Special Issue},
  issn = {0003-4916},
  doi = {10.1016/j.aop.2010.09.012},
  url = {https://www.sciencedirect.com/science/article/pii/S0003491610001752},
  author = {Schollw\"{o}ck, Ulrich}
}

@article{POLYAK19641,
  title = {Some methods of speeding up the convergence of iteration methods},
  journal = {USSR Computational Mathematics and Mathematical Physics},
  volume = {4},
  number = {5},
  pages = {1-17},
  year = {1964},
  issn = {0041-5553},
  doi = {10.1016/0041-5553(64)90137-5},
  url = {https://www.sciencedirect.com/science/article/pii/0041555364901375},
  author = {Polyak, B. T.}
}

@article{rumelhart1986learning,
  title = {Learning representations by back-propagating errors},
  author = {Rumelhart, David E. and Hinton, Geoffrey E. and Williams, Ronald J.},
  journal = {Nature},
  volume = {323},
  number = {6088},
  pages = {533-536},
  year = {1986},
  publisher = {Nature Publishing Group UK London},
  doi = {10.1038/323533a0},
  url = {https://doi.org/10.1038/323533a0}
}

@book{ben2006generalized,
  title = {Generalized inverses: theory and applications},
  author = {Ben-Israel, Adi and Greville, Thomas N. E.},
  year = {2006},
  publisher = {Springer Science \& Business Media},
  doi = {10.1007/b97366},
  url = {https://link.springer.com/book/10.1007/b97366}
}

@article{yasuda2025dataset,
  title = {Dataset-free weight-initialization on restricted Boltzmann machine},
  author = {Yasuda, Muneki and Maeno, Ryosuke and Takahashi, Chako},
  journal = {Neural Networks},
  volume = {187},
  pages = {107297},
  year = {2025},
  publisher = {Elsevier},
  doi = {10.1016/j.neunet.2025.107297}
}

@InProceedings{pmlr-v9-glorot10a,
  title = {Understanding the difficulty of training deep feedforward neural networks},
  author = {Glorot, Xavier and Bengio, Yoshua},
  booktitle = {Proceedings of the Thirteenth International Conference on Artificial Intelligence and Statistics},
  pages = {249-256},
  year = {2010},
  editor = {Teh, Yee Whye and Titterington, Mike},
  volume = {9},
  series = {Proceedings of Machine Learning Research},
  address = {Chia Laguna Resort, Sardinia, Italy},
  month = {13-15 May},
  publisher = {PMLR},
  url = {https://proceedings.mlr.press/v9/glorot10a.html}
}

@inproceedings{optuna_2019,
  title = {Optuna: A Next-generation Hyperparameter Optimization Framework},
  author = {Akiba, Takuya and Sano, Shotaro and Yanase, Toshihiko and Ohta, Takeru and Koyama, Masanori},
  booktitle = {Proceedings of the 25th {ACM} {SIGKDD} International Conference on Knowledge Discovery and Data Mining},
  year = {2019},
  doi = {10.1145/3292500.3330701}
}

@article{hansen2016cma,
  title         = {The CMA Evolution Strategy: A Tutorial},
  author        = {Hansen, Nikolaus},
  year          = {2016},
  journal       = {arXiv preprint arXiv:1604.00772},
  eprint        = {1604.00772},
  archivePrefix = {arXiv},
  primaryClass  = {cs.LG},
  doi           = {10.48550/arXiv.1604.00772}
}

@article{bergstra2011algorithms,
  title = {Algorithms for hyper-parameter optimization},
  author = {Bergstra, James and Bardenet, R{\'e}mi and Bengio, Yoshua and K{\'e}gl, Bal{\'a}zs},
  journal = {Advances in Neural Information Processing Systems},
  volume = {24},
  year = {2011}
}

@misc{chen2025convolutionaltransformerwavefunctions,
  title = {Convolutional transformer wave functions},
  author = {Chen, Ao and Naik, Vighnesh Dattatraya and Heyl, Markus},
  year = {2025},
  eprint = {2503.10462},
  archivePrefix = {arXiv},
  primaryClass = {cond-mat.dis-nn},
  doi = {10.48550/arXiv.2503.10462},
  url = {https://arxiv.org/abs/2503.10462}
}

@article{parisi1995number,
  title = {On the number of metastable states in spin glasses},
  author = {Parisi, Giorgio and Potters, Marc},
  journal = {Europhysics Letters},
  volume = {32},
  number = {1},
  pages = {13-17},
  year = {1995},
  publisher = {IOP Publishing},
  doi = {10.1209/0295-5075/32/1/003}
}

@article{RevModPhys.58.801,
  title = {Spin glasses: Experimental facts, theoretical concepts, and open questions},
  author = {Binder, K. and Young, A. P.},
  journal = {Rev. Mod. Phys.},
  volume = {58},
  issue = {4},
  pages = {801-976},
  year = {1986},
  month = {Oct},
  publisher = {American Physical Society},
  doi = {10.1103/RevModPhys.58.801},
  url = {https://link.aps.org/doi/10.1103/RevModPhys.58.801}
}

@article{doi:10.1126/sciadv.abl6850,
  author = {Schmitt, Markus and Rams, Marek M. and Dziarmaga, Jacek and Heyl, Markus and Zurek, Wojciech H.},
  title = {Quantum phase transition dynamics in the two-dimensional transverse-field Ising model},
  journal = {Science Advances},
  volume = {8},
  number = {37},
  pages = {eabl6850},
  year = {2022},
  doi = {10.1126/sciadv.abl6850},
  URL = {https://www.science.org/doi/abs/10.1126/sciadv.abl6850}
}

@misc{schmitt2025simulatingdynamicscorrelatedmatter,
  title = {Simulating dynamics of correlated matter with neural quantum states},
  author = {Schmitt, Markus and Heyl, Markus},
  year = {2025},
  eprint = {2506.03124},
  archivePrefix = {arXiv},
  primaryClass = {quant-ph},
  doi = {10.48550/arXiv.2506.03124},
  url = {https://arxiv.org/abs/2506.03124}
}

@misc{pascanu2014revisitingnaturalgradientdeep,
  title = {Revisiting Natural Gradient for Deep Networks},
  author = {Pascanu, Razvan and Bengio, Yoshua},
  year = {2014},
  eprint = {1301.3584},
  archivePrefix = {arXiv},
  primaryClass = {cs.LG},
  doi = {10.48550/arXiv.1301.3584},
  url = {https://arxiv.org/abs/1301.3584}
}

@misc{brodoloni2025spinglassquantumphasetransition,
  title = {Spin-glass quantum phase transition in amorphous arrays of Rydberg atoms},
  author = {Brodoloni, L. and Vovrosh, J. and Juli\`{a}-Farr\'{e}, S. and Dauphin, A. and Pilati, S.},
  year = {2025},
  eprint = {2505.05117},
  archivePrefix = {arXiv},
  primaryClass = {cond-mat.dis-nn},
  doi = {10.48550/arXiv.2505.05117},
  url = {https://arxiv.org/abs/2505.05117}
}

@misc{li2025acceleratednaturalgradientmethod,
  title = {Accelerated Natural Gradient Method for Parametric Manifold Optimization},
  author = {Li, Chenyi and Zhu, Shuchen and Xie, Zhonglin and Wen, Zaiwen},
  year = {2025},
  eprint = {2504.05753},
  archivePrefix = {arXiv},
  primaryClass = {math.OC},
  doi = {10.48550/arXiv.2504.05753},
  url = {https://arxiv.org/abs/2504.05753}
}

@misc{khan2018fastscalablebayesiandeep,
  title = {Fast and Scalable Bayesian Deep Learning by Weight-Perturbation in Adam},
  author = {Khan, Mohammad Emtiyaz and Nielsen, Didrik and Tangkaratt, Voot and Lin, Wu and Gal, Yarin and Srivastava, Akash},
  year = {2018},
  eprint = {1806.04854},
  archivePrefix = {arXiv},
  primaryClass = {stat.ML},
  doi = {10.48550/arXiv.1806.04854},
  url = {https://arxiv.org/abs/1806.04854}
}

@misc{martens2020optimizingneuralnetworkskroneckerfactored,
  title = {Optimizing Neural Networks with Kronecker-factored Approximate Curvature},
  author = {Martens, James and Grosse, Roger},
  year = {2020},
  eprint = {1503.05671},
  archivePrefix = {arXiv},
  primaryClass = {cs.LG},
  doi = {10.48550/arXiv.1503.05671},
  url = {https://arxiv.org/abs/1503.05671}
}

@inproceedings{10.5555/3042817.3043064,
  author = {Sutskever, Ilya and Martens, James and Dahl, George and Hinton, Geoffrey},
  title = {On the importance of initialization and momentum in deep learning},
  year = {2013},
  publisher = {JMLR.org},
  booktitle = {Proceedings of the 30th International Conference on Machine Learning (ICML)},
  pages = {1139-1147},
  location = {Atlanta, GA, USA},
  series = {ICML'13},
  doi = {10.5555/3042817.3043064},
  url = {https://dl.acm.org/doi/10.5555/3042817.3043064}
}

@article{goh2017why,
  author = {Goh, Gabriel},
  title = {Why Momentum Really Works},
  journal = {Distill},
  year = {2017},
  doi = {10.23915/distill.00006},
  url = {http://distill.pub/2017/momentum}
}

@article{PhysRevX.14.021029,
  title = {Wave-Function Network Description and Kolmogorov Complexity of Quantum Many-Body Systems},
  author = {Mendes-Santos, T. and Schmitt, M. and Angelone, A. and Rodriguez, A. and Scholl, P. and Williams, H. J. and Barredo, D. and Lahaye, T. and Browaeys, A. and Heyl, M. and Dalmonte, M.},
  journal = {Phys. Rev. X},
  volume = {14},
  issue = {2},
  pages = {021029},
  numpages = {21},
  year = {2024},
  month = {May},
  publisher = {American Physical Society},
  doi = {10.1103/PhysRevX.14.021029},
  url = {https://link.aps.org/doi/10.1103/PhysRevX.14.021029}
}

@article{torta2025quantum,
  title = {Quantum computing for space applications: a selective review and perspectives},
  author = {Torta, Pietro and Casati, Rebecca and Bruni, Stefano and Mandarino, Antonio and Prati, Enrico},
  journal = {EPJ Quantum Technology},
  volume = {12},
  number = {1},
  pages = {66},
  year = {2025},
  publisher = {Springer},
  doi = {10.1140/epjqt/s40507-025-00369-8}
}

@article{gao2017efficient,
  title = {Efficient representation of quantum many-body states with deep neural networks},
  author = {Gao, Xun and Duan, Lu-Ming},
  journal = {Nature Communications},
  volume = {8},
  number = {1},
  pages = {662},
  year = {2017},
  publisher = {Nature Publishing Group UK London},
  doi = {10.1038/s41467-017-00705-2}
}

@misc{ft_nqs,
year = {n.a.},
note = {We denote the Pauli-Z eigenvalues with $x_i=\pm 1$ for notational consistency with the following sections.}}

@misc{ft_rbm_1,
year = {n.a.},
note = {To be consistent with the standard notation in quantum mechanics, we use spin variables with values $\pm1$ rather than conventional binary variables. The two conventions are equivalent up to an affine transformation.}}

@misc{ft_rbm_2,
year = {n.a.},
note = {To simplify the notation, we omit from now on the explicit dependence on $\boldsymbol{\theta}$ of the partition function $Z$.}}

@misc{ft_zero_energy,
year = {n.a.},
note = {The zero-energy value is actually $-\gamma NKT$, due to the constant offset introduced by the third term in the cost function.}
}

@software{Nishimura_OpenJij,
  author = {Nishimura, Kohji and Sakamoto, Yoshiki and Shimizu, Taro and Suzuki, Kohei and Yamashiro, Yu},
  title = {{OpenJij}}
}

@misc{sg_server,
  howpublished = {\url{http://spinglass.uni-bonn.de/}}
}

@inproceedings{lorch2016visualizing,
  title = {Visualizing deep network training trajectories with PCA},
  author = {Lorch, Eliana},
  booktitle = {ICML Workshop on Visualization for Deep Learning},
  year = {2016},
  url={https://api.semanticscholar.org/CorpusID:211560470}
}

@incollection{jolliffe2011principal,
  title = {Principal component analysis},
  author = {Jolliffe, Ian},
  booktitle = {International Encyclopedia of Statistical Science},
  pages = {1094-1096},
  year = {2011},
  publisher = {Springer},
  doi = {10.1007/978-3-642-04898-2_455}
}

@article{hopfield1982neural,
  author = {Hopfield, J. J.},
  title = {Neural networks and physical systems with emergent collective computational abilities},
  journal = {Proceedings of the National Academy of Sciences},
  volume = {79},
  number = {8},
  pages = {2554-2558},
  year = {1982},
  doi = {10.1073/pnas.79.8.2554},
  url = {https://www.pnas.org/doi/abs/10.1073/pnas.79.8.2554}
}

@inproceedings{gabor2020insights,
  author = {Gabor, Thomas and Feld, Sebastian and Safi, Hila and Phan, Thomy and Linnhoff-Popien, Claudia},
  title = {Insights on Training Neural Networks for QUBO Tasks},
  year = {2020},
  isbn = {9781450379632},
  publisher = {Association for Computing Machinery},
  address = {New York, NY, USA},
  url = {https://doi.org/10.1145/3387940.3391470},
  doi = {10.1145/3387940.3391470},
  booktitle = {Proceedings of the IEEE/ACM 42nd International Conference on Software Engineering Workshops},
  pages = {436-441},
  numpages = {6},
  location = {Seoul, Republic of Korea}
}

@misc{bello2016neural,
  title = {Neural Combinatorial Optimization with Reinforcement Learning},
  author = {Bello, Irwan and Pham, Hieu and Le, Quoc V. and Norouzi, Mohammad and Bengio, Samy},
  year = {2017},
  eprint = {1611.09940},
  archivePrefix = {arXiv},
  primaryClass = {cs.AI},
  doi = {10.48550/arXiv.1611.09940},
  url = {https://arxiv.org/abs/1611.09940}
}

@inproceedings{khalil2017learning,
  author = {Khalil, Elias and Dai, Hanjun and Zhang, Yuyu and Dilkina, Bistra and Song, Le},
  booktitle = {Advances in Neural Information Processing Systems},
  editor = {Guyon, I. and Von Luxburg, U. and Bengio, S. and Wallach, H. and Fergus, R. and Vishwanathan, S. and Garnett, R.},
  title = {Learning Combinatorial Optimization Algorithms over Graphs},
  volume = {30},
  year = {2017},
  url = {https://proceedings.neurips.cc/paper/2017/file/d9896106ca98d3d05b8cbdf4fd8b13a1-Paper.pdf}
}

@inproceedings{he2024quantum,
  author = {He, Haoqi},
  editor = {Ghosh, Smita and Zhang, Zhao},
  title = {Quantum Annealing and GNN for Solving TSP with QUBO},
  booktitle = {Algorithmic Aspects in Information and Management},
  year = {2024},
  publisher = {Springer Nature Singapore},
  address = {Singapore},
  pages = {134-145},
  doi = {10.1007/978-981-97-7801-0_12},
  url = {https://link.springer.com/chapter/10.1007/978-981-97-7801-0_12}
}

@article{PhysRev.98.1479,
  title = {Many-Body Problem with Strong Forces},
  author = {Jastrow, Robert},
  journal = {Phys. Rev.},
  volume = {98},
  issue = {5},
  pages = {1479-1484},
  year = {1955},
  month = {Jun},
  publisher = {American Physical Society},
  doi = {10.1103/PhysRev.98.1479},
  url = {https://link.aps.org/doi/10.1103/PhysRev.98.1479}
}

@book{becca2017quantum,
  title = {Quantum Monte Carlo Approaches for Correlated Systems},
  author = {Becca, F. and Sorella, S.},
  isbn = {9781108547314},
  year = {2017},
  publisher = {Cambridge University Press},
  doi = {10.1017/9781316417041},
  url = {https://books.google.de/books?id=4phADwAAQBAJ}
}

@software{JAX,
  author = {Bradbury, James and Frostig, Roy and Hawkins, Peter and Johnson, Matthew James and Leary, Chris and Maclaurin, Dougal and Necula, George and Paszke, Adam and Vander{P}las, Jake and Wanderman-{M}ilne, Skye and Zhang, Qiao},
  title = {{JAX}: composable transformations of {P}ython+{N}um{P}y programs},
  url = {http://github.com/jax-ml/jax},
  version = {0.3.13},
  year = {2018}
}

@misc{nqa_server_github,
  howpublished = {\url{https://github.com/lucaleonect/nqa_server}}
}

@misc{Venturelli2015JSSP,
  title = {Quantum Annealing Implementation of Job-Shop Scheduling},
  author = {Venturelli, Davide and Marchand, Dominic J. J. and Rojo, Galo},
  year = {2016},
  eprint = {1506.08479},
  archivePrefix = {arXiv},
  primaryClass = {quant-ph},
  doi = {10.48550/arXiv.1506.08479},
  url = {https://arxiv.org/abs/1506.08479}
}

@INPROCEEDINGS{casati2025iterative,
  author={Casati, Rebecca and Giudici, Thomas and Torta, Pietro and Prati, Enrico},
  booktitle={2025 IEEE International Conference on Quantum Computing and Engineering (QCE)}, 
  title={Iterative Quantum Annealing for Job Shop Scheduling with a Directed Graph Representation}, 
  year={2025},
  volume={01},
  number={},
  pages={2161-2169},
  doi={10.1109/QCE65121.2025.00236}}

@article{PhysRevB.54.3328,
  title = {Griffiths singularities in the disordered phase of a quantum Ising spin glass},
  author = {Rieger, H. and Young, A. P.},
  journal = {Phys. Rev. B},
  volume = {54},
  issue = {5},
  pages = {3328--3335},
  numpages = {0},
  year = {1996},
  month = {Aug},
  publisher = {American Physical Society},
  doi = {10.1103/PhysRevB.54.3328},
  url = {https://link.aps.org/doi/10.1103/PhysRevB.54.3328}
}

@inbook{Sachdev_2011, place={Cambridge}, title={Quantum spin glasses}, booktitle={Quantum Phase Transitions}, publisher={Cambridge University Press}, author={Sachdev, Subir}, year={2011}, pages={463–478},doi={10.1017/CBO9780511973765}}

@article{PhysRevB.105.L020201,
  title = {Glassy quantum dynamics of disordered Ising spins},
  author = {Schultzen, P. and Franz, T. and Geier, S. and Salzinger, A. and Tebben, A. and Hainaut, C. and Z\"urn, G. and Weidem\"uller, M. and G\"arttner, M.},
  journal = {Phys. Rev. B},
  volume = {105},
  issue = {2},
  pages = {L020201},
  numpages = {6},
  year = {2022},
  month = {Jan},
  publisher = {American Physical Society},
  doi = {10.1103/PhysRevB.105.L020201},
  url = {https://link.aps.org/doi/10.1103/PhysRevB.105.L020201}
}

@misc{cugliandolo2022quantumglassesreview,
      title={Quantum Glasses -- a review}, 
      author={L. F. Cugliandolo and M. Mueller},
      year={2022},
      eprint={2208.05417},
      archivePrefix={arXiv},
      primaryClass={cond-mat.dis-nn},
      doi={10.48550/arXiv.2208.05417},
      url={https://arxiv.org/abs/2208.05417}, 
}

@book{MPV1987,
  title        = {Spin Glass Theory and Beyond: An Introduction to the Replica Method and Its Applications},
  author       = {M{\'e}zard, Marc and Parisi, Giorgio and Virasoro, M. A.},
  year         = {1987},
  publisher    = {World Scientific},
  address      = {Singapore},
  doi          = {10.1142/0271}
}

@book{Young1998,
  title        = {Spin Glasses and Random Fields},
  editor       = {Young, A. P.},
  year         = {1998},
  publisher    = {World Scientific},
  address      = {Singapore},
  doi          = {10.1142/3517}
}

@article{doi:10.1137/1033004,
author = {Padberg, Manfred and Rinaldi, Giovanni},
title = {A Branch-and-Cut Algorithm for the Resolution of Large-Scale Symmetric Traveling Salesman Problems},
journal = {SIAM Review},
volume = {33},
number = {1},
pages = {60-100},
year = {1991},
doi = {10.1137/1033004},

URL = { 
    
        https://doi.org/10.1137/1033004
    
    

},
eprint = { 
    
        https://doi.org/10.1137/1033004
    
    

}
}

@article{PhysRevLett.58.86,
  title = {Nonuniversal critical dynamics in Monte Carlo simulations},
  author = {Swendsen, Robert H. and Wang, Jian-Sheng},
  journal = {Phys. Rev. Lett.},
  volume = {58},
  issue = {2},
  pages = {86--88},
  numpages = {0},
  year = {1987},
  month = {Jan},
  publisher = {American Physical Society},
  doi = {10.1103/PhysRevLett.58.86},
  url = {https://link.aps.org/doi/10.1103/PhysRevLett.58.86}
}

@article{PhysRevLett.62.361,
  title = {Collective Monte Carlo Updating for Spin Systems},
  author = {Wolff, Ulli},
  journal = {Phys. Rev. Lett.},
  volume = {62},
  issue = {4},
  pages = {361--364},
  numpages = {0},
  year = {1989},
  month = {Jan},
  publisher = {American Physical Society},
  doi = {10.1103/PhysRevLett.62.361},
  url = {https://link.aps.org/doi/10.1103/PhysRevLett.62.361}
}

@article{Marinari_1992,
doi = {10.1209/0295-5075/19/6/002},
url = {https://doi.org/10.1209/0295-5075/19/6/002},
year = {1992},
month = {jul},
publisher = {},
volume = {19},
number = {6},
pages = {451},
author = {E. Marinari and G. Parisi},
title = {Simulated Tempering: A New Monte Carlo Scheme},
journal = {Europhysics Letters}
}

@article{doi:10.1143/JPSJ.65.1604,
author = {Hukushima ,Koji and Nemoto ,Koji},
title = {Exchange Monte Carlo Method and  Applications to Spin Glass Simulations},
journal = {Journal of the Physical Society of Japan},
volume = {65},
number = {6},
pages = {1604-1608},
year = {1996},
doi = {10.1143/JPSJ.65.1604},

URL = { 
    
        https://doi.org/10.1143/JPSJ.65.1604
    
    

},
eprint = { 
    
        https://doi.org/10.1143/JPSJ.65.1604
    
    

}
}

@article{ben2018spectral,
   title={{Spectral Gap Estimates in Mean Field Spin Glasses}},
   volume={361},
   ISSN={1432-0916},
   url={http://dx.doi.org/10.1007/s00220-018-3152-6},
   DOI={10.1007/s00220-018-3152-6},
   number={1},
   journal={Communications in Mathematical Physics},
   publisher={Springer Science and Business Media LLC},
   author={Ben Arous, Gérard and Jagannath, Aukosh},
   year={2018},
   month=may, pages={1–52} }

@article{Bernaschi_2024,
   title={The quantum transition of the two-dimensional Ising spin glass},
   volume={631},
   ISSN={1476-4687},
   url={http://dx.doi.org/10.1038/s41586-024-07647-y},
   DOI={10.1038/s41586-024-07647-y},
   number={8022},
   journal={Nature},
   publisher={Springer Science and Business Media LLC},
   author={Bernaschi, Massimo and González-Adalid Pemartín, Isidoro and Martín-Mayor, Víctor and Parisi, Giorgio},
   year={2024},
   month=jul, pages={749–754}, }

@article{PhysRevLett.126.090506,
  title = {Tropical Tensor Network for Ground States of Spin Glasses},
  author = {Liu, Jin-Guo and Wang, Lei and Zhang, Pan},
  journal = {Phys. Rev. Lett.},
  volume = {126},
  issue = {9},
  pages = {090506},
  numpages = {7},
  year = {2021},
  month = {Mar},
  publisher = {American Physical Society},
  doi = {10.1103/PhysRevLett.126.090506},
  url = {https://link.aps.org/doi/10.1103/PhysRevLett.126.090506},
}

@InProceedings{10.1007/978-3-642-83154-6_17,
author="Ishii, H.
and Yamamoto, T.",
editor="Suzuki, Masuo",
title="Monte Carlo Study of the Sherrington-Kirkpatrick Spin Glass Model in a Transverse Field",
booktitle="Quantum Monte Carlo Methods in Equilibrium and Nonequilibrium Systems",
year="1987",
publisher="Springer Berlin Heidelberg",
address="Berlin, Heidelberg",
pages="176--185",
isbn="978-3-642-83154-6"
}

@misc{oliveira2008complexityquantumspinsystems,
      title={The complexity of quantum spin systems on a two-dimensional square lattice}, 
      author={Roberto Oliveira and Barbara M. Terhal},
      year={2008},
      eprint={quant-ph/0504050},
      archivePrefix={arXiv},
      primaryClass={quant-ph},
      url={https://arxiv.org/abs/quant-ph/0504050}, 
}

@misc{viteritti2025quantumspinglasstwodimensional,
      title={Quantum Spin Glass in the Two-Dimensional Disordered Heisenberg Model via Foundation Neural-Network Quantum States}, 
      author={Luciano Loris Viteritti and Riccardo Rende and Giacomo Bracci Testasecca and Jacopo Niedda and Roderich Moessner and Giuseppe Carleo and Antonello Scardicchio},
      year={2025},
      eprint={2507.05073},
      archivePrefix={arXiv},
      primaryClass={cond-mat.dis-nn},
      url={https://arxiv.org/abs/2507.05073}, 
}

@article{PhysRevB.41.4858,
  title = {Solvable model of the quantum spin glass in a transverse field},
  author = {Goldschmidt, Yadin Y.},
  journal = {Phys. Rev. B},
  volume = {41},
  issue = {7},
  pages = {4858--4861},
  numpages = {0},
  year = {1990},
  month = {Mar},
  publisher = {American Physical Society},
  doi = {10.1103/PhysRevB.41.4858},
  url = {https://link.aps.org/doi/10.1103/PhysRevB.41.4858}
}

@article{PhysRevB.52.384,
  title = {Landau theory of quantum spin glasses of rotors and Ising spins},
  author = {Read, N. and Sachdev, Subir and Ye, J.},
  journal = {Phys. Rev. B},
  volume = {52},
  issue = {1},
  pages = {384--410},
  numpages = {0},
  year = {1995},
  month = {Jul},
  publisher = {American Physical Society},
  doi = {10.1103/PhysRevB.52.384},
  url = {https://link.aps.org/doi/10.1103/PhysRevB.52.384}
}

@inbook{Rieger,
   title={Quantum spin glasses},
   ISBN={9783540691235},
   url={http://dx.doi.org/10.1007/BFb0104832},
   DOI={10.1007/bfb0104832},
   booktitle={Complex Behaviour of Glassy Systems},
   publisher={Springer Berlin Heidelberg},
   author={Rieger, Heiko and Young, A. Peter},
   pages={256–265} }

@book{10.5555/3159044, author = {Tanaka, Shu and Tamura, Ryo and Chakrabarti, Bikas K.}, url={https://dl.acm.org/doi/10.5555/3159044}, title = {Quantum Spin Glasses, Annealing and Computation}, year = {2017}, isbn = {1107113199}, publisher = {Cambridge University Press}, address = {USA}, edition = {1st} }

@software{zenodo_repo,
  author    = {Leone, L. and Dutta, A. and Heyl, M. and Prati, E. and Torta, P.},
  title     = {Data relative to the paper "Solving Classical and Quantum Spin Glasses with Deep Boltzmann Quantum States"},
  year      = 2026,
  publisher = {Zenodo},
  doi       = {https://doi.org/10.5281/zenodo.18351403},
  url       = {https://zenodo.org/records/18351403}
}

@article{Nomura_2021,
doi = {10.1088/1361-648X/abe268},
url = {https://doi.org/10.1088/1361-648X/abe268},
year = {2021},
month = {apr},
publisher = {IOP Publishing},
volume = {33},
number = {17},
pages = {174003},
author = {Nomura, Yusuke},
title = {Helping restricted Boltzmann machines with quantum-state representation by restoring symmetry},
journal = {Journal of Physics: Condensed Matter},
}

\end{document}